\documentclass[a4paper,fleqn,usenatbib]{mnras}

\usepackage{newtxtext,newtxmath}

\usepackage[T1]{fontenc}
\usepackage{ae,aecompl}

\usepackage{graphicx}	

\newcommand{\gapprox}{\,\rlap{\lower 2.5pt 
\hbox{$\sim$}}\raise 1.5pt\hbox{$>$}\,}
\newcommand{\gsim}{\,\rlap{\lower 2.5pt 
\hbox{$\sim$}}\raise 1.5pt\hbox{$>$}\,}
\newcommand{\lapprox}{\,\rlap{\lower 2.5pt 
\hbox{$\sim$}}\raise 1.5pt\hbox{$<$}\,}
\newcommand{\lsim}{\,\rlap{\lower 2.5pt 
\hbox{$\sim$}}\raise 1.5pt\hbox{$<$}\,}

\def\eeq{\end{equation}}
\def\beq{\begin{equation}}

\title[SDSS-IV/MaStar stellar population models]{Stellar population models based on the SDSS-IV MaStar library of stellar spectra. I. Intermediate-age/old models.}
\author[C. Maraston et al.]{
C. Maraston$^{1}$\thanks{E-mail:claudia.maraston@port.ac.uk}, 
L. Hill$^{1}$,
D. Thomas$^{1}$, 
R. Yan$^{2}$, 
Y. Chen$^{3}$,
J. Lian$^{1,13}$,
T. Parikh$^{1,14}$,
\newauthor
J. Neumann$^{1}$,
S. Meneses-Goytia$^{1}$, 
M. Bershady$^{4,15,16}$,
N. Drory$^{5}$,
D. Bizyaev$^{6,7}$,
\newauthor
A. Concas$^{8,9}$,
J. Brownstein$^{10}$,
D. Lazarz$^{2}$, 
G. Stringfellow$^{11}$,
K. Stassun$^{12}$
\\
$^{1}$Institute of Cosmology, University of Portsmouth, Burnaby Road, Portsmouth PO1 3FX, UK \\
$^{2}$Department of Physics and Astronomy, University of Kentucky, 505 Rose St., Lexington, KY 40506-0057, USA
Department, Institution, Street Address\\
$^{3}$ New York University Abu Dhabi, Abu Dhabi, P.O. Box 129188, United Arab Emirates\\
$^{4}$ University of Winsconsin-Madison, Department of Astronomy, 475 N.Charter Street, Madison, WI 53706-1582, USA\\
$^{5}$ McDonald Observatory, The University of Texas at Austin, 1 University Station, Austin, TX 78712, USA\\
$^{6}$ Apache Point Observatory and New Mexico State University, P.O. Box 59, Sunspot, NM 88349, USA\\
$^{7}$ Sternberg Astronomical Institute, Moscow State University, Universitetskij pr. 13, Moscow, Russia\\
$^{8}$ Cavendish Laboratory, University of Cambridge, 19 J. J. Thomson Ave., Cambridge CB3 0HE, UK\\
$^{9}$ Kavli Institute for Cosmology, University of Cambridge, Madingley Road, Cambridge CB3 0HA, UK\\
$^{10}$ The University of Utah
115 S. 1400 E.
James Fletcher Bldg
Salt Lake City, UT 84112-0830\\
$^{11}$ Center for Astrophysics and Space Astronomy, Department of Astrophysical and Planetary Sciences, University of Colorado, 389 UCB,Boulder, CO 80309-0389, USA\\
$^{12}$ Vanderbilt University,
Dept. of Physics and Astronomy,
VU Station B 1807,
Nashville, TN 37235\\
$^{13}$ Department of Physics \& Astronomy, University of Utah, Salt Lake City, UT, 84112, USA\\
$^{14}$ Max-Planck-Institut f\"ur extraterrestrische Physik, Giessenbachstrasse 1, 85748 Garching bei M\"unchen, Germany\\
$^{15}$ South African Astronomical Observatory, PO Box 9, Observatory 7935, Cape Town, South Africa\\
$^{16}$ Department of Astronomy, University of Cape Town, Private Bag X3, Rondebosch 7701, South Africa}

\date{Accepted XXX. Received YYY; in original form ZZZ}

\pubyear{2020}

\begin{document}
\label{firstpage}
\pagerange{\pageref{firstpage}--\pageref{lastpage}}
\maketitle

\begin{abstract}
We use the first release of the SDSS/MaStar stellar library comprising $\sim 9000$, high $S/N$~spectra, to calculate integrated spectra of stellar population models. The models extend over the wavelength range 0.36-1.03~$\mu$ and share the same spectral resolution ($\rm R\sim1800$) and flux calibration as the SDSS-IV/MaNGA galaxy data. The parameter space covered by the stellar spectra collected thus far allows the calculation of models with ages and chemical composition in the range $\rm {t>200 Myr, -2 <= [Z/H] <= + 0.35}$, which will be extended as MaStar proceeds. Notably, the models include spectra for dwarf Main Sequence stars close to the core H-burning limit, as well spectra for cold, metal-rich giants. Both stellar types are crucial for modelling $\lambda>0.7 \mu$ absorption spectra. Moreover, a better parameter coverage at low metallicity allows the calculation of models as young as 500 Myr and the full account of the Blue Horizontal Branch phase of old populations. We present models adopting two independent sets of stellar parameters ($T_{eff}, logg, [Z/H]$). In a novel approach, their reliability is tested 'on the fly' using the stellar population models themselves. We perform tests with Milky Way and Magellanic Clouds globular clusters, finding that the new models recover their ages and metallicities remarkably well, with systematics as low as a few per cent for homogeneous calibration sets. We also fit a MaNGA galaxy spectrum, finding residuals of the order of a few per cent comparable to the state-of-art models, but now over a wider wavelength range. 
\end{abstract}



\section{Introduction}
Stellar population models \citep{tinsley_1972, bruzual_1983, bruzual_and_charlot_2003, leitherer_et_al_1999, maraston_1998,maraston_2005,maraston_and_stromback_2011, vazdekis_etal_1996,vazdekis_etal_2010, vazdekis_etal_2012,fioc_and_rocca_1997,conroy_etal_2009, thomas_etal_2003, thomas_etal_2011} are instrumental to the investigation of the fundamental properties of stellar systems such as stellar ages, chemical composition, initial mass function (IMF), mass in stars, star formation rate and history (i.e.\ exponentially-declining, constant, etc.), and redshift, be they encoded in observational data or predicted by galaxy formation and evolution models \citep[e.g.][]{kauffmann_etal_1993,baugh_2006}. Due to their widespread use in astrophysics and cosmology, there is always scope to improve and extend these models beyond the already high standard which has been reached \citep[e.g.][for a review]{conroy_2013}. 

Stellar population models combine energy and timescales of evolutionary phases from stellar evolution calculations with stellar spectral energy distributions. In this work we focus on the 'stellar spectrum' component.  Stellar spectra can either be theoretical \citep[from model atmosphere calculations, e.g.][]{kurucz_1979} or empirical \citep[from real stars, e.g. ][see below]{miles}. Both carry advantages and disadvantages, as discussed in e.g. \citet{thomas_etal_2003},\citet{korn_etal_2005},\citet{maraston_and_stromback_2011}. In brief, theoretical spectra can be calculated for well-defined, arbitrary stellar parameters, spectral resolution and wavelength coverage, and are free from observational problems. As a downside, element line-lists are known to be incomplete. On the other hand, empirical spectra contain all lines of the elements as they occur in Nature. However, they are not exempt from problems. For example, spectral libraries for early-type stars have several observational issues, such as Balmer lines contaminated by nebular emission or strong dust reddening. Most importantly, for observed spectra stellar parameters, the accuracy of which affects population models, need to be derived. It is therefore useful to consider both theoretical and empirical spectra when attacking the problem of calculating the integrated spectra of stellar systems.
This comprehensive approach is taken in \citet[][hereafter M11]{maraston_and_stromback_2011}, who calculated stellar population models as a function of major available empirical libraries, namely the Pickles library \citep{pickles_1998}, the STELIB library \citep{stelib}, the ELODIE library \citep{elodie} and the MILES library \citep[Medium-resolution Isaac Newton Telescope library of empirical spectra][]{miles} plus the high spectral resolution fully theoretical library MARCS \citep {gustafsson_etal_2008}.  
M11 found discrepancies in stellar population models due to different empirical libraries (see their Figure~12) which affect the derived results. For this reason \citet{wilkinson_etal_2017} (hereafter W17, {\it www.port.ac.uk/firefly}) and Comparat et al.\ (2018) performed fits to all M11 model flavours for $\sim$~2M galaxy spectra from SDSS. 

There are two limitations affecting our and similar analyses, which we shall hope to alleviate with the project described here. Firstly, stellar population models spanning as wide a wavelength range as the SDSS data ($0.36$-$\sim1\mu$m) with a wide range of ages and chemical composition cannot be calculated with current empirical stellar libraries (see discussion in M11). Existing models with large wavelength coverage \citep{vazdekis_etal_2016,conroy_etal_2018} are a patchwork of different libraries. While the latter approach is hard to avoid in several cases, it would still be beneficial to extend the wavelength range of the input spectra as much as possible using the same instrument. Equally important is the fact that crucial stellar types are missing in current empirical stellar libraries, in particular: the coolest end of the Main Sequence which is needed to assess the initial mass function \citep[e.g.][]{conroy_and_van_dokkum_2012}; Carbon-type stars which are important for the spectral modelling of $\sim1$~Gyr populations, as in high-redshift galaxies or local star forming galaxies \citep{maraston_2005}; and metal-rich, $\alpha$-enhanced stars which dominate the most massive galaxies \citep{thomas_etal_2005}.

Clearly, a new empirical spectral library is needed for improving population synthesis modelling. In this spirit we initiated a new effort within the SDSS-IV project \citep{blanton_etal_2017,york_etal_2000}, called MaStar \citep[i.e. MaNGA Stellar Library,][]{yan_etal_2019}. MaStar uses the BOSS spectrographs \citep{smee_etal_2013} with the new fiber integral field units in SDSS-IV \citep{drory_etal_2015} to obtain stellar spectra for the whole duration of the mission (2014-2020), enabling the acquisition of an unprecedentedly large number of spectra (of order 30,000 at completion) that will comprise the ultimate empirical spectral stellar library for years to come. 

In this paper we present the first MaStar-based stellar population models, enabled by the first release of MaStar spectra \citep{yan_etal_2019}. This release includes 8646 spectra (corresponding to 3321 individual Milky Way stars) and -- already at this stage -- it is an order of magnitude larger than the state-of-the-art empirical stellar library for population synthesis models, MILES \citet{miles}, comprising $\sim 900$ spectra. At completion (end 2020), the observed database will include of the order 30,000 spectra and will be by far the largest spectral stellar library for years to come. The stellar population models resulting from our project will overcome several limitations of past efforts, by means of: i) a wide contiguous wavelength range (0.36-1$\mu$m), which includes unique spectral features for the understanding of galaxy physics; ii) the same spectral resolution and flux calibration of the SDSS-IV MaNGA galaxy data and similar to data from the legacy surveys\footnote{Strictly speaking the single-fiber legacy Sloan surveys have slightly different spectral resolution and coverage, but the present models will be suitable to those data as well.}; iii) a better coverage of stellar phases, including a large number of low-mass Main Sequence dwarfs for the first time \citep[cfr. Table 1 in][]{yan_etal_2019}  and small statistical errors due to the large number of spectra. These models will leverage the analysis of spectroscopic galaxy and star clusters data bases and usher in a new era of galaxy evolution studies.

The initial set of models we release in this paper covers the age range $t \gapprox 100-300~\rm Myr$\footnote{The exact range depends on the chemical composition, see Table~1.}~and a wide range in chemical composition. Note that the age limit is set by the coverage of the parameter space appropriate to stellar phases for various population ages, as we will extensively demonstrate (see Table~1 for the details of the grid). In particular, these models include dwarf MS star spectra at or close to, the core-$H$~ burning limit ($0.1-0.3~M_{\odot}$, depending on metallicity). This will allow the investigation of IMF-sensitive spectral features \citep[e.g.][and references therein]{parikh_etal_2018}. 

Besides presenting the models, in this paper we also demonstrate the effect of different assumptions regarding the adopted stellar parameters on the final model results, and provide models for two fully independent sets of stellar parameters, one based on previous empirical spectra from MILES, the other on theoretical spectra from model atmospheres. 

In both cases, we employ the population model calculation as a {\it goal} as well as a {\it mean} to check the soundness of stellar parameters 'on the fly'. Indeed, incorrect parameter assignment for energetically-relevant stellar sub-phases produces erroneous 
integrated model spectra. We perform this analysis by comparing the spectra of MaStar population models as a function of parameters for each evolutionary phase, to our previous calculations based on the same energetics. This novel approach also allows us to identify any missing parameter coverage for future target selection efforts. 
Both sets of parameters are further checked against GAIA data for breaking the insidious degeneracy between cool dwarfs and giants.

To assure ourselves of the overall correctness of model calculations, we test on globular cluster (GCs) spectra as in our standard approach and previous papers \citep[e.g.][]{maraston_and_stromback_2011,wilkinson_etal_2017}. In particular, we quantify the accuracy in age and metallicity determinations that can be achieved with the present models.

It should be noted that the current MaStar release does not include Carbon-type and Oxygen-rich-type stars in sufficient numbers/variety for properly describing the Thermally-Pulsating Asymptotic Giant Branch (TP-AGB) phase of stellar evolution, which, in the Maraston models, contributes to ages $0.2\lapprox t/Gyr\lapprox2$. Hence, the MaStar-based models in this release do not include the TP-AGB contribution. However, we provide a version of them where we use lower-resolution TP-AGB empirical spectra as in \citet{maraston_2005}, which we re-bin to the MaStar wavelength vector \citep[following the same procedure as in][]{maraston_and_stromback_2011}. Furthermore, this version of MaStar-based models do not explicitly account for the variation in  element abundance ratios, which is the matter of a future effort.

This paper is organised as follows. In Section~\ref{sec:eps} we recapitulate the main features of the evolutionary population synthesis code and in Section~\ref{sec:mastar}, those of the MaStar empirical stellar spectral library. In the same Section~\ref{sec:mastar}, we also describe the calculations of the two sets of stellar parameters we adopt and the various quality cuts we apply (e.g.\ in S/N). In Section~\ref{sec:ssp} we describe the implementation of the empirical stellar spectra in the population synthesis code and the model calculation procedure. The MaStar stellar population models are presented and extensively discussed in Section~\ref{sec:results} including a comparison with our previous models, while in Section~\ref{sec:testing} we test the models with globular cluster data. Section~\ref{sec:summary} summarises and discusses the work.
\section{Summary of the evolutionary population synthesis code}
\label{sec:eps}
The evolutionary population synthesis we adopted is comprehensively described in \citet[hereafter M98 and M05]{maraston_1998,maraston_2005} and here we will only repeat the most relevant features. 

\begin{figure*}
	\includegraphics[width=0.33\textwidth]{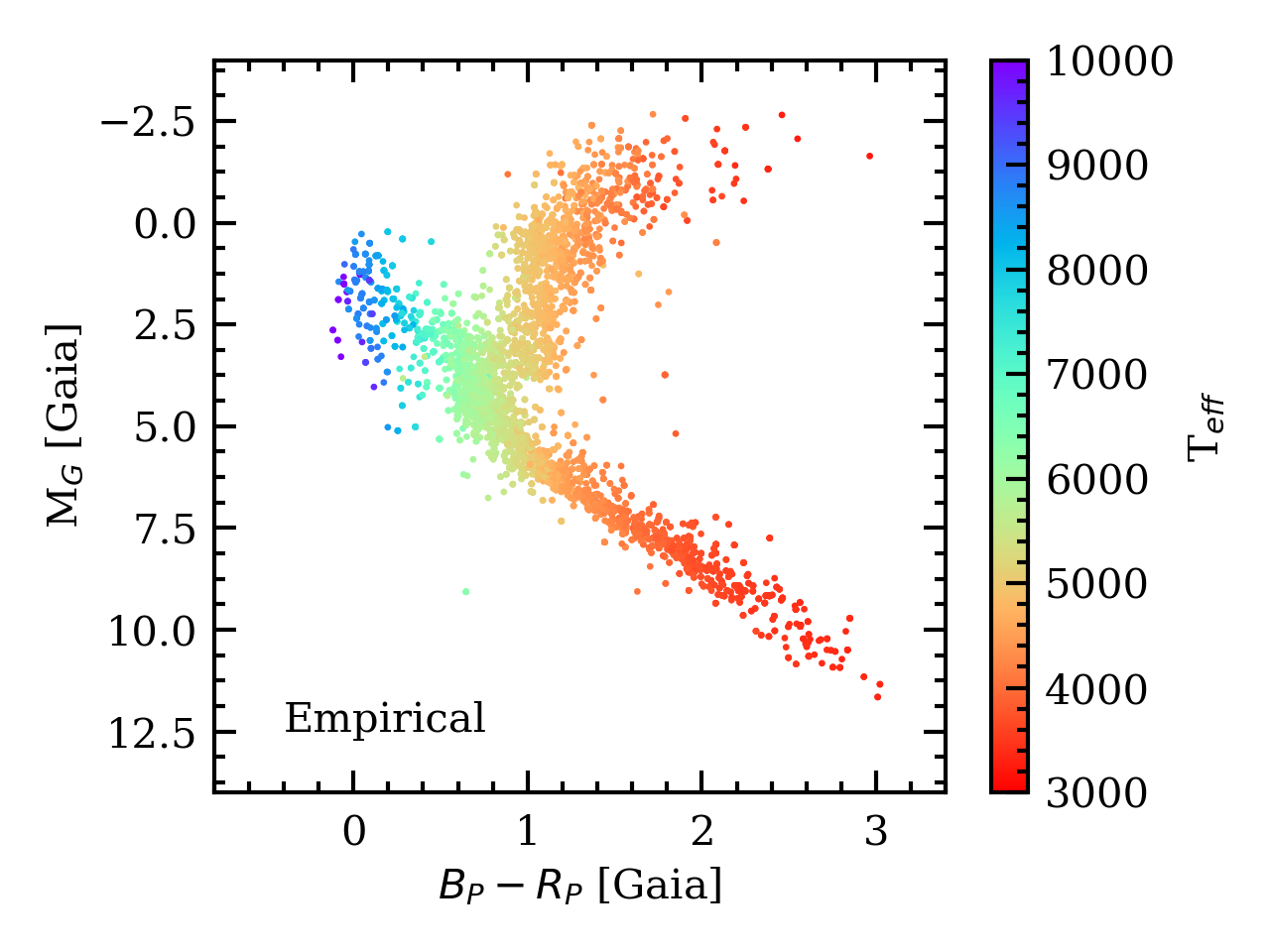}
		\includegraphics[width=0.33\textwidth]{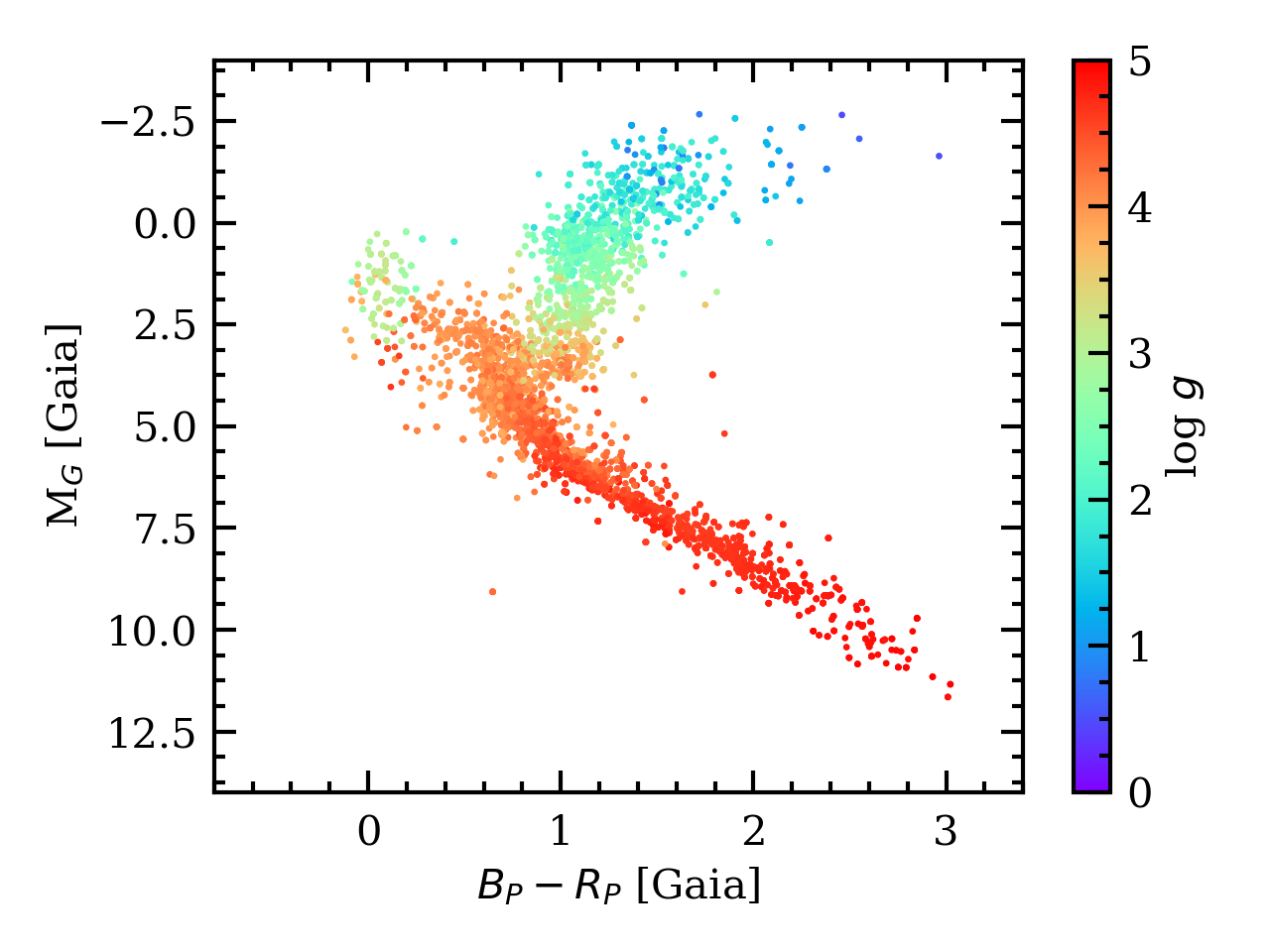}
	\includegraphics[width=0.33\textwidth]{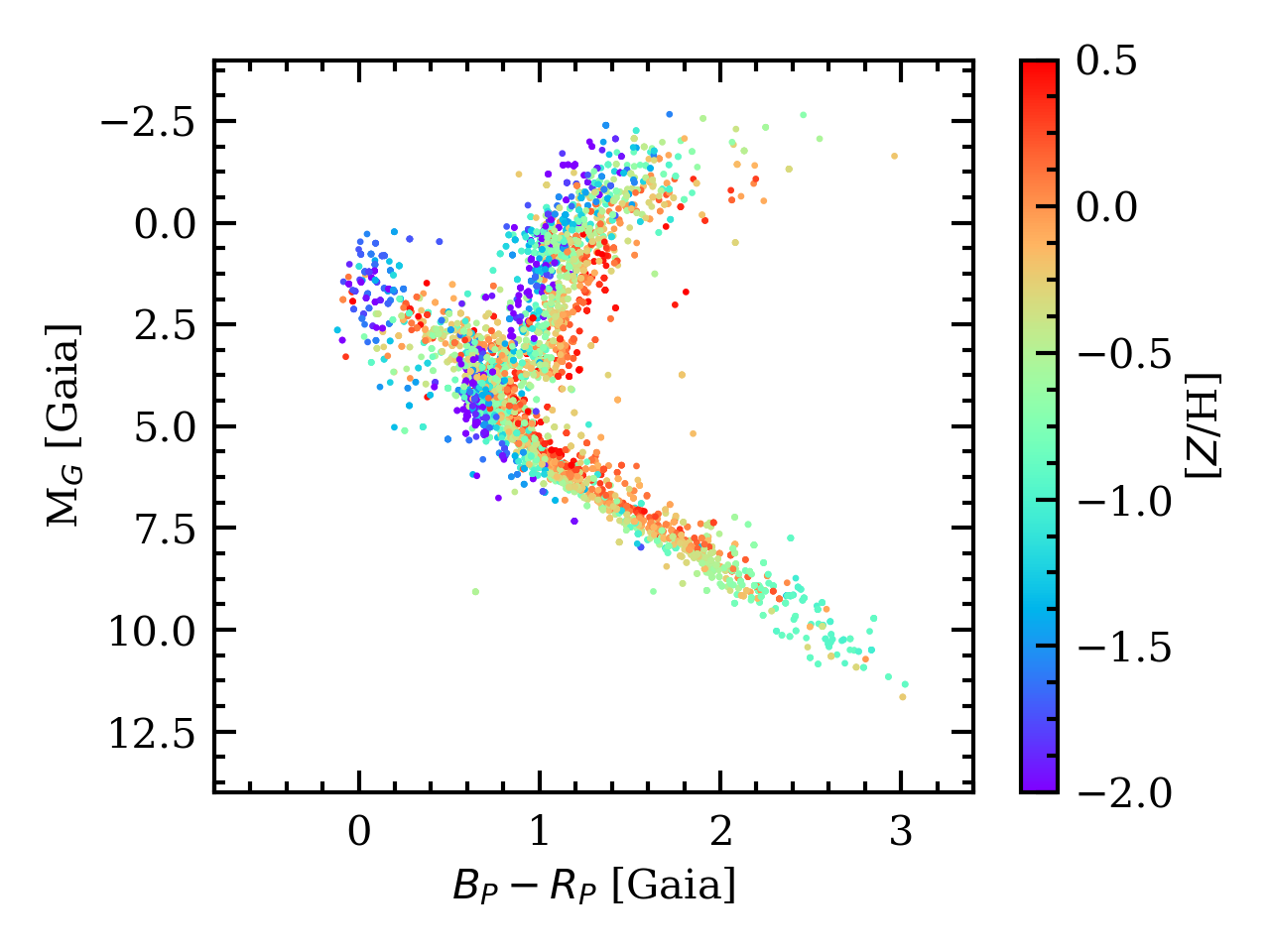}
		\includegraphics[width=0.33\textwidth]{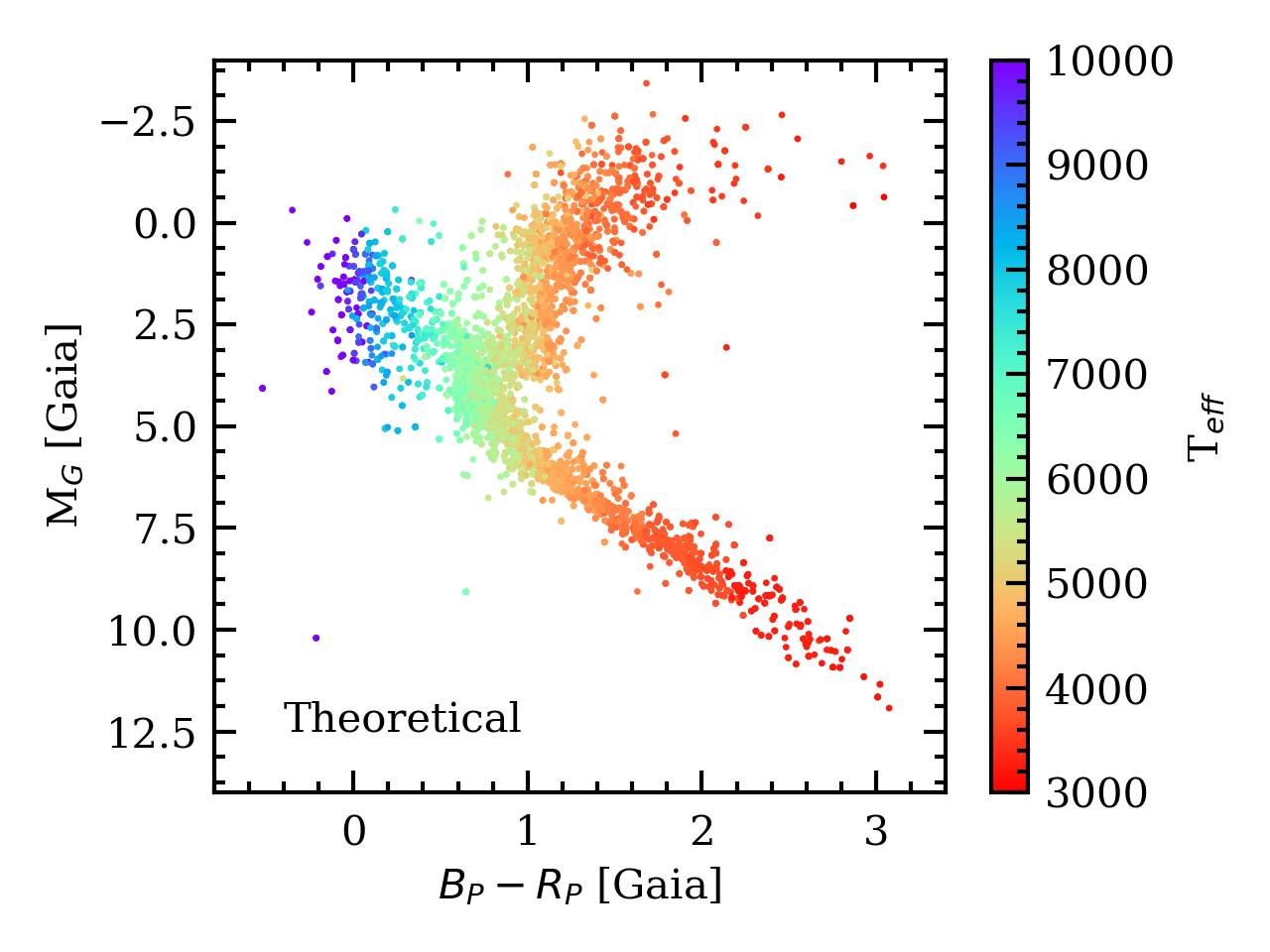}
		\includegraphics[width=0.33\textwidth]{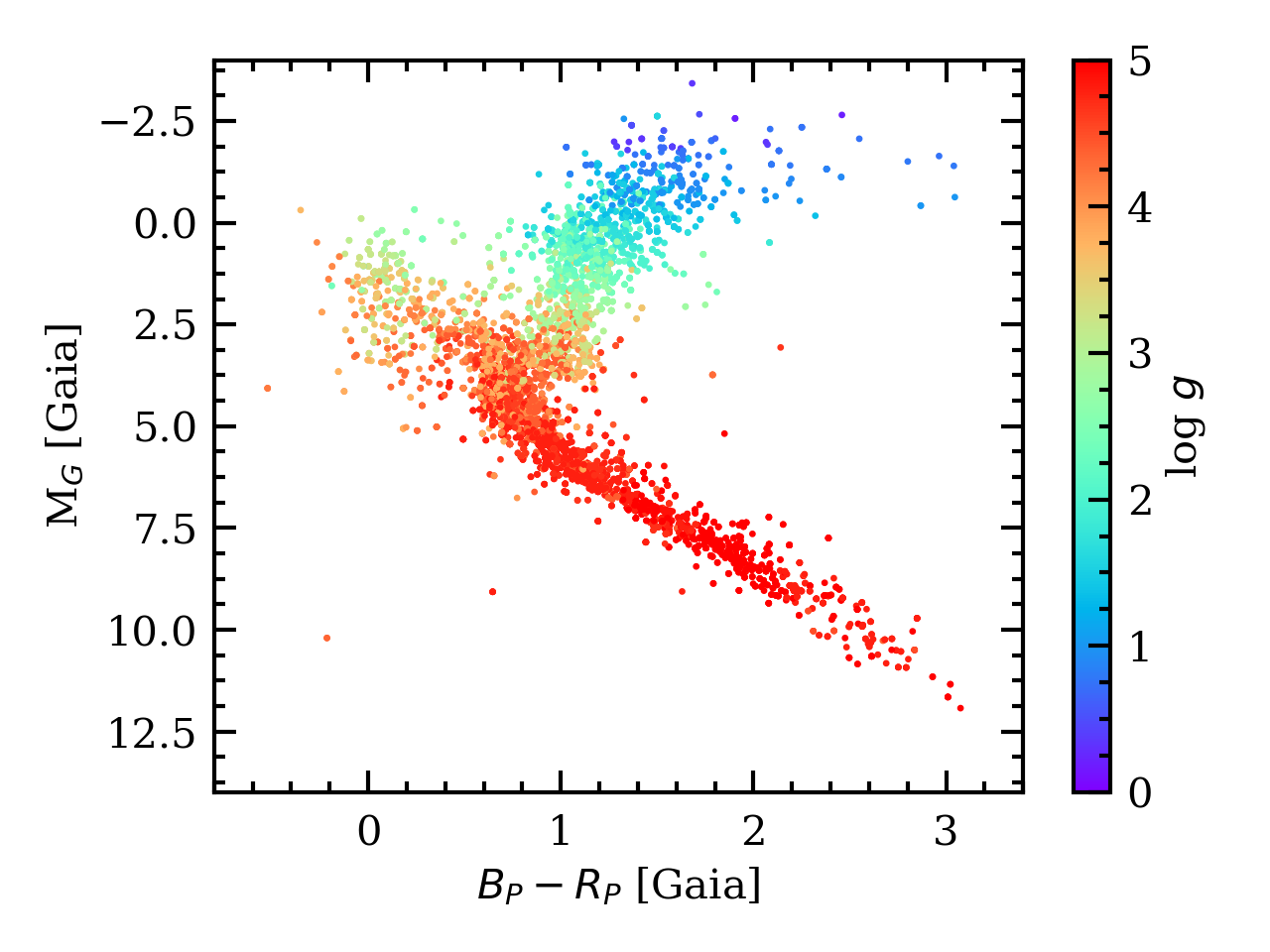}
	\includegraphics[width=0.33\textwidth]{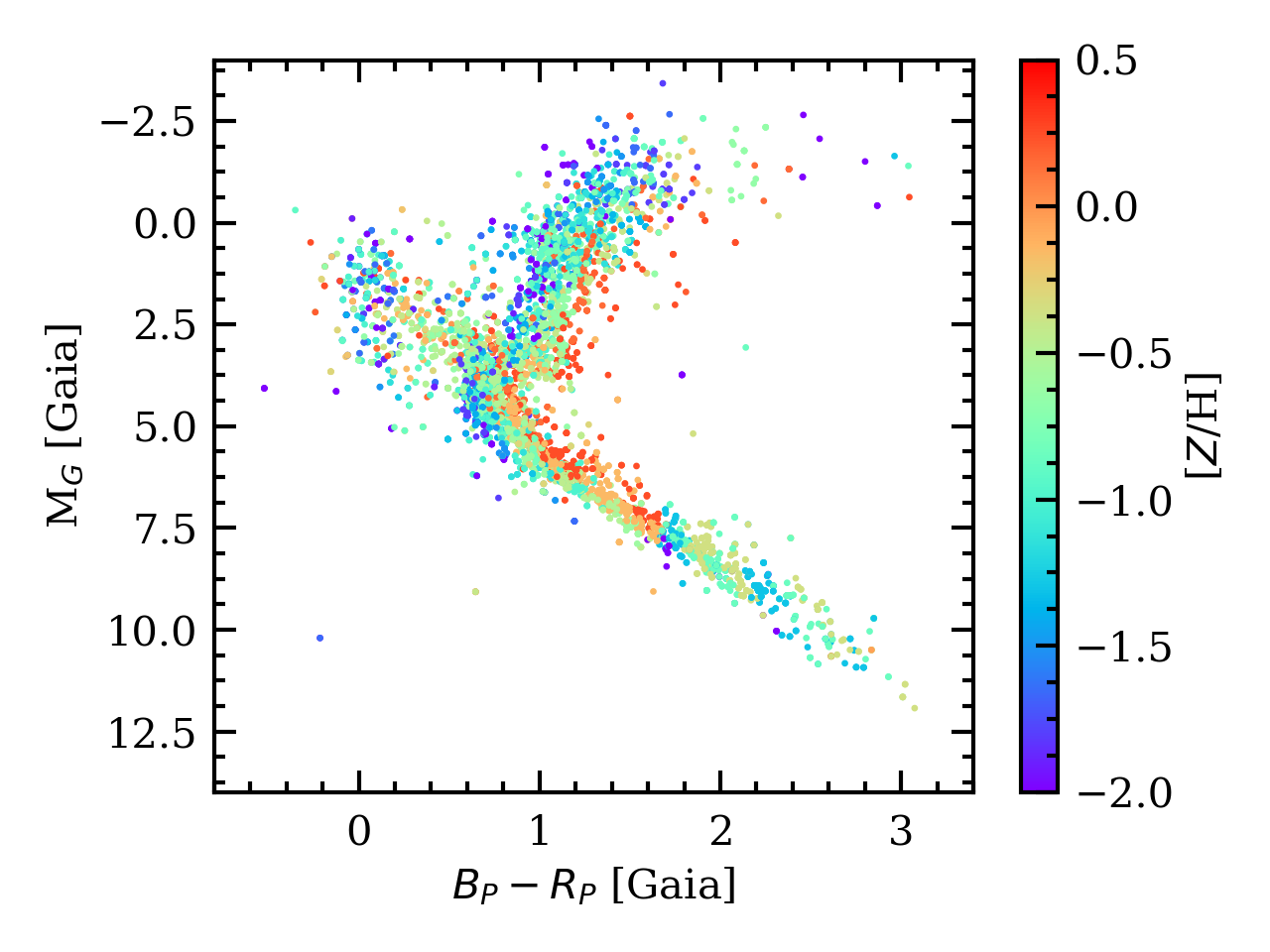}
	\caption{Extinction-corrected colour-magnitude diagrams (CMD) in GAIA filters $M_{G}$ vs. ($B_{\rm p}-R_{\rm p}$) for MaStar spectra, colour-coded by effective temperature $T_{eff}$, surface gravity $log g$ and total metallicity [Z/H], from left to right, obtained from the E-set and the Th-set (upper and lower rows, Sections~\ref{sec:emparam}) and ~\ref{sec:thparam}, respectively). These diagrams have already been 'cleaned' by ill-assigned parameters using GAIA-based information, as is described in Section~\ref{sec:thparam}.}
    \label{fig:cmd}
\end{figure*}

Evolutionary population synthesis models assume a stellar evolution prescription, which in this case consists of the isochrones and stellar tracks by
 \citet{cassisi_etal_1997} for ages larger than $\sim~30$~Myr and by \citet{genevatracks} for younger populations. Sets of models are also computed with the widely-used Padova stellar evolutionary models \citep{girardi_etal_2000}. The fuel consumption for the TP-AGB phase is calibrated with observational data. Reimers-type mass-loss was applied to the Red Giant Branch track in order to generate blue and intermediate Horizontal Branch (HB) morphologies as observed in Milky Way globular clusters, and also blue HBs at high metallicity \citep[see][for details]{maraston_2005}.
 
The synthesis technique exploits the fuel consumption approach~\citep[see][and references therein]{maraston_1998} for calculating the energetics of post Main Sequence phases, i.e.\ the amount of fuel available for nuclear burning using the evolutionary track of the turnoff mass. In the M05 models, the fuel is used as an integration variable in the post Main Sequence. For the Main Sequence, the standard isochrone synthesis technique is applied where the integration variable is the mass. 

We note that the fuel consumption approach is particularly useful when performing population synthesis using empirical stellar libraries in which regions of parameter space might not be evenly populated because of short evolutionary timescales (e.g.\ the tip-RGB). An accurate weighting of the energetics in under-populated phases becomes even more important.

As for the stellar spectra, in M05 the theoretical spectral library by \citet{lejeune_etal_1997} \citep[based on][]{kurucz_1979} is employed for all evolutionary phases but the TP-AGB for which empirical Carbon-rich and Oxygen-rich averaged stellar spectra are used \citep{lancon_and_mouhcine_2002}. In M11, population models based on several empirical libraries were calculated and comprehensively inter-compared and also confronted with the literature. Here, we will not repeat that analysis. 

In this paper, we adopt the first release of the MaStar stellar library (Section~3) to calculate stellar population models. The code structure is modular \citep[see][]{maraston_1998}, meaning the three main inputs of the model -
the energetics, the atmospheric parameters temperature and gravity, and the individual stellar spectra - are stored in independent matrices, which allows one to easily study their differential impact on the final models. This structure is particularly useful to compute the models of this paper for which we only vary the spectral transformation matrix (i.e.\ the stellar spectra and their stellar parameters, see Section~\ref{sec:parameters}).

\section{The MaStar library of empirical stellar spectra}
\label{sec:mastar}
\citet{yan_etal_2019} present the first release of the MANGA Stellar library of Milky Way stellar spectra (MaStar). Here we summarise its main features with particular attention to stellar population model calculations.

MaStar is a large, well-calibrated, high-quality empirical library aiming at observing more than 10,000 individual stars, whose spectra cover the wavelength range of $3,622--10,354$\AA\ at a resolution of $\sim~1800$. The spectra were obtained with the same instrument as used by the Mapping Nearby Galaxies at Apache Point Observatory galaxy survey project \citep[MaNGA][]{bundy_etal_2016,yan_etal_2016}, by piggybacking on SDSS-IV/APOGEE-2N observations \citep{apogee,nidever_etal_2015}. A sophisticated target selection strategy, based on stellar parameter catalogs available in the literature, is employed in order to ensure an unprecedented coverage of stellar parameters especially in the cool dwarf regime, low-metallicity regime and for acquiring a variety of element abundance ratios. The first version of the library published by \citet{yan_etal_2019}, which we use here, contains 8646 high quality per-visit spectra for 3321 unique stars distributed accross the HR diagram (see Section~\ref{sec:parameters}). Stars are selected to be brighter than 17.5 in either the $g$- or the $i$-band. This ensures a $S/N$~larger than 50 with 8 15-minute exposures \citep[cfr. Figure~5 in][]{yan_etal_2019}. Compared to photometry, the relative flux calibration of the library is accurate to 3.9\% in $g-r$, 2.7\% in $r-i$, and 2.2\% in $i-z$. For comparison \citet{miles} quote an accuracy of 2.5\% in the $B-V$~colour with a systematic offset of 1.5\% for the MILES library (c.f.\ their Figure 8). 

The spectra comprising the library are marked with a series of quality flags fully described in \citet{yan_etal_2019}. For population synthesis purposes, we apply further selections as will be described in Section~\ref{sec:flags}.

Note that the current release does not yet include Carbon- and Oxygen-rich type spectra in sufficient number as to properly model the TP-AGB phase of stellar evolution. For the latter phase, we shall use the same empirical spectra by \citet{lancon_and_wood_2000} as in the M05 and M11 models, but also provide a version of our MaStar population models not including the TP-AGB phase. 

\subsection{Golden MaStar spectral tables for population synthesis.}
\label{sec:flags}

Not all spectra can be used for population synthesis. Observed spectra can be corrupted, have an incorrect flux calibration, exist for the same star for a range of S/N ('visits' per star), or be affected by other problems that are flagged during visual inspection \citep[cfr.][]{yan_etal_2019}. Hence we apply a series of quality selections in order to construct our 'golden sample' of MaStar spectra. The large number of available spectra (8646) allows us to set relatively strict selection criteria.

We set the MJDQUAL bitmask to {\em exclude} the following spectra:
\begin{itemize}
\item Spectra with bad sky subtraction (bit 1).
\item Spectra where the PSF-covering fraction by fiber is smaller than 10\%\ (bit 4).
\item Spectra flux calibration affected by incorrect extinction estimates (bit 5).
\item Spectra with large scatter in radial velocity among multiple exposures (bit 6)\footnote{Two are the reasons for radial velocity scatter. In the reduction for DR15, the radial velocity was fitted using a limited set of model spectra. Thus,  there are cases (mostly white dwarfs or very hot stars) for which no models produce good fitting to any of the lines and the results can have large scatter in the derived radial velocity. In addition, radial velocity variations maybe caused by binary stars, a fraction of which is inevitable in large data set.}
\item Spectra flagged as problematic by visual inspection (bit 7).
\item Spectra with median S/N per pixel less than or equal to 15 (bit 9)\footnote{We emphasise again that the vast majority of spectra have a much larger $S/N$, as shown in Figure~5 of Yan et al.\ 2019.}. 

Among the spectra effectively used for our models, the S/N ranges from 15 to 400, with a median of 150, with only 100 stars (corresponding to 3.5\% of the sample) having a S/N between 15 and 30.

\end{itemize}

In addition to the selection through this quality bitmask, we apply the same criteria as \citet{yan_etal_2019} to select stars with reliable extinction correction. These criteria include stars either with low total extinction, or at a vertical distance $|z|>300\;$pc, or at a distance $d<100\;$pc. These quality flags are the same as those applied by Chen et al.\ (2020, {\it in prep.}). Additionally, there are some quality flags we set after or as part of the spectral fitting procedure for parameter determination (see below).

Finally, during the procedure of population model calculations we have further removed as much as possible stellar spectra if they were corrupted by bad pixels by visually inspecting the spectra again. Indeed the visual inspection described in \citet{yan_etal_2019} has allowed spectra with bad pixels as it was thought that the good portion of those spectra can still be useful.

\subsection{Stellar parameters}
\label{sec:parameters}
In order to associate an empirical spectrum to a stellar evolutionary phase, we need to determine the following parameters: the effective temperature $T_{\rm eff}$, the surface gravity $log~g$ and the chemical composition. The latter can be in the form of total fractional abundance of heavy elements ([Z/H]), of metals including Helium ([M/H]), or of the Iron abundance [Fe/H]. These three quantities ($T_{\rm eff}$, $log~g$~and chemical composition) will be referred to as {\it stellar parameters}.  

In this work, we use two independent sets of stellar parameters for the MaStar spectra and calculate stellar population models for each of them, which will allow us to assess their effect on the final models.  The two sets of stellar parameters are described in the following sections.
\subsubsection{Stellar parameters based on empirical spectra - E-set.}
\label{sec:emparam}
This set of stellar parameters is adopted in \citet{yan_etal_2019} to describe the first release of the library. As it is mostly (though not only, see below) based on parameters derived from a MILES-based interpolator (see below) we dub this set as 'E' (standing for 'empirical'). Here we provide a concise description of the procedure and refer to \citet{yan_etal_2019} and Chen et al.\ (2020, {\it in prep.}) for details. 

Chen et al.\ use a publicly-available interpolation tool \citep{prugniel_etal_2011} to calculate stellar parameters for the MaStar sample. This tool adopts parameters re-calculated for the MILES stellar spectra by \citet{wu_etal_2011}\footnote{As discussed in \citet{prugniel_etal_2011}, the original stellar parameters for the MILES library were a non-homogeneous combination of literature values based on both spectroscopy and photometry. For this reason they have re-evaluated them. On the other hand, as shown in Chen et al.\ original and re-calculated parameters compare quite well.} as a training set to assign parameters to any arbitrary stellar spectrum via interpolation within the range of MILES parameters. 
The tool uses full-spectrum-fitting \citep[ULySS][]{koleva_etal_2009} which matches MILES spectra to the spectrum for which parameters are to be evaluated - MaStar in our case - and finds the best solution as a linear combination of MILES spectra. The parameter assigned to each MaStar spectrum are then similarly linear combinations of the parameters of each MILES spectrum entering the best-fit linear combination.
\begin{figure}
	\includegraphics[width=0.49\textwidth]{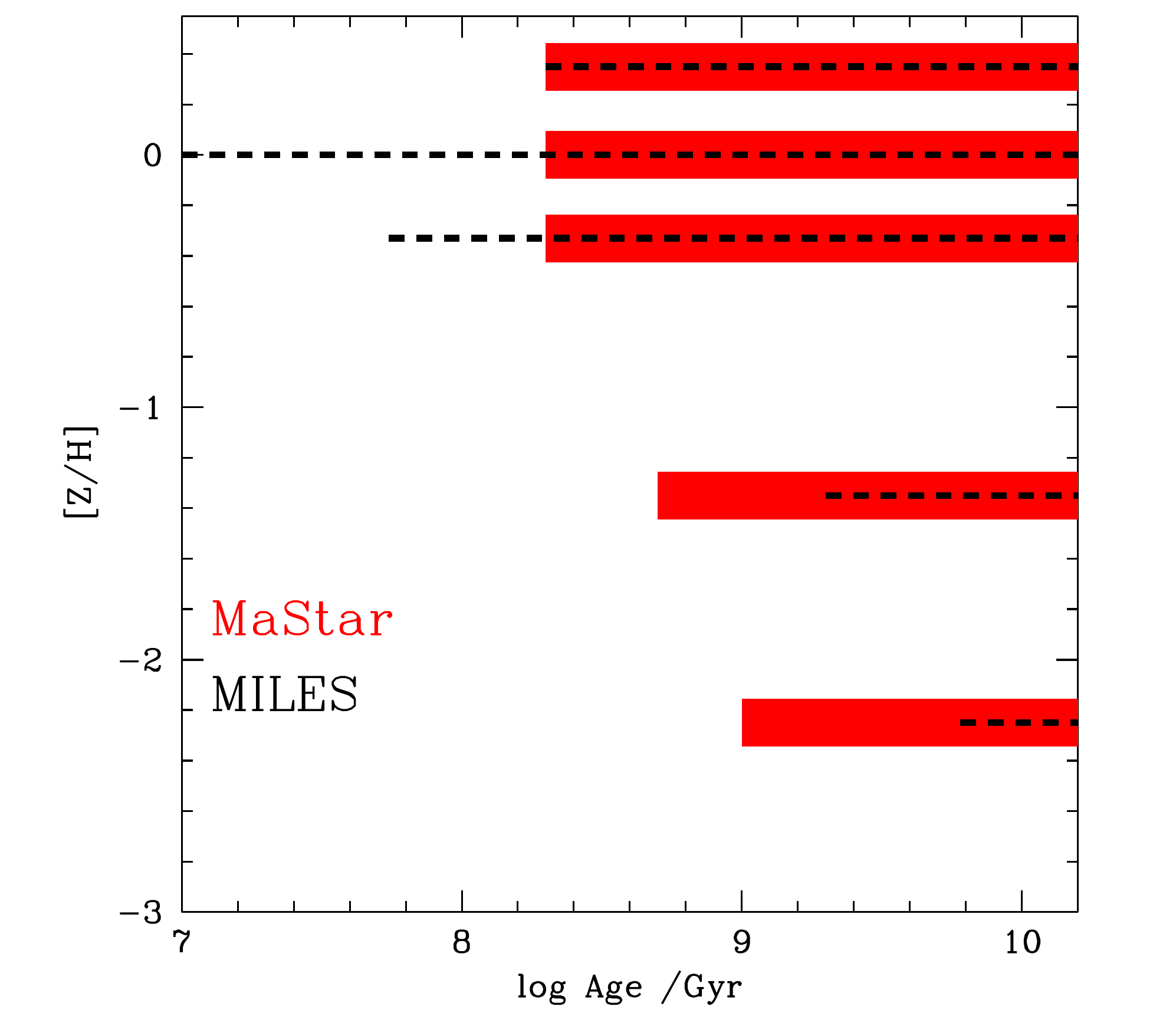}
	\caption{Coverage in age at the given chemical composition pillars, for the new MaStar models (red) and for M11-MILES (black dashed). At sub-solar metallicity MaStar allows the calculation of a wider age grid, whereas around solar metallicity MILES-based models reach younger ages.}
    \label{fig:grid}
\end{figure}
Chen et al.\ also provide a flag for the quality of the spectral fit, which we use to exclude spectra with uncertain parameter determination based on the quality of the fit. Finally, we apply constraints on $T_{\rm eff}$ and $\log g$ derived from matching MaStar with GAIA that will be described in Section~\ref{sec:thparam}.
\begin{table*}
\centering
{\footnotesize 
\begin{tabular}{ l | cccccc |}
\hline\hline
Model & Wavelength coverage & Age coverage & Age grid & Metallicity & HB morphology\\
 & (min -- max) / \AA & (min -- max) / Gyr & N ages & [Z/H] \\ \hline \hline
 Th-MaStar & 3621.6 -- 10352.3 & 1 -- 15 & 15 & [$-2.3$] & red \\ 
  &  & 1 -- 15 & 15 & [$-2.3$] & blue\\
            & & 0.5 -- 15 & 20 & [$-1.3$] & red \\ 
            &  & 0.5 -- 15 & 20 & [$-1.3$] & blue \\ 
            &  & 0.2 -- 15 & 23 & [$-0.3$] & red \\
            &  & 0.2 -- 15 & 24 & [$+0.0$] & red \\ 
             &  & 0.2 -- 15 & 25 & [$+0.3$] & red \\ \hline 
 E-MaStar & 3621.6 -- 10352.3 & 6 -- 15 & 10 & [$-2.3$] & red \\ 
  &  & 6 -- 15 & 10 & [$-2.3$] & blue\\
            & & 0.5 -- 15 & 13 & [$-1.3$] & red \\ 
            &  & 0.5 -- 15 & 20 & [$-1.3$] & blue \\ 
            &  & 0.4 -- 15 & 21 & [$-0.3$] & red \\
            &  & 0.2 -- 15 & 24 & [$+0.0$] & red \\ 
             &  & 0.2 -- 15 & 25 & [$+0.3$] & red \\ \hline 
 M11-MILES & 3500 -- 7429 & 5 -- 15 & 11 & [$-2.3$] & red\\ 
  &  & 5 -- 12 & 8 & [$-2.3$] & blue\\
            & & 2 -- 15 & 14 & [$-1.3$] & red \\ 
            &  & 2 -- 15 & 14 & [$-1.3$] & blue \\ 
            &  & 0.055 -- 15 & 34 & [$-0.3$] & red \\
            &  & 0.0065 -- 15 & 50 & [$+0.0$] & red \\ 
             &  & 0.1 -- 15 & 25 & [$+0.3$] & red \\ \hline 
 \end{tabular}
\caption[Parameter space covered by the stellar population models calculated in this paper.]{Population parameter coverage of the MaStar-based stellar population models released in this paper (plus those referred to M11-MILES models for comparison.). The set denoted with 'Th' refers to the one based on stellar parameters calculated from theoretical spectra  (Section~\ref{sec:thparam}). The set denoted with 'E' refers to the one based on empirical MILES spectra (Section~\ref{sec:emparam}). Grids are identical for any assumed IMF. Models are available at \protect\url{www.icg.port.ac.uk/MaStar/}. 
\label{tab:grid}}
} 
\end{table*}
The main features of the E-parameter set are then: i) it is limited to the parameter space covered by the MILES empirical spectra library; ii) it is composed by combinations of parameters from individual fits.

There are two caveats to the above procedure which motivated us to calculate a further set of parameters. Firstly, the MaStar library aims at covering a wider range of parameters compared to the MILES library \citep[cfr.][Figure~13]{yan_etal_2019} and  including spectral types - such as cool dwarfs and Carbon stars - which were not included in MILES. However, parameters interpolated within the MILES range will naturally squeeze any newly determined parameter to lie within the MILES range. Hence, the E-set imposes the MILES-parameter-dimension prior into MaStar.

Furthermore, stars unlike galaxies, are individual objects and parameters obtained by linearly combining different individual fits carry the risk at altering the real nature of the stellar spectrum. In fact, the search for the minimum $\chi^2$~will include spectra that can improve the fit which not necessarily respond to actual physics rather they might just result from missing certain features.
Due to the above caveats we decided to calculate another set of parameters as will be described in the next section. 

A further important comment to the E-set needs to be made. \citet{yan_etal_2019} find that metallicity determinations from other sources (see below) perform better than metallicities from the MILES-based interpolator when used to construct colour-magnitude diagrams for MaStar spectra, as the Red Giant branch dependence on metallicity is better recovered \citep[Figure~12 in][]{yan_etal_2019}. These sources are, in the order of preference set by \citet{yan_etal_2019}: the APOGEE Stellar Parameter and Chemical Abundance Pipeline (ASPCAP) catalogs \citep{holtzman_etal_2015,garcia_etal_2016,holtzman_etal_2018}; the Stellar Parameter Pipeline (SSPP) catalog \citep{lee_etal_2008a,lee_etal_2008b,allendeprieto_etal_2008,allendeprieto_etal_2014}; the Large Sky Area Multi-Object Fiber Spectroscopic
Telescope \citep[LAMOST,][]{lamost1, lamost2,lamost3} Experiment for Galactic Understanding and Exploration (LEGUE) Data Release 2 AFGK catalog \citep{lamost4}.

Hence, when available we use the metallicity from those other sources as compiled by \citet{yan_etal_2019} and referred to as 'input parameters'. These are available for $2/3$~of the spectra.

Figure~\ref{fig:cmd} illustrates the Colour Magnitude diagrams (CMD) we obtain for the MaStar spectra after applying quality flags (Section~\ref{sec:flags}\footnote{The parameter space of the {\it removed} spectra after application of all flags is shown in Appendix A.}) as displayed in the GAIA filters of absolute magnitude $M_{G}$ vs. the colour $B_{\rm p}-R_{\rm p}$. Spectra are colour-coded by their associated stellar parameters effective temperature $T_{eff}$, surface gravity $log g$ and total metallicity [Z/H], from left to right. The upper row refers to the E-set, while the lower row to the Th-set that will be explained in the next subsection~\ref{sec:thparam}. 
The E-set offers good coverage of parameters with range in gravity and metallicity in agreement with stellar evolution. In order to achieve such a sensible display of gravities and break insidious spectral degeneracies between cool dwarf and giants, we have used GAIA distance information to check whether the determined parameters would not erroneously place a giant star along the dwarf sequence and vice versa. This test will be discussed in the next subsection.
Note that the dependance of the Red Giant Branch (RGB) on metallicity, where higher metallicity stars have redder (i.e.\ lower temperature) RGBs is recovered by the E-set. 

Note also that the lower end of the Main Sequence (MS) as well as the upper part of the RGB seem to be mostly populated by sub-solar metallicity spectra. This impacts on the integrated population models we are able to calculate.

CMDs as such are not able to inform us on which model ages we are able to compute. For that we need a comparison with stellar evolution models. 
\begin{figure*}
	\includegraphics[width=0.9\textwidth]{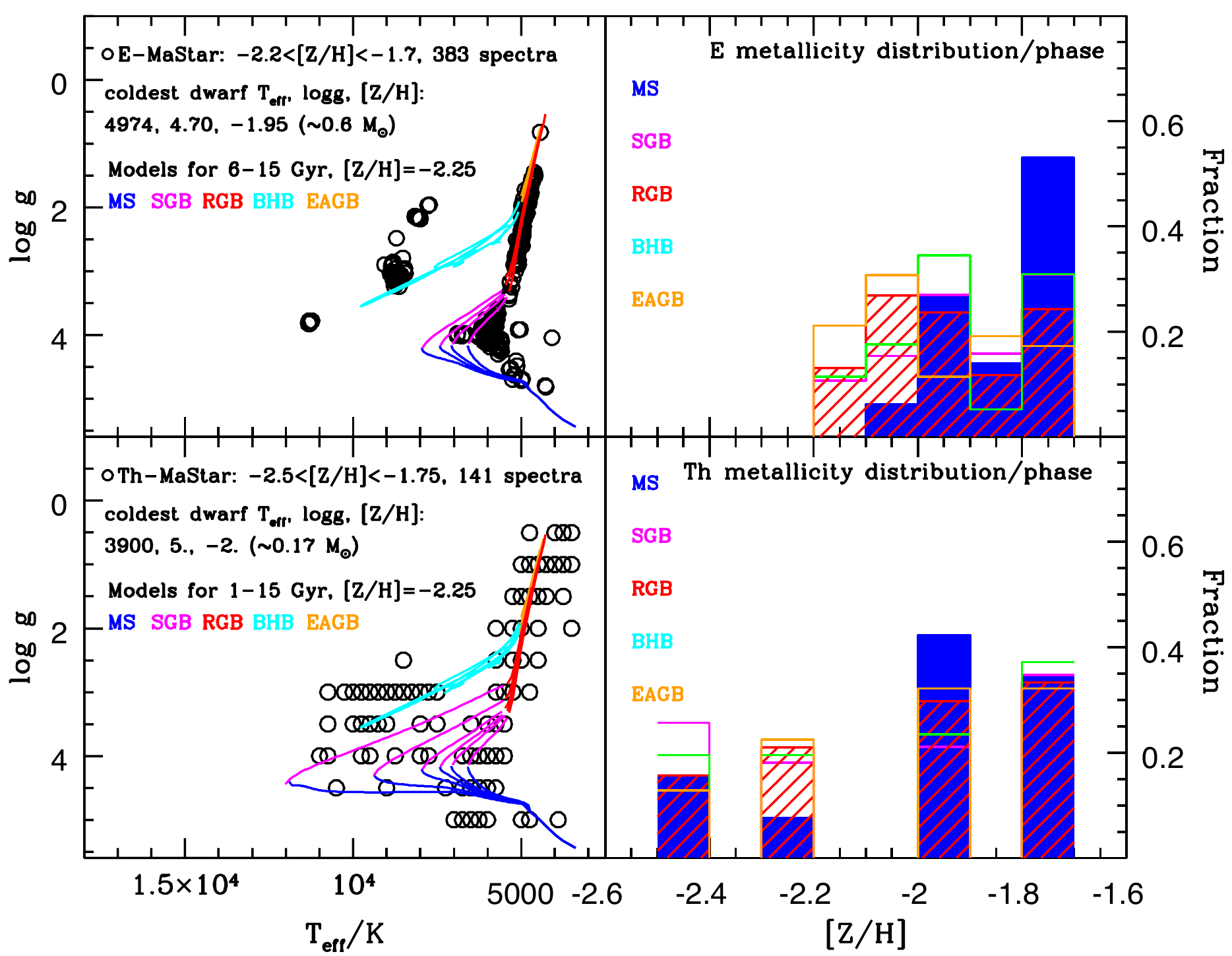}
	\caption{{\it Left-hand panels}: Theoretical HR diagram of effective temperature vs log gravity displaying the input stellar models to the evolutionary synthesis, namely isochrones up to the turnoff and the evolutionary track correspondent to the turnoff mass for post Main Sequence phases (coloured lines split by stellar phase) - and the parameters from MaStar spectra (circles), according to the theoretical set (Th-set, Section \ref{sec:thparam}) and to the empirical set (E-set, Section \ref{sec:emparam}). The line up of parameters for the Th-case reflects the binning of the model atmosphere grid. The E-set is based on interpolation instead. Th-based temperatures extend to hotter values with respect to the E- ones, allowing the calculation of model ages down to $\sim 1$~Gyr, while for the E-case the minimum age is 6 Gyr (cfr. Table~1). Worth noticing is the population of blue horizontal branch (BHB) stars - around $T_{eff}\sim9000~K$~and $logg\sim3$ - which is found consistently with both parameter estimators, allowing a proper modelling of metal-poor, old stellar populations with BHBs. The RGB slope is well recovered by both sets. We also label the parameters of the coldest MS dwarf and we quote the approximate evolutionary mass to which they correspond. The Th set allows us to nearly reach the lowest mass MS dwarf. {\it Right-hand panels}: metallicity distribution of MaStar spectra split by phases for the two parameter sets.}
    \label{fig:hr_verypoor}
\end{figure*}

The E-parameters as well as the Th-parameters that we shall discuss next are input to our population models (Section~\ref{sec:ssp}). 
\subsubsection{Theoretical stellar parameters - Th-set}
\label{sec:thparam}

The other set of stellar parameters is obtained through full spectral fitting of theoretical spectra with model atmospheres (Hill et al.\ {\it in prep.}). We refer to this set as 'theoretical' set (Th).

Prior to fitting, the observed MaStar spectra are de-reddened using the standard  \citet{schlegel_etal_1998} Milky Way reddening maps. The model fit is then performed using the penalized pixel-fitting (pPXF) method by \citet{ppxf}. This method uses Gauss-Hermite parameterization to describe the line of sight velocity distributions in the spectra hence it accounts for dispersion effects in the observed spectra. The main benefit is to ensure that individual spectral lines match the templates. Furthermore, by fitting the observed continuum with multiplicative polynomials, inaccuracies in spectral calibration or in reddening effects by dust are minimized.

The backbone of the method is straightforward. Individual theoretical spectra with known stellar parameters are fitted to the observed spectra and the model that best matches (in terms of minimum $\chi^2$) gives the stellar parameters of the observed spectrum. 
As fitting templates we use a combination of two theoretical spectral libraries, namely BOSZ-ATLAS9 \citep{meszaros_etal_2012,bohlin_etal_2017} and MARCS \citep{gustafsson_etal_2008}. By combining both libraries, we obtain a full coverage of stellar parameters, with the BOSZ-ATLAS9 models covering O, B, A, F, and G stars whilst MARCS models complement F and G stars, and  provide state-of-art models for the coldest K and M stars. The combined libraries allow for the following parameter coverage: $T_{\rm eff}$ (2400 - 34800~K), $log~g$ (-0.5 - 5.5 dex), [Z/H] (-2.5 - 0.5). More precisely, the two models overlap in the $T_{\rm eff}$~region of 3500 to
8000~K (with MARCS extending to higher surface gravity values, namely $log~g=5.5$~dex). In the common region we leave the fit choosing the best match. We find that $\sim 88\%$~of the observed spectra are best-matched by a BOSZ model, with the rest $12\%$~selecting instead a MARCS model. Finally, it should be noted that the original binning of the models in the atmospheric parameters $T_{\rm eff}$, $log~g$~and $[Z/H]$ is retained. This is due to the fact that the evolutionary synthesis code performs interpolation for assigning a spectrum to a stellar track point. In future work based on the full MaStar library we shall also experiment on the effect of the model atmosphere grid size and binning. Details will be given in Hill et al. {\it in prep.}.

As the full spectral fitting of 4563 flux points in each spectrum to a large grid of models (8149) is computationally intensive, we first narrow the theoretical grid by adopting priors based on photometry. We have verified that the result without the application of these priors is the same, but the calculation time is much larger. The use of photometric priors has the additional benefit that the stellar parameters derived from spectroscopy are consistent with the star's position in the CMD.

To estimate $T_{\rm eff}$, we define a function that describes the relationship between the colour ($g - i$ magnitude) and $T_{\rm eff}$ of the theoretical models. As this is sampled from a discrete grid, a one-dimensional interpolation function is used to match our observed colour to an effective temperature. The $g - i$ magnitudes are derived from the MaStar spectra for each observation and are provided to the interpolation function which returns the corresponding $T_{\rm eff}$, according to the theoretical models. This estimate is then used to determine a prior range, whereby an estimate of $T_{\rm eff}$>12000~K has a range of $\pm2000~K$ and an estimate of $T_{\rm eff}$<12000~K has a range of $\pm1000~K$. The prior range is larger for high temperature stars as estimates are less accurate at such temperatures and the theoretical grid of parameters becomes sparse in this domain.

The $\log g$ priors are obtained by matching MaStar photometry to GAIA DR2 isochrones with ages of 1-15 Gyr (GAIA Collaboration 2016, GAIA Collaboration 2018, \footnote{\url{https://gea.esac.esa.int/archive/}} GAIA DR2). To this end we overlay GAIA DR2 isochrones on the CMD of our data. A horizontal bound of width $\pm0.1$ mag is assigned to each data point on the CMD. We then calculate the average $\log g$ of all isochrones that fall within this horizontal bound and use this as the prior, with a width of $\pm 0.8 \log g$. No prior constraint is set for the metallicity parameter, which is left as a free parameter.

As will be discussed in Lewis et al., GAIA information helps in removing outliers and adds in reliability. It also helps in reducing the number of input templates hence improving the speed of calculation.

\section{Stellar Population model calculation.}
\label{sec:ssp}
In this Section we describe the steps taken to calculate stellar population models based on the empirical MaStar library in addition to our standard population synthesis procedure. These steps are identical for both sets of input parameters.  
\subsection{Scaling the empirical stellar spectra}
\label{sec:scaling}
An important difference between theoretical and empirical stellar spectra is the flux units. While theoretical spectra are preferentially provided in absolute units (e.g.\ luminosity as ergs s$^{-1}$ {\AA}$^{-1}$) which makes the population synthesis straightforward, things are more 
complicated with empirical spectra. Fluxes are either given as they were 
measured and thus depend on the 
star distance or they 
have been normalised to unity at some wavelength. In both cases, the true relative energy scale
between the spectra is removed. It is important to note that expressing the stellar population models on a real energy scale is crucial for evaluating any model output connected to luminosity, e.g.\ integrated magnitudes, redshift evolution and mass-to-light ratios.  

Here we adopt the same procedure as in M11 which has the advantage of offering a consistent approach that can be used for any empirical library -- meaning the stellar population models based on the re-normalised libraries are readily comparable. The procedure is as follows. First, all empirical spectra are (re-)normalized to the average flux in a 100~{\AA} passband around 5550 {\AA} \citep[as in][] {pickles_1998}. This choice is independent of the specific resolution of the library because we use the average flux within the band. The spectra are 
then scaled with the average luminosity calculated within a 100~{\AA} 
passband centred at 5550 {\AA} of theoretical spectra with the same stellar 
atmospheric parameters (interpolated as necessary). For consistency with M11, we used the same models \citep[from][]{lejeune_etal_1997} that were the input of the M05 models. M11 studied the impact of the choice of width and wavelength (for the MILES case), by recalculating the scaling as a function of these parameters, including a much larger $\lambda$~range as adopted by other authors. As discussed in M11, their adopted scaling allows them to keep the continuum {\it shape} of the empirical spectra intact, which is not the case for other procedures adopted in the literature (see discussion in M11).
\subsection{Sorting Mastar spectra by metallicity.}
\label{sec:sorting}
The M05 population synthesis code performs calculations for stellar evolution models defined by their total metallicity [Z/H]. Hence we sort the flag-selected and re-normalised MaStar spectra into bins appropriate to each theoretical metallicity value. In other words, when we assign a spectrum to a theoretical ($T_{eff},logg$) location we do {\it not} also interpolate in metallicity. 

The choice of metallicity ranges has been experimented with until we found the best compromise between number of stars and range width. The large number of stars offered by our library allows us to be relatively strict and our metallicity ranges are typically $\pm 0.2-0.3$~dex wide, with the exception of the lowest bin where in order to gain a coverage of all evolutionary phases we adopt a larger bin. Around solar metallicity instead we are able to use only spectra with a best-fit solar metallicity $[Z/H]=0$.
The adopted metallicity binning is comparable to the metalllcity grid size in M05/M11 (the smallest being $\pm 0.3$). One should also note that metallicity determinations for real stars are affected by observational errors which are comparable to our range size \citep[e.g. $\pm 0.2 \rm dex$,][]{allendeprieto_etal_2008}. It should be stressed that the partition in metallicity is a key procedure as its accuracy will impact on the capacity of the models to determine the metallicity of real stellar populations of star clusters and galaxies. Anticipating the result of model testing with GC spectra (Section~\ref{sec:testing}), our metallicity partition is of good quality. Precise ranges and stellar numbers will be specified in Section~\ref{sec:results} where we present the models for each metallicity bin. 

Note also that when assembling the tables of MaStar spectra per metallicity, we also check whether there are spectra which have identical stellar parameters, an event that can occur for the Th-set whose parameters are based on individual fits on a discrete model grid\footnote{This event never happens for the E-set due to the linear combination procedure.}. If this is the case, we select the spectrum with the highest $S/N$~ratio. Again, the large number of spectra collected by MaStar allows us to always have a sufficient number of spectra in each metallicity bin.  
\subsection{Implementation of MaStar spectra in the Maraston code}
\label{sec:interp}
In the M05 code, when using theoretical spectra the spectrum is assigned to each [T$_{eff}$, $logg$]-bin 
by means of a quadratic interpolation in T$_{eff}$ and
$logg$~at given metallicity. This approach is generally 
not feasible when empirical libraries are used, due to their coarser sampling 
of the parameter space.
\begin{figure*}
\includegraphics[width=1.\textwidth]{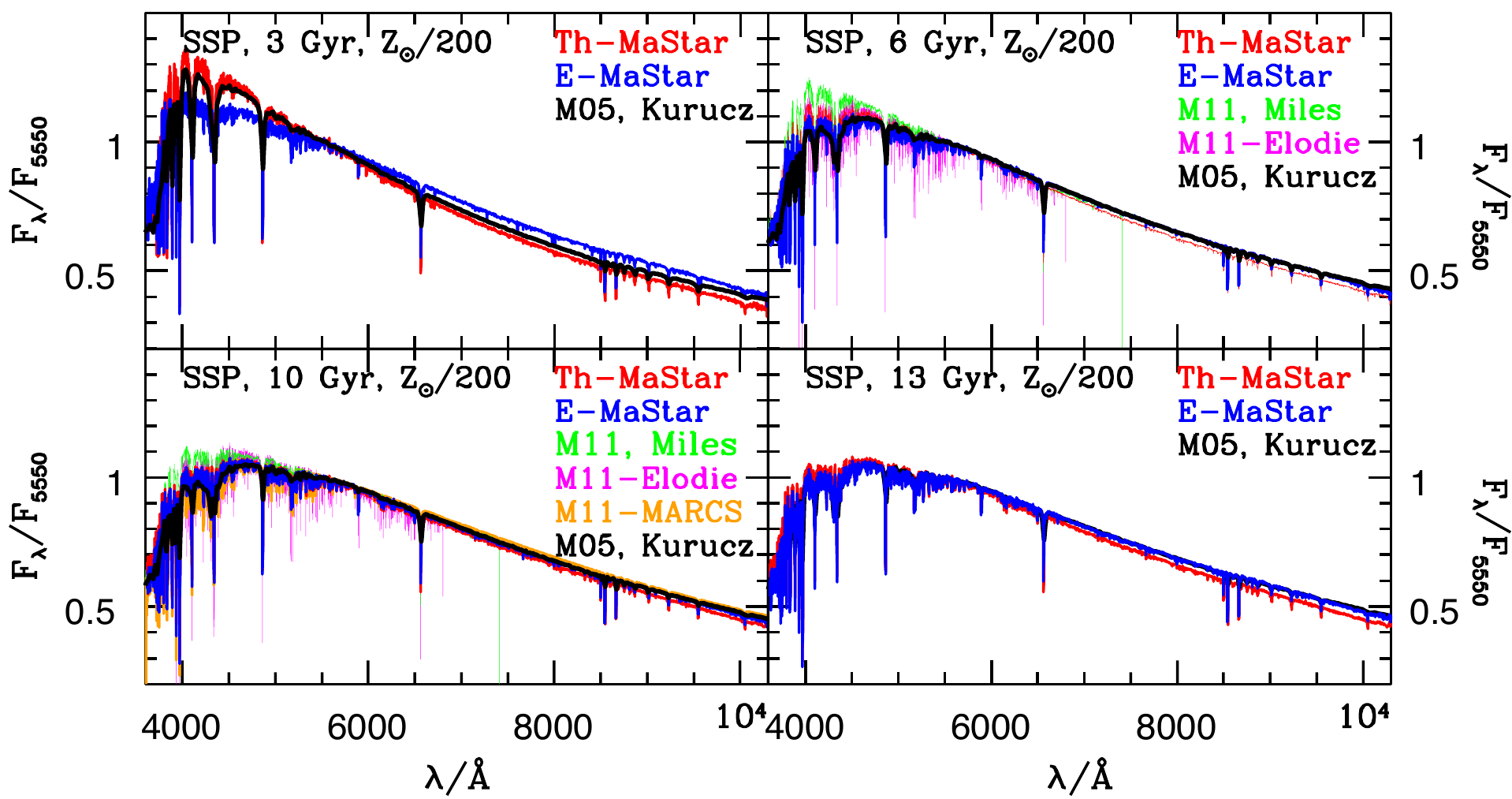}
	\caption{Integrated SEDs of MaStar-based stellar population models, for four representative ages, ordered as increasing age from  top-left to  bottom right. Models refer to a BHB morphology and a Kroupa IMF. Red and blue colours refer to MaStar-based models adopting the Th and the E sets of stellar parameters, respectively. Also shown are the SEDs of M05 models based on Kurucz model atmospheres (black), and of the M11-MILES and M11-ELODIE models (green and magenta, respectively). The fully theoretical M11-MARCS models, based on MARCS model atmospheres are shown in orange. The absence of a certain M11-type model means that no model with the chosen parameters is available (cfr. M11, Table~1). Models have different spectral resolutions, namely (referred to the FWHM at 5500 \AA\ ): 0.55 \AA\ (M11-Elodie), 2.54 \AA\ (M11-Miles and M11-MARCS), 3.05 \AA\ (MaStar, see Figure~\ref{fig:resolution}), 3.4 \AA\ (M11-Stelib), 10--20 \AA\ (M05-Kurucz).}
	  \label{fig:ssp_vp}
    \end{figure*}
As in M11, for each empirical library we sort stellar spectra, corresponding to the different evolutionary phases, i.e. MS, SGB, RGB, HB, 
E-AGB \footnote{M11 also identify spectra for the supergiant phase of massive stars. We do not need this further division as this first Mastar release does not contain these stellar types \citep[cfr.][]{yan_etal_2019}, but this will be the case in the next model release}. The categorisation is performed based primarily on the surface gravity, as in M11, but now we also explicitly use the effective temperature when necessary (e.g.\ for intermediate-phases on the HR diagram such as the SGB). Furthermore, we also make the cuts metallicity dependent. For example, in M11 the cut for stars in the MS was $logg>$4.0, which is maintained here, but enlarged to 3.5 for metal-rich models as these values are found on the corresponding isochrones. For uncertain cases, we also resort to other criteria, such as further visual inspection of the spectra. Note that the ranges of adjacent phases may overlap and therefore spectra be permitted to be present in both corresponding tables. 
We are able to apply stringent limits to the individual stellar phase parameters as we have many more spectra than M11 did. As we shall see in the next Section though, the partition done by M11 was already sufficient as the stellar population models calculated there and here are generally in good agreement. 
Within one specific evolutionary phase, the actual representative spectrum for each theoretical temperature/gravity location is then calculated  via linear interpolation of the logarithmic fluxes in log(T$_{eff}$), and when the number of stars allows it, by means of quadratic interpolation in log(T$_{eff}$) and $logg$ as in M05.
\begin{figure*}
  \includegraphics[width=0.9\textwidth]{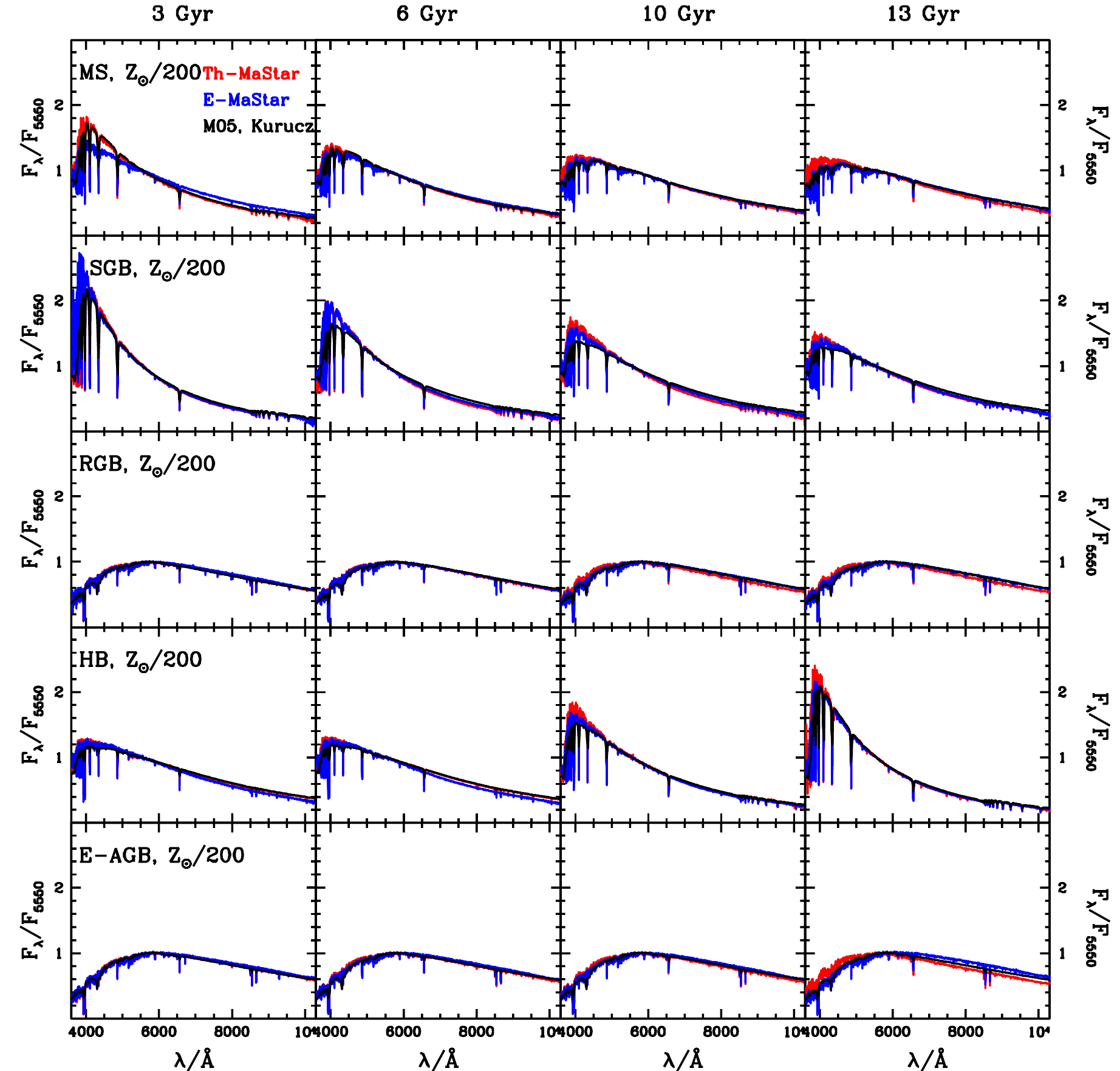}
  \caption{Integrated SEDs of individual stellar phases, for Th and E-MaStar models (red and blue) for four representative ages, ordered as increasing age from left to right. Each row refers to an individual stellar phase, namely from top to bottom: Main Sequence (MS), Sub Giant Branch (SGB), Red Giant Branch (RGB), Horizontal Branch (HB) and Early Asymptotic Giant Branch (E-AGB). The corresponding SEDs for M05 models are shown in black.}
  \label{fig:phasesvpoor}
\end{figure*}
The number of stars assigned to each phase is never smaller than 10 (in case of the E-parameters is actually much larger, reaching hundred of spectra per phase and metallicity). We check the stability of the procedure by using fewer or more stars and fix the procedure when the results are found to be stable. Note that the whole procedure is also checked at the bottom of the calculations by comparing the integrated models with our previous calculations based on theoretical stellar spectra (see Section~\ref{sec:results}).

\section{Results.}
\label{sec:results}
In this section we present the integrated spectra of simple stellar populations (SSP) models obtained by including the first release of MaStar empirical spectra with the two sets of stellar parameters E- and Th-, into the Maraston population synthesis code. Each subsection refers to one metallicity of the M05 grid, where we first present the details of the MaStar stellar parameters entering the models in the form of a theoretical HR diagram, followed by the model spectra, integrated over the whole population and over individual stellar phases. We also compare the models to our previous calculations based on either empirical or theoretical stellar libraries. As all models share the same energetics, these comparisons highlight the effect of the input library. 

As in M11, we also show plots of integrated line-strengths, which are a powerful tool for inferring the physical properties of stellar systems. In order to allow a direct comparison to previous models, we consider the widely used optical indices in the Lick system. In addition we discuss a few near-IR absorption features that became available thanks to the wavelength extension of MaStar (see Section~\ref{sec:nearirindices}).

The comparison of integrated spectra and line-strengths among various models constitutes a first sanity check on the models. The models will then be more quantitatively tested in their power to recover independently determined metallicity and age of star clusters (Section~\ref{sec:testing}). 

The MaStar model grids are given in Table~\ref{tab:grid} \citep[similar to Table~1 in][] {wilkinson_etal_2017}, where for comparison we also report the M11-MILES model grid. We note that the youngest ages in MaStar models lie around 0.1-0.2 Gyr, not yet comparable to the youngest models M11 was able to publish based on MILES (or other libraries) around solar metallicity. On the other hand, MaStar models are able to reach younger ages than M11-MILES at low metallicity, which can be useful to analyse the stellar populations of dwarf galaxies or high-$z$~galaxies. Moreover, we can model old, metal-poor populations with blue horizontal branches due to a larger populations of those stars in MaStar with respect to MILES (see Section~\ref{sec:vpoor}). For a quick visualisation Figure~\ref{fig:grid} depicts the coverage in age at the given chemical composition pillars, for the new MaStar models (red) and for M11-MILES (black dashed). 
\begin{figure*}
	\includegraphics[width=0.9\textwidth]{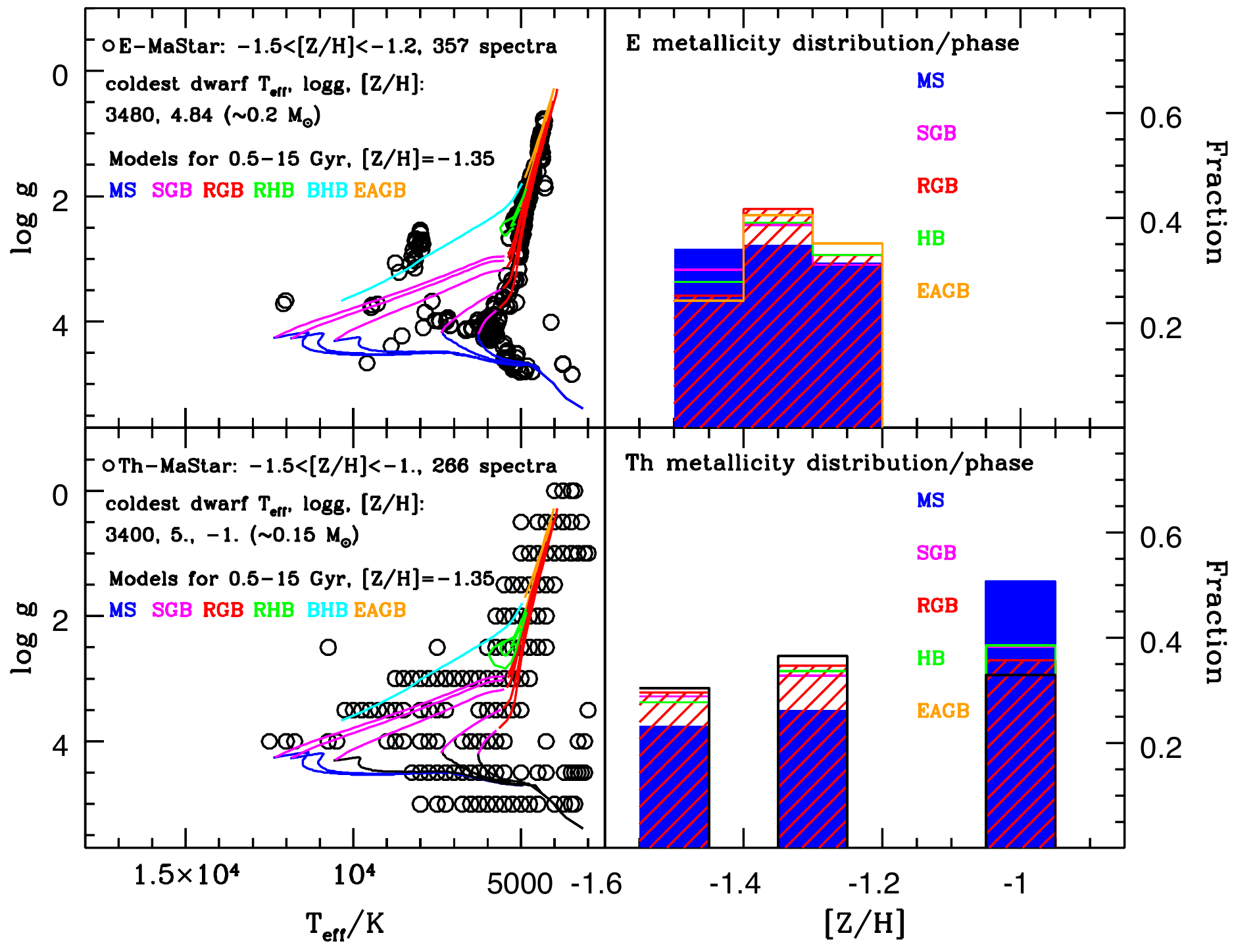}
	\caption{Theoretical HR diagram of $T_{\rm eff}$ vs logg of MaStar stellar parameters as in Figure~\ref{fig:hr_verypoor} here for $[Z/H]=-1.35$. We are able to push the calculation of models as young as 0.5 Gyr for the first time using empirical spectra at this low metallicity (cfr. Table 1). Also in this metallicity range we could secure BHB stars, whose parameters nicely match the theoretical HB location (shown for both 'red' as well as 'blue' morphologies, green and cyan lines, respectively). Both parameter sets reach near the bottom of the dwarf MS. The metallicity distribution per stellar phase (right-hand panels) reflects the model atmosphere grid employed to derive the Th-set.}
    \label{fig:hr_poor}
\end{figure*}
\subsection{Very metal-poor models $[Z/H]=-2.25$.}
\label{sec:vpoor}
Figure~\ref{fig:hr_verypoor} displays - in the theoretical HR diagram of effective temperature vs.\ log gravity - the isochrones (for the MS) and stellar tracks (for the post-MS) which are the input of the M05 evolutionary synthesis (coloured lines split by evolutionary phase, see Section~\ref{sec:eps}) and the available parameters from MaStar spectra (circles), according to the theoretical set (Th set, Section \ref{sec:thparam}) and to the empirical set (E set, Section \ref{sec:emparam}) in the left-hand panels, respectively. The available number of spectra is also quoted. The right-hand panels show the distribution in metallicity of the available MaStar spectra split by phases for the two sets. 

We use plots of this kind here and in the next subsections in order to illustrate the span of models in terms of age and chemical composition we are able to calculate with this first MaStar release. The exact age grid is given in Table~1.

We will first comment on a few features that are in common to all chemical composition bins. Firstly, the E-parameters allow for more spectra (in a given metallicity bin) due to being calculated via interpolation, which implies that there are never two spectra having identical parameters. On the other hand, the Th parameters which result from individual template fitting onto a wide grid without any a priori restriction, usually provide a wider coverage of the theoretical HR diagram. In this metallicity bin, the Th set allows us to calculate models with turnoff ages down to 1 Gyr (blue lines, lower left panel), whereas the E-set allows the calculation of robust models only for ages larger than 6 Gyr (upper panel).
\begin{figure*}
		\includegraphics[width=1.\textwidth]{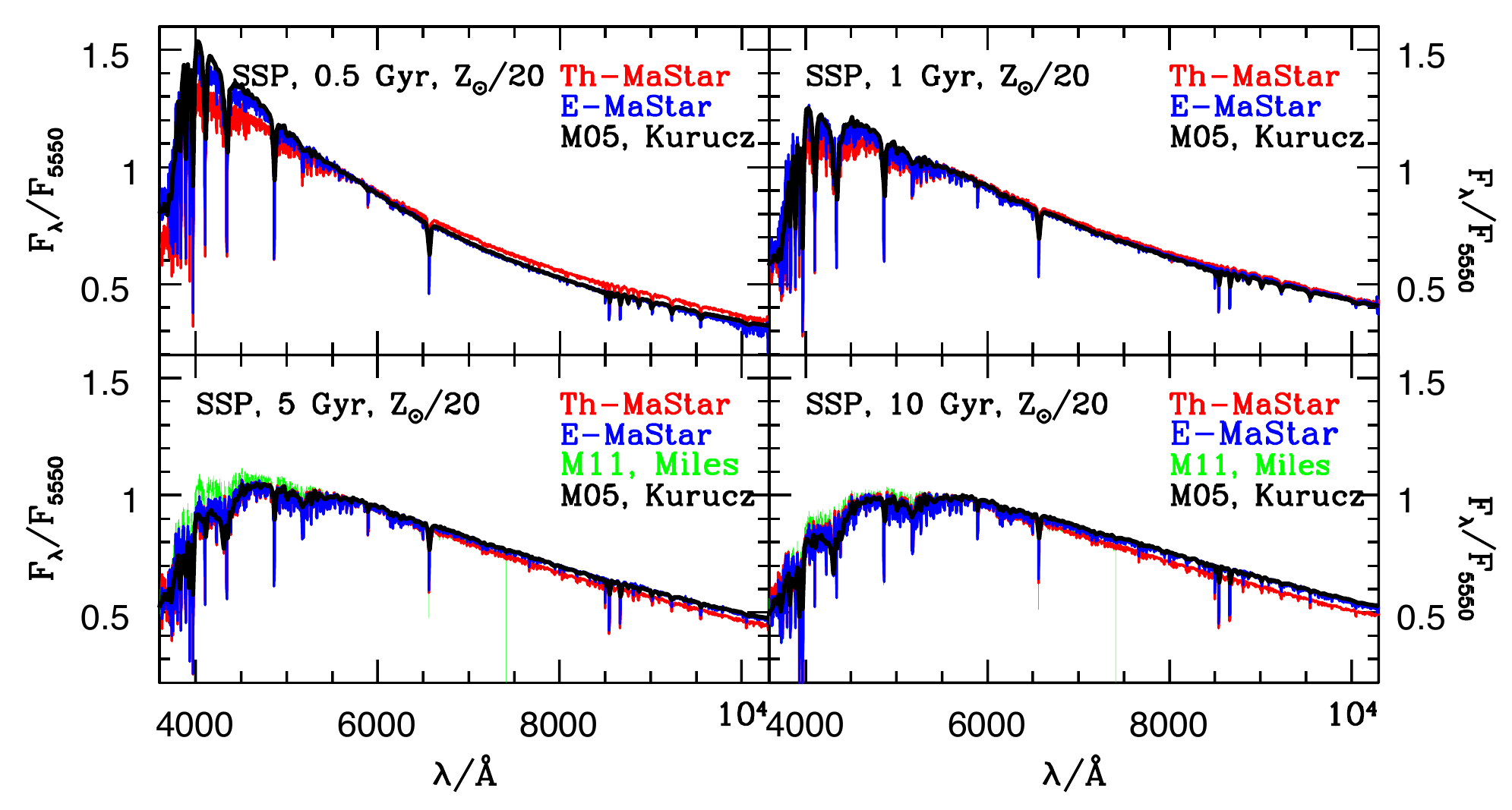}
	\caption{Integrated SEDs of MaStar-based stellar population models for $[Z/H]=-1.35$ as in Figure~\ref{fig:ssp_vp}.}
    \label{fig:ssp_poor}
\end{figure*}
Both Th- and E-sets cover the Red Giant Branch (RGB, red lines) well, showing a morphology which is remarkably consistent with the theoretical slope. Note that this is a non trivial result as the two parameter sets are calculated independently from the M05 adopted stellar evolution and also independently from each other. 
 
 Another excellent achievement is the acquisition of spectra appropriate to describe the Blue Horizontal Branch (BHB, cyan lines) in its full temperature extension down to the oldest model age (15 Gyr). This was not possible using the MILES library, which is why M11 could only calculate metal-poor models with BHB down to 12 Gyr (cfr Table~1).
 
 In Figure~\ref{fig:hr_verypoor} we also label the parameters of the coldest MS dwarf spectrum we obtain with each set and quote the approximate evolutionary mass that we are able to model. The Th set allows us to nearly reach the lowest mass MS dwarf down to the Hydrogen-burning limit, which is a remarkable result. Note that when we lack MaStar spectra with parameters appropriate to describe stellar masses below a certain value down to the bottom of the MS of $0.1~M_{\odot}$, we use the coldest, highest gravity available spectrum for all masses below the limiting value.
 
For the non-extragalactic reader, we recall that the merit of obtaining spectra for cold dwarfs and giants lies in the effect they have on the population models that are used for interpreting integrated galaxy spectra. 
Spectra for cold dwarfs and giants exhibit lines (e.g. the Sodium doublet at 8200\AA, hereafter  NaI0.82, the FeH band, the TiO indices) that are sensitive to the star surface gravity hence those lines integrated over a stellar generation are in principle sensitive to the giant-to-dwarf stellar ratio, which in turn sets constraints to the IMF.  By comparing galaxy data for near-IR spectral indices (e.g. the NaI0.82) with population models, \citet{conroy_and_van_dokkum_2012} (and several others afterwards) argued for a steepening of the IMF exponent over the Salpeter one, for stellar masses below $0.3~M_{\odot}$, in the most massive galaxies. Key to this finding are the stellar spectra employed in the models, for describing the coldest dwarfs (down to the H-burning limit) and giants. We shall come back to this point in Section~\ref{sec:nearirindices}.
 
The metallicities of MaStar spectra associated to the various stellar phases according to the Th and E sets are shown in the right-hand panels of Figure~\ref{fig:hr_verypoor}. 
 In case of the E-set we had to include stars up to $[Z/H]=-1.7$~in order to achieve enough cool giant and dwarfs for population model calculations. For consistency we kept the same upper limit for the Th-set. For the latter we included stars down to -2.5 as we could gain SGB and MS stars. This was not the case for the E-set, hence the metallicity distribution for the E-set is narrower.
For the E-set the metallicity of the MaStar spectra lies around -2 rather than the -2.25 of the input models. However, the latter value refers to the proper definition of $[Z/H]$~for stellar models (see M05, Table~1) where $X$ (the Hydrogen abundance $H$), $Y$ (the Helium abundance) and $Z$ (the abundance of heavier elements) are clearly defined whereas the 'metallicity' parameter assigned to empirical spectra is less strictly defined and has its own uncertainties hence we do not regard this slight mismatch as a major worry compared to other uncertainties. Most of all, we do not have enough stars in all evolutionary phases at lower values of metallicity. 

The integrated SEDs of our new stellar population models based on MaStar are shown in Figure~\ref{fig:ssp_vp}, for four representative ages, a Kroupa IMF and a Blue Horizontal Branch (BHB) morphology. Red and blue colours refer to MaStar-based models adopting the Th and the E sets of parameters, respectively. Also shown are the SEDs of M05 models based on Kurucz model atmospheres (black), and of the M11-MILES, M11-ELODIE and M11-MARCS models (green, magenta and orange). 

In examining comparisons between models based on different libraries the following should be noted. Firstly, models may not exist for all ages due to the sparsiness of empirical libraries (cfr M11). 
Secondly, models have different spectral resolution. The resolution of the MaStar models need to be calculated as the models are based on a mixture of spectra taken under various seeing conditions and the resolution of the SDSS spectrograph is wavelength-dependent. In this work we provide the resolution of the MaStar population models for a best comparison of these models to data (see Section~\ref{sec:specres}). In particular, MaStar models have a FWHM of 3.05 \AA\ at 5500 \AA\. For the other models the correspondent values are (from Table~1 of M11): 0.55, 2.54 and 3.4 \AA\ for M11-Elodie, M11-Miles and M11-Stelib, respectively. M11-MARCS models with the much higher resolution of 0.25 \AA\ have been smoothed to the MILES resolution. M05-Kurucz have a resolution ranging between 10 \AA\ and 20 \AA\ in the region of interest.
\begin{figure*}
	\includegraphics[width=0.9\textwidth]{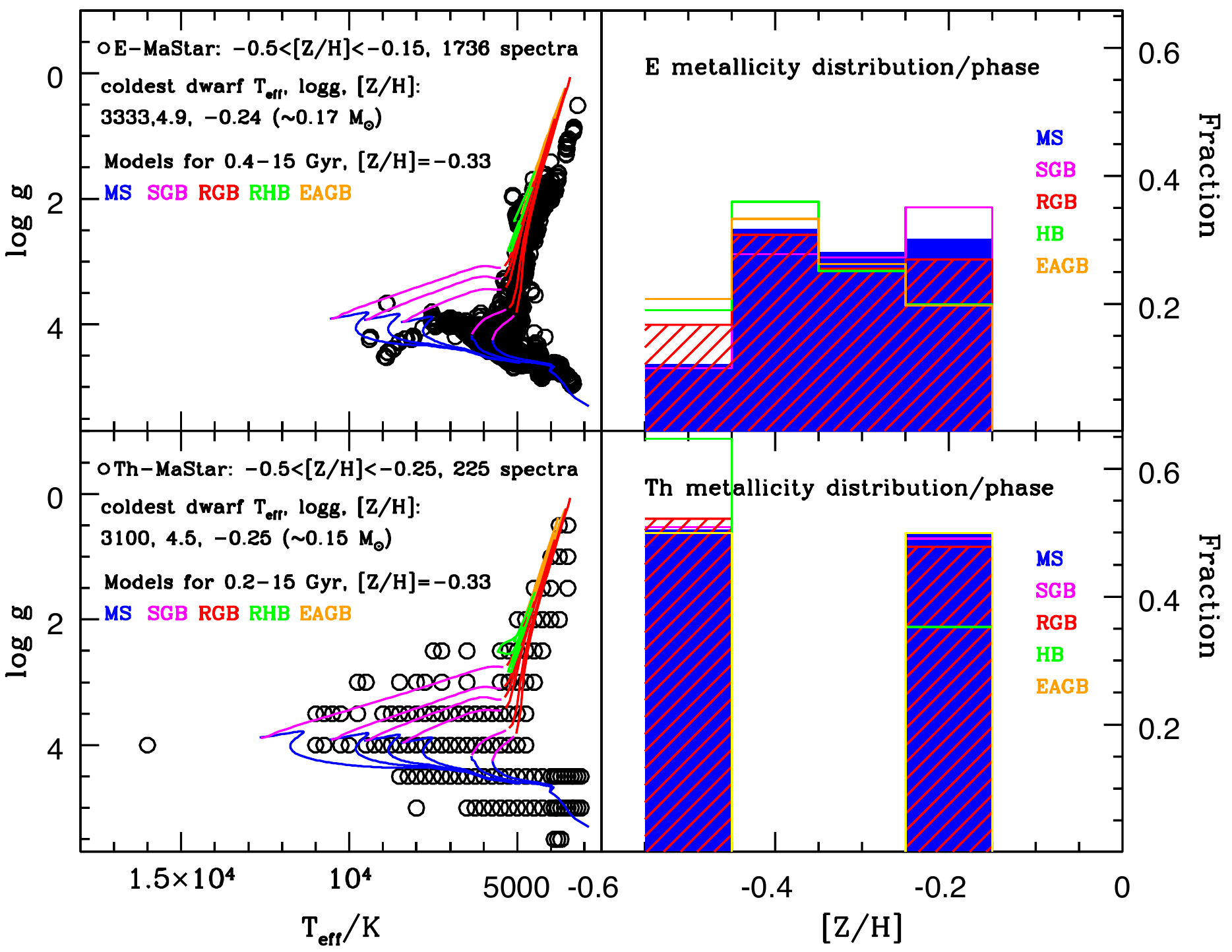}
	\caption{Theoretical HR diagram of $T_{\rm eff}$ vs $logg$ of MaStar stellar parameters as in Figure~\ref{fig:hr_verypoor} here for half-solar metallicity $[Z/H]=-0.33$. Th and E parameters allow the calculation of models down to $\sim 0.1$~Gyr. The assumed theoretical RGB slope is well matched by Th-parameters, whereas E-parameters shift to colder temperatures around the RGB bump up to the RGB-tip. The right-hand panels show the metallicity distributions per stellar phase.}
    \label{fig:hr_half}
\end{figure*}

Before proceeding, we explain how we treat such comparisons which we have used in order to improve the models and the stellar parameters during the whole development of this work, for which we show only the final outcome. 
Firstly, the comparison with the M05 models or with M11-MARCS based on theoretical atmospheres informs us on issues which may affect empirical spectra such as flux calibration and mis-assignments of stellar parameters. For the understanding of these issues, a comparison with the integrated SSPs is not sufficient since we also need to inspect each evolutionary phase (see below). This is because subtle mismatches may affect different phases in a way that compensate each other, e.g.\ a too red MS may couple with a too blue RGB, with these errors cancelling out in the total SED, but having potentially catastrophic effects on line-strengths sensitive to the dwarf-to-giant ratio. The comparison with M05 should not be quantitative because: i) line-strengths are going to be different because of known (and unknown) shortcomings in model atmospheres and because of spectral resolution; ii) we are not seeking to obtain identical models, rather we wish to improve on previous models.

Secondly, the comparison with M11 models based on empirical libraries such as e.g.\ M11-MILES is instructive as MaStar is the new generation empirical stellar library which aims at improving over previous empirical libraries (see discussion in M11). Moreover, population models based on empirical libraries have been extensively used for galaxy evolution investigations. 

Referring back to Figure~\ref{fig:ssp_vp}, a good overall agreement between the models is evident. Note that at 3~Gyr the E-models are somewhat too red due to the lack of hot stars required to model the MS turnoff at this age. We show them for illustration, but we do not advise their use at ages lower than 6 Gyr (cfr.\ Table~1). At the oldest ages the Th SEDs are somewhat bluer than the other models due to the input stellar SEDs for BHB stars (cfr.\ Figure~\ref{fig:phasesvpoor}). 

Figure~\ref{fig:phasesvpoor} illustrates the model SEDs integrated over each stellar phase, i.e.\ Main Sequence (MS), Sub Giant Branch (SGB), Red Giant Branch (RGB), Horizontal Branch (HB), Early Asymptotic Giant Branch (E-AGB), from top to bottom right, for the same model ages as in Figure~\ref{fig:ssp_vp} (one per column). We find good agreement overall between MaStar-based and M05 models in each evolutionary phase; no unexplained offset is apparent. Similar plots for other metallicities can be found in Appendix~\ref{app:phases}, unless we wish to emphasise a certain result, as will be the case for metal-rich models, see Section~\ref{sec:supersolar}.

{In Appendix~\ref{app:lick} we also show Lick absorption line strengths calculated on the new models and compare them with previous calculations.}.
\begin{figure*}
		\includegraphics[width=1.\textwidth]{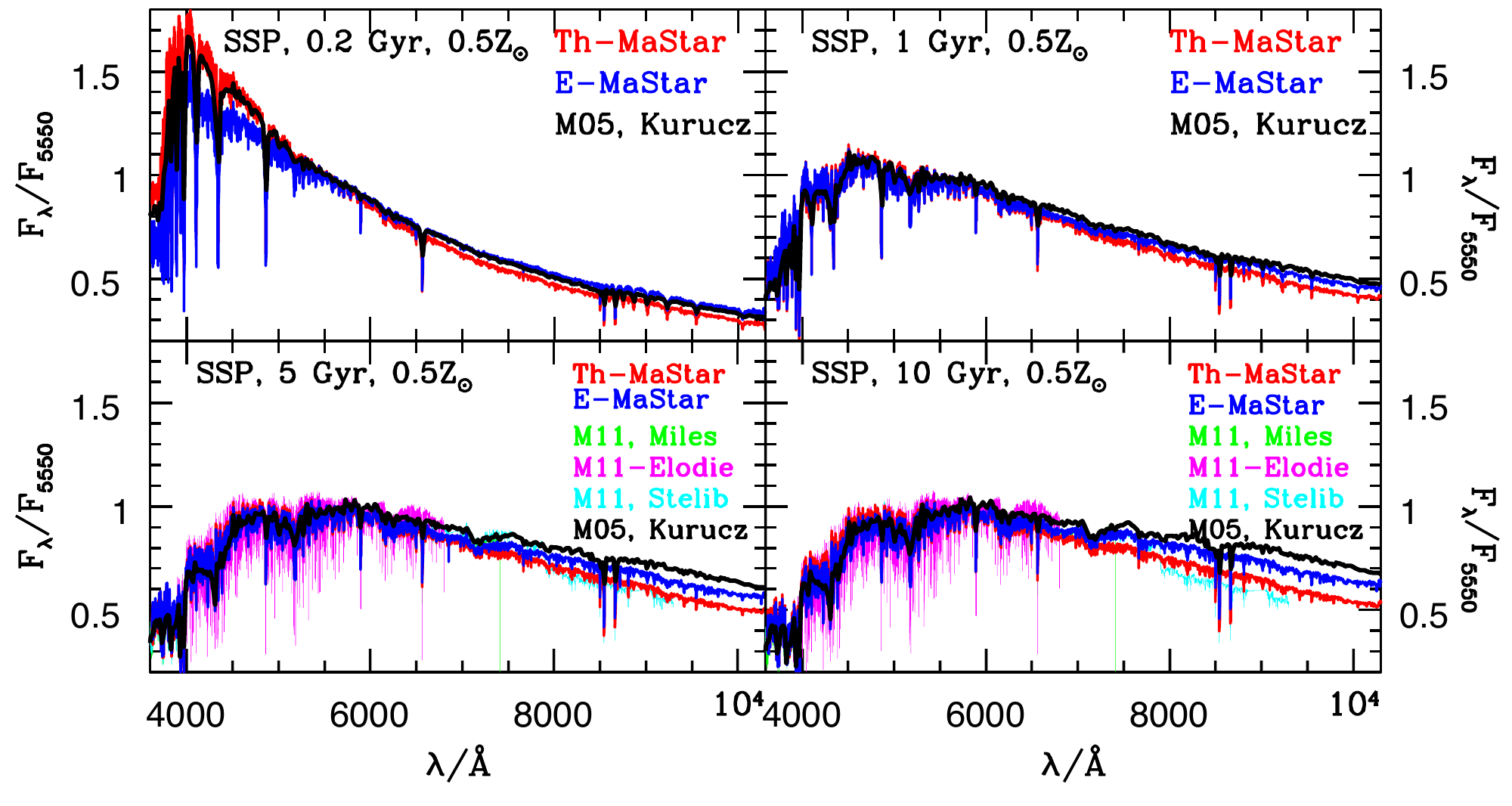}
	\caption{Integrated SEDs of MaStar-based stellar population models for $[Z/H]=-0.33$ as in Figure~\ref{fig:ssp_vp}. Also shown here with respect to previous similar figures are M11-Stelib models (cyan).}
    \label{fig:ssp_half}
\end{figure*}
\subsection{Metal-poor models $[Z/H]=-1.35$.}
\label{sec:poor}
In this section we focus on model results for a chemical composition with $[Z/H]=-1.35$.

\begin{figure*}
	\includegraphics[width=0.9\textwidth]{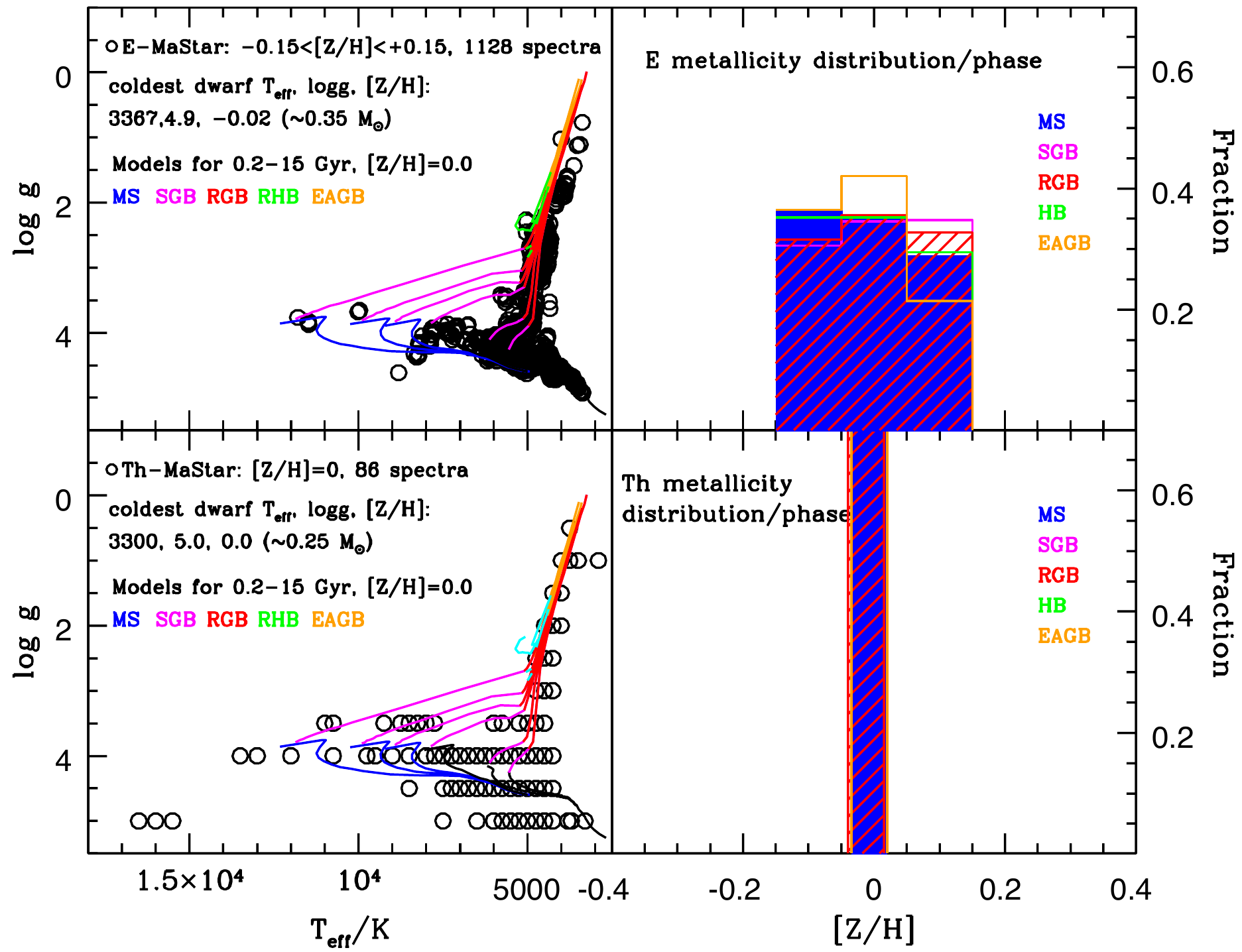}
	\caption{Theoretical HR diagram of $T_{\rm eff}$ vs $logg$ of MaStar stellar parameters as in Figure~\ref{fig:hr_verypoor} here for solar metallicity $[Z/H]=0$. Th and E parameters allow the calculation of models down to $\sim 0.3$~Gyr. The right-hand panels show the metallicity distributions per stellar phase. In case of the Th-set, all spectra have a nominal solar metallicity from the model atmosphere fitting.}
    \label{fig:hr_sol}
\end{figure*}

Figure~\ref{fig:hr_poor} similarly to Figure~\ref{fig:hr_verypoor} displays in the theoretical HR diagram of effective temperature vs.\ log gravity, the parameters of MaStar spectra (circles), according to the theoretical and empirical sets (Th and E sets, as explained in Sections \ref{sec:thparam} and \ref{sec:emparam}) in the left-hand panels, respectively. Lines show the stellar models which are the input of the evolutionary synthesis (coloured lines split by evolutionary phase). The available number of spectra is also given as well as the parameters of the coldest MS spectrum and the corresponding stellar mass. The right-hand panels show the distribution in metallicity of the available MaStar spectra split by phases for the two sets. 

Similar comments to those made for the lowest metallicity regime hold at this metallicity, too. The E method allows the gain of more spectra due to unique parameters obtained via interpolation, but the Th method allows the coverage of a wider parameter range. At this metallicity we could calculate models with turnoff ages down to 0.5 Gyr for both parameter sets, which is a new achievement for models based on empirical libraries (see discussion in M11 and Table~1). For both parameter sets the RGB slope is well matched (red lines) and the BHB location is well described. At this metallicity, the MaStar parameters allow us to model stars as small as $0.12-0.2~M_{\odot}$, which is an important step forward with respect to previous empirical libraries (see discussion in M11).\\
The right-hand panels show the metallicity distributions per stellar phase obtained with the two parameter sets. The one corresponding to the Th-set just reflects the entry grid of model atmospheres.\\
As in the previous subsection, we show the integrated SEDs of MaStar-based population models obtained using the E and Th sets of stellar parameters (Figure~\ref{fig:ssp_poor}). Also shown are the SEDs of M05 models based on Kurucz model atmospheres (black) and of the M11-MILES models (green). Also at this metallicity, a good consistency is found between the two MaStar-based models and the M05 and M11-MILES models. The SEDs of evolutionary phases and model Lick indices are shown and discussed in Appendices \ref{app:phases} and \ref{app:lick}.

\subsection{Half-solar metallicity models $[Z/H]=-0.33$.}
\label{sec:half}
It is around half-solar metallicity where most MaStar spectra seem to be located according to both parameter sets, in particular the E- one, which gives over 1700 spectra in the $[Z/H]$~range -0.5 to -0.15 (Figure~\ref{fig:hr_half}). The theoretical parameter space is well populated in all evolutionary phases for both sets, allowing the calculation of integrated models 
down to ages of 0.2 and 0.4~Gyr, for the Th and E sets, respectively. Stellar spectra with parameters appropriate for the bottom of the MS around 0.15-0.17~$M_{\odot}$ are obtained with both sets.

We note that the slope of the RGB as traced by the E parameter set lies at colder temperatures with respect to the theoretical one, with the offset increasing towards the RGB-tip. This effect is much less pronounced in case of the Th set. The RGB slope depends on the assumed mixing-length and other input physics and it is shown to vary among different stellar evolution calculations (see M05, Figure 9), with models from the Padova stellar evolutionary code \citep{girardi_etal_2000} having colder RGBs than e.g.\ the stellar tracks input to M05 models. 
As the Th parameters are obtained fully independently of stellar tracks, it is tempting to conclude that the actual RGB slope found in Nature is closer to the M05 input and that too cold RGBs are not proper to this metallicity. This conclusion is in line with the results by \citet{tayar_etal_2017} which are based on a much more comprehensive and quantitative study. We are unable to trace back any stellar track influence in case of the E-parameters.

The right-hand panels show the metallicity distributions per stellar phase that are input of our models. We note that while the E parameters distribute themselves homogeneously around the nominal track metallicity of $-0.3$, the Th metallicities are bimodally distributed at the grid values around the nominal track metallicity, again due to the input grid. No bias is observed in the distribution per phase, which is important for the synthesis procedure.

The integrated SEDs of MaStar-based population models for half-solar metallicity as obtained with the E and Th sets of stellar parameters are shown in Figure~\ref{fig:ssp_half}, along with the M05 models plus other population models published in M11 for different input empirical libraries (M11-MILES, M11-ELODIE and M11-STELIB as green, magenta and cyan line colours, respectively). 
Models generally agree well up to $\lambda \sim 7000$~Angstrom, after which the input stellar spectra and parameters make the two MaStar models deviate from each other. In particular, M05-Kurucz is the reddest model and Th-MaStar the bluest, with E-MaStar and M11-MARCS lying in between. Th-MaStar models seem consistent with M11-STELIB.  

The split of the integrated model SEDs per evolutionary phase shown in Figure~\ref{fig:phaseshalf} helps clarifying these discrepancies. The origin of the differences lies in the modelling of the giant phases and not in the Main Sequence (see Appendix~\ref{app:phases} for details).
As usual, a plot showing the model Lick indices for this metallicity is found in Appendix~\ref{app:lick}.

\subsection{Solar metallicity models $[Z/H]=0.0$.}
\label{sec:solar}

Figure~\ref{fig:hr_sol} shows the theoretical parameter space for population models with solar metallicity. The parameter space is well populated in all evolutionary phases for both sets, allowing the calculation of integrated models down to ages of approx 0.3~Gyr. We note that the lowest stellar mass on the MS we are able to model is $\sim 0.25-0.35 \rm M_{\odot}$. As before, the same (coldest available) empirical spectrum is used down to $\sim 0.1 \rm M_{\odot}$.    
As in the half-solar metallicity bin, though less markedly here, the E-based RGB slope lies at slightly colder temperatures with respect to the theoretical one, towards the RGB-tip. 

The right-hand panels show the metallicity distributions per stellar phase that are input of our models. In this metallicity bin, after the inspection of the HR diagram revealed we could cover all phases of stellar evolution, we decided to use just only spectra with an exact solar metallicity from the atmospheric parameter fitting. 
\begin{figure*}
		\includegraphics[width=1.\textwidth]{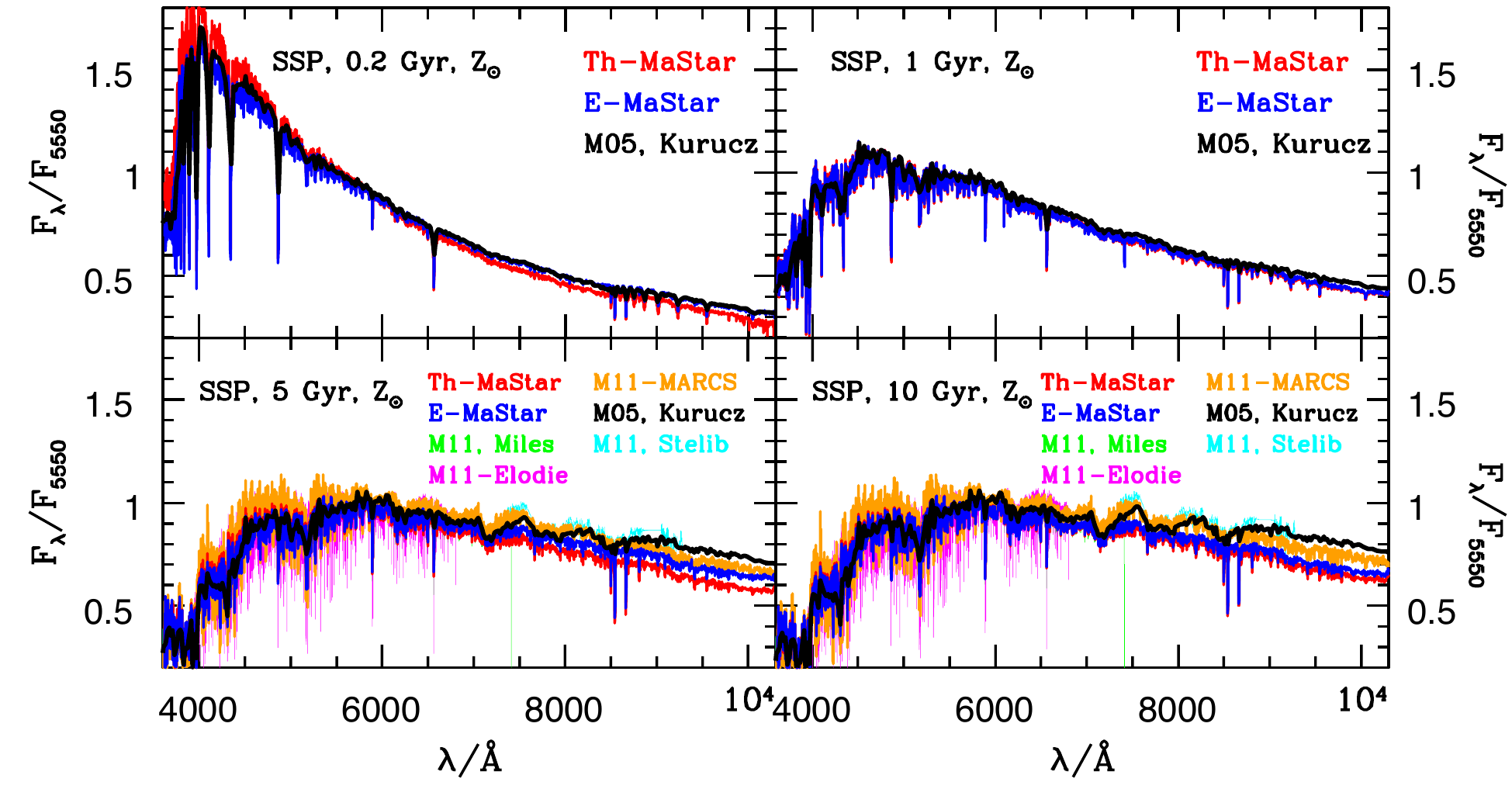}
	\caption{Integrated SEDs of MaStar-based stellar population models with solar metallicity $[Z/H]=0.$ as in Figure~\ref{fig:ssp_vp}.}
    \label{fig:ssp_sun}
\end{figure*}
The integrated SEDs of MaStar-based population models for solar metallicity are shown in Figure~\ref{fig:ssp_sun}, along with the other population models published in M11 for different input empirical (M11-MILES, M11-ELODIE and M11-STELIB as green, magenta and cyan line colours, respectively) and theoretical stellar libraries (M05 based on Kurucz and M11-MARCS, based on MARCS atmosphere models, as black and orange linestyles respectively).
As in the half-solar metallicity case, models better agree up to $\lambda \sim 7500$~Angstrom, after which some deviations are noted. As before, M05-Kurucz is the reddest model, but at this metallicity MaStar-models agree better especially at old ages and M11-MARCS lies in between. 

Again, the split of the integrated model SEDs per evolutionary phase shown in Figure~\ref{fig:phasessun} clarifies these discrepancies.

\begin{figure*}
  \includegraphics[width=\textwidth]{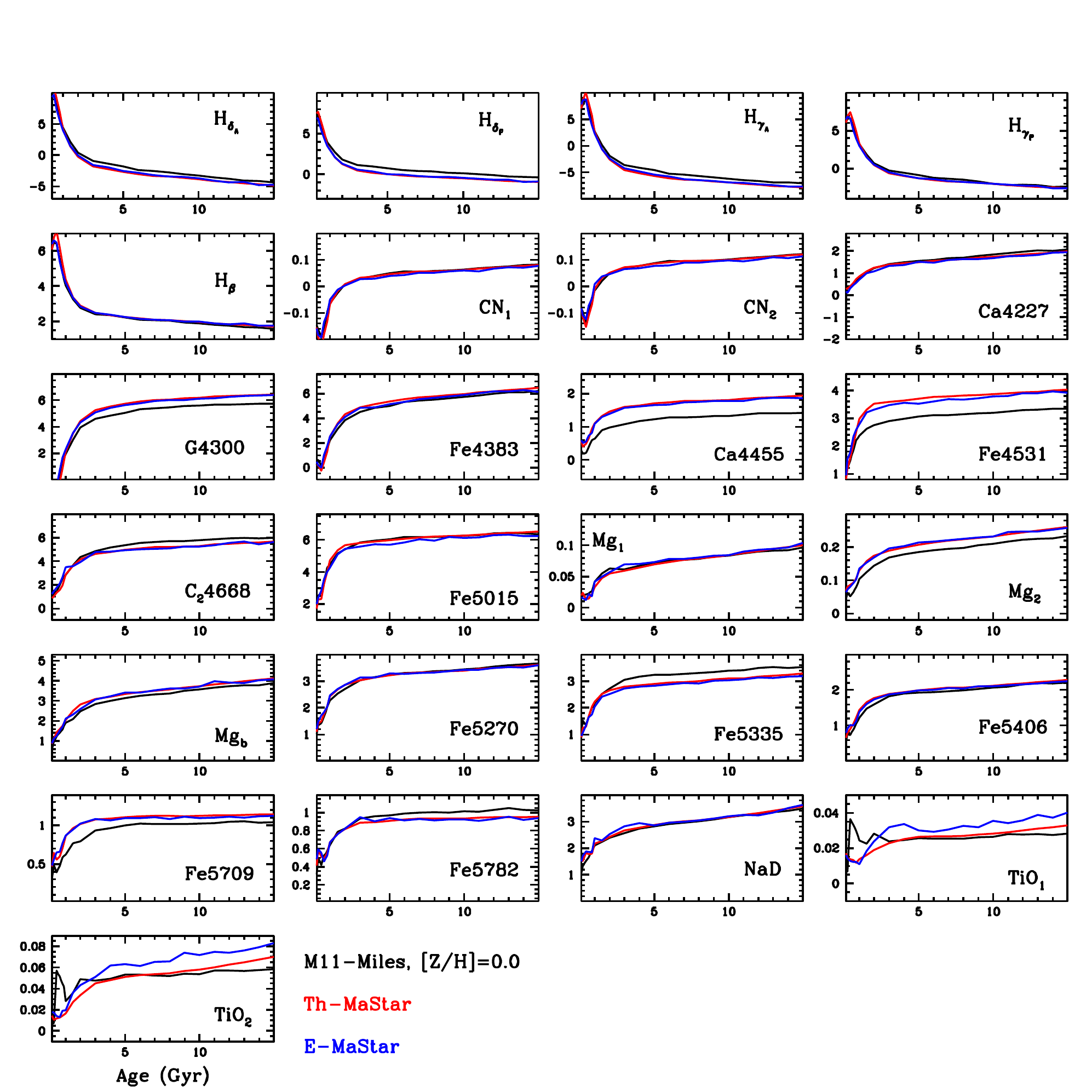}
  \caption{Comparison between the Lick line indices calculated on the two flavours of MaStar-based models (Th- and E-, depicted in red and blue as in previous plots) and those for the M11-MILES models (black), for solar metallicity $[Z/H]=0$. See Appendix~C for for the same plots for other metallicities.}
  \label{fig:licksun}
\end{figure*}

In Figure~\ref{fig:licksun} we compare the full set of optical Lick indices \citep{worthey_etal_1994,trager_etal_1998} calculated on the two flavours of MaStar-based models (Th and E) to those for the M11-MILES models (black). Overall there is a very good agreement between the two MaStar models for the vast majority of indices and a good agreement with those from M11-MILES, with the exception of the TiO indices. The TiO lines for the new MaStar models are stronger than the M11-MILES and also show a stronger age dependence. This effects is more significant at super-solar metallicity (see Figure~\ref{fig:lick_2sun} in Appendix~A). This is a good outcome as the model TiO indices were found to be too weak with respect to those of galaxies (i.e. high-metallicity systems) by \citet{thomas_etal_2003} and \citet{thomas_etal_2011} and due to this mismatch \citet{parikh_etal_2018} did not include these indices in their analysis of metallicity and IMF radial gradients in galaxies. It will be interesting to explore whether the new models will help the analysis of galaxy data.

\subsection{Super solar metallicity models $[Z/H]=+0.35$.}
\label{sec:supersolar}
Stellar populations at metallicity above solar are notoriously complicated to model due to the lack of local calibrators for the stellar evolution and model atmospheres. Yet, this is a crucial regime for studying the most massive (elliptical) galaxies, whose absorption lines point to super-solar metallicity either when using chemical modelling \citep[e.g.][]{thomas_etal_2005} as well as when just empirically comparing their absorption line-strengths to those of solar-metallicity Milky Way globular clusters \citep[][ Figure~1]{maraston_etal_2003}. In our Galaxy the most metal-rich stars are located into the Bulge, which is difficult to observe in the optical due to strong extinction. The SDSS telescope has little access to the MW Bulge due to its northern location.

\begin{figure*}
	\includegraphics[width=0.9\textwidth]{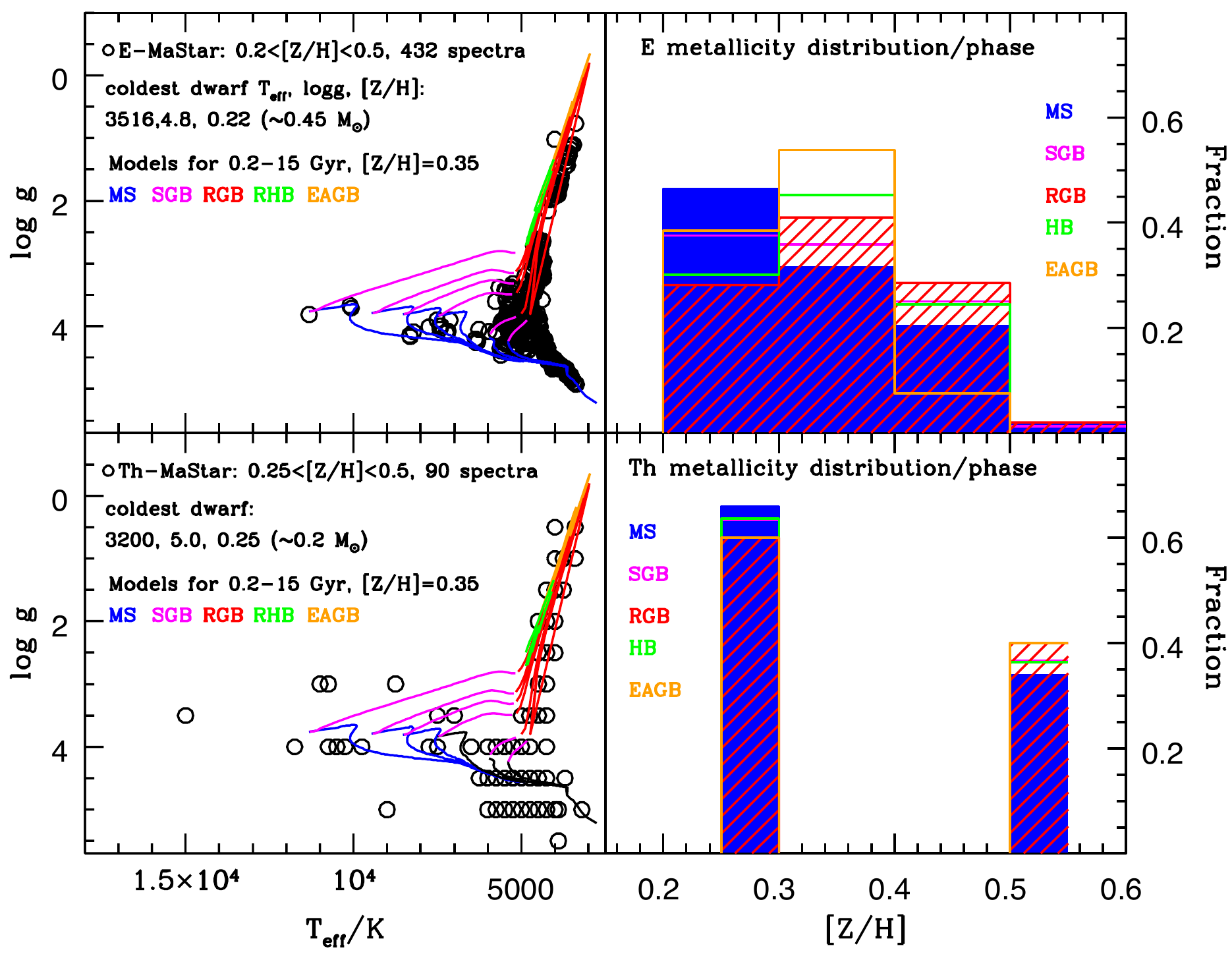}
	\caption{Theoretical HR diagram of $T_{\rm eff}$ vs $logg$ of MaStar stellar parameters as in Figure~\ref{fig:hr_verypoor} here for super solar metallicity $[Z/H]=0.35$. Th and E parameters allow the calculation of models down to $\sim 0.2 $~Gyr. In this metallicity regime, the coldest MS dwarf we cover is $\sim 0.2 \rm M_{\odot}$ with the Th set and $\sim 0.45 \rm M_{\odot}$ with the E set. Tip-RGB stars are missing, especially in the E-case. The right-hand panels show the metallicity distributions per stellar phase.}
    \label{fig:hr_2sun}
\end{figure*}

In spite of these complications, we are able to cover a wide range of parameters also at super-solar metallicity, allowing the calculation of population models down to $\sim0.2~\rm Gyr$, which is basically the range offered by MILES-based models (cfr. Table~1).
The distributions of MaStar stellar parameters in the theoretical HR diagram for super-solar metallicity is shown in Figure~\ref{fig:hr_2sun}.

Note how well both sets follow the lines defined by the theoretical models from the MS along the SGB and up to the RGB. This is a remarkable finding. No MaStar spectra is found with parameters lying to the right of the theoretical evolutionary sequences. The E set instead finds a large number of spectra with parameters lying to the right of the theoretical MS/SGB. Whether these low temperature values are real cannot yet be easily established, but future work on stellar parameters as well as the addition of more MaStar spectra will hopefully shed light on this problem.

The MaStar spectra metallicity distributions in the various stellar phases (right-hand panels in the same figure) are similar to previous cases, with the Th-one being by-modal by construction and the E-set stretching around the nominal value. 

Regarding the bottom end of the MS, the Th set allows the modelling down to $\sim 0.25 \rm M_{\odot}$ while the E set down to $\sim 0.45 \rm M_{\odot}$. Although we do not (yet) reach the core-H-burning limit of $0.1 \rm M_{\odot}$, these values are impressive and testify the innovative parameter coverage MaStar is able to offer.
\begin{figure*}
		\includegraphics[width=1.\textwidth]{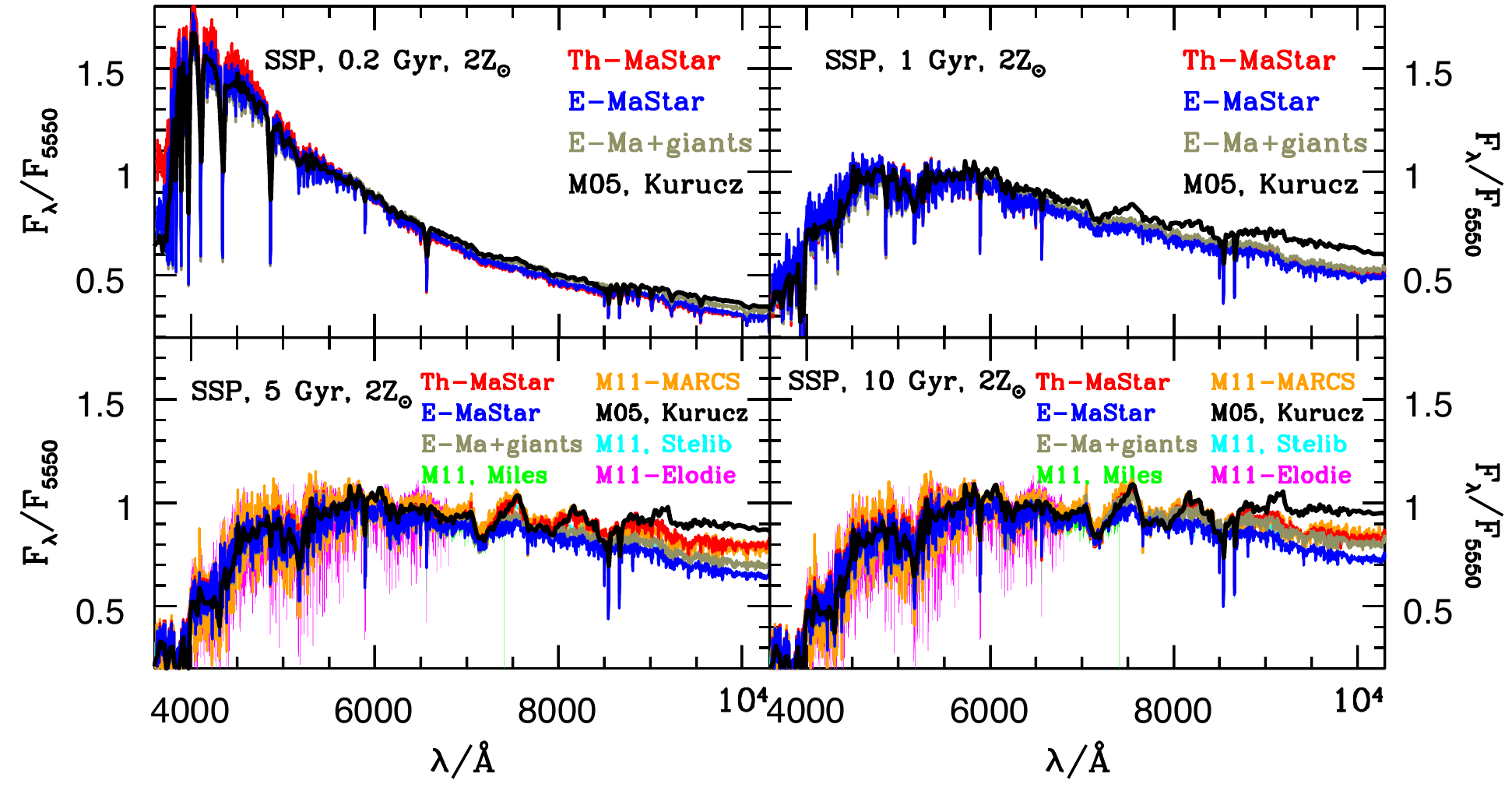}
	\caption{Integrated SEDs of MaStar-based stellar population models with super solar metallicity $[Z/H]=0.35$ as in previous figures (e.g. Figure~\ref{fig:ssp_sun}). Note the strong near-IR absorption bands visible in the Th-MaStar models (red lines), which are absent in the E-MaStar models (blue). To probe the cause of the difference, we calculated a version of E-MaStar models with additional RGB tip spectra - E-MaStar + giants models (grey). These models display more reasonable spectra.}
    \label{fig:ssp_2sun}
\end{figure*}
In spite of the different metallicity distributions of the two explored stellar parameter sets, the resulting stellar population model SEDs are consistent except in the old range regime ($t>1$~Gyr) where Th-Mastar models display strong absorption bands in the near-IR, very similar to the theoretical models M05-Kurucz and M11-MARCS (Figure~\ref{fig:ssp_2sun}). E-MaStar models instead are quite featureless at the same wavelengths (blue lines). In order to improve the E-MaStar model spectra, we perform the experiment of adding cool upper RGB spectra to the E-parameter set. The results are the grey models in Figure~\ref{fig:ssp_2sun}, which show that it was indeed the missing upper RGB spectra in E-MaStar the cause for the lower flux. This experiment highlights the importance of ensuring a full coverage of stellar phases including the extreme regions of the evolutionary diagram, for evolutionary population synthesis.
\begin{figure*}
\includegraphics[width=1.\textwidth]{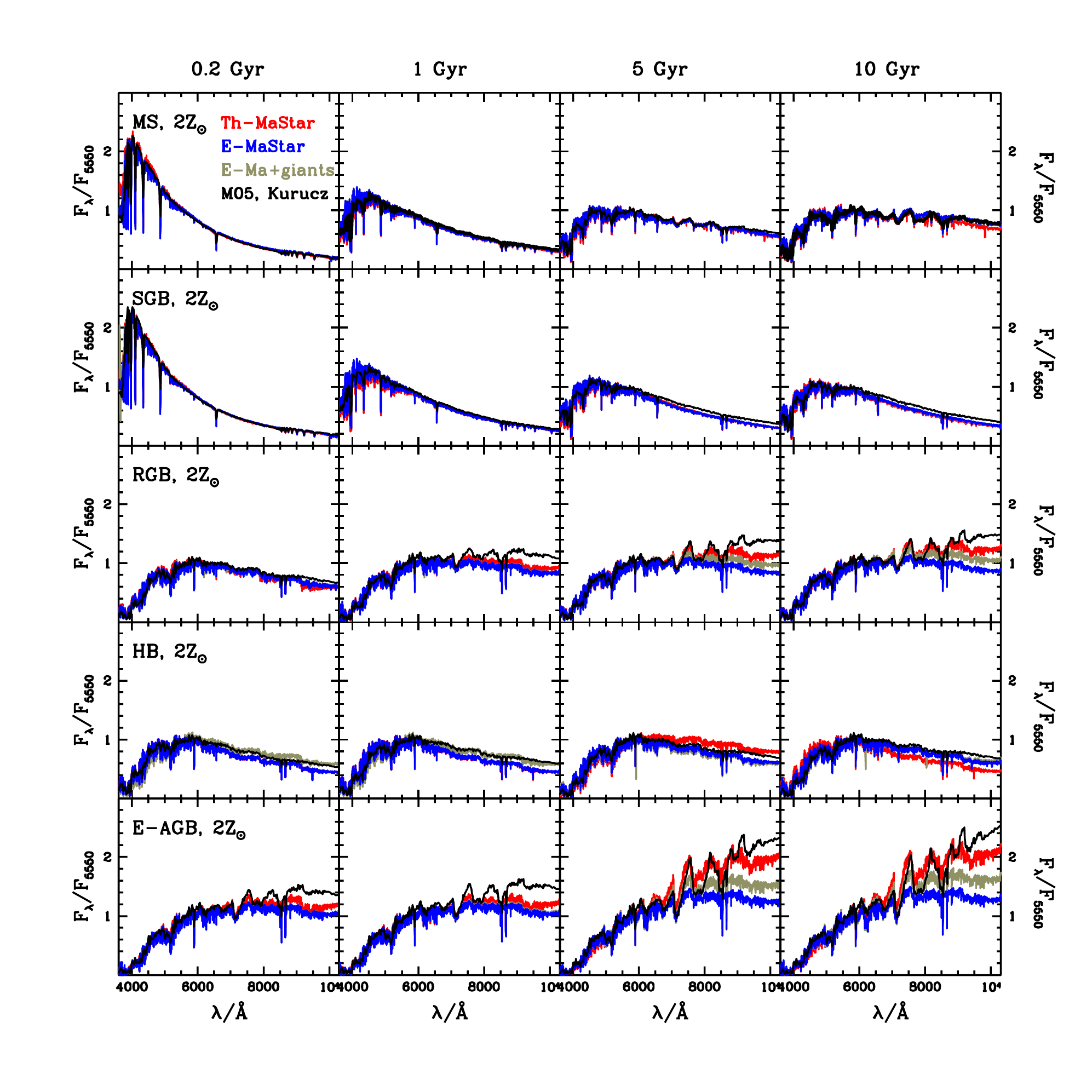}
  \caption{Integrated SEDs of individual stellar phases MaStar-models, as in Figure~\ref{fig:phasesvpoor} here for $2 Z_{\odot}$. In case of Th-MaStar, the SEDs for the giant phases RGB and E-AGB agree remarkably well with those for theoretical models based on MARCS or Kurucz atmospheres. E-MaStar models (blue lines) display a lower flux due to missing spectra with RGB-tip parameters. In the E-MaStar+giants models (grey lines) we added missing giant spectra and the result is a better agreement with Th-MaStar and other models.}
\label{fig:phases2sun}
\end{figure*}

The detailed comparison of the SEDs of stellar phases (Figure~\ref{fig:phases2sun}) confirms that the difference in the integrated model SEDs is due to the SEDs of the giant phases. Th-Mastar models display strong absorption bands of cold giants, which look amazingly similar to those from theoretical models. As MaStar is based on empirical stellar SEDs, this consistency is remarkable and informs us on the correctness of model atmosphere in the cold regime. E-MaStar models (blue lines) display a lower flux due to missing the coldest RGB-tip stars. E-MaStar+giants models (grey lines) where we added cool giants display a more reasonable spectrum. Note that from now onwards we shall use E-MaStar+giants as default E-MaStar models at supersolar metallicity. 

Finally, the modelling of the MS is instead consistent among the two MaStar models.

\begin{figure*}
  \includegraphics[width=\textwidth]{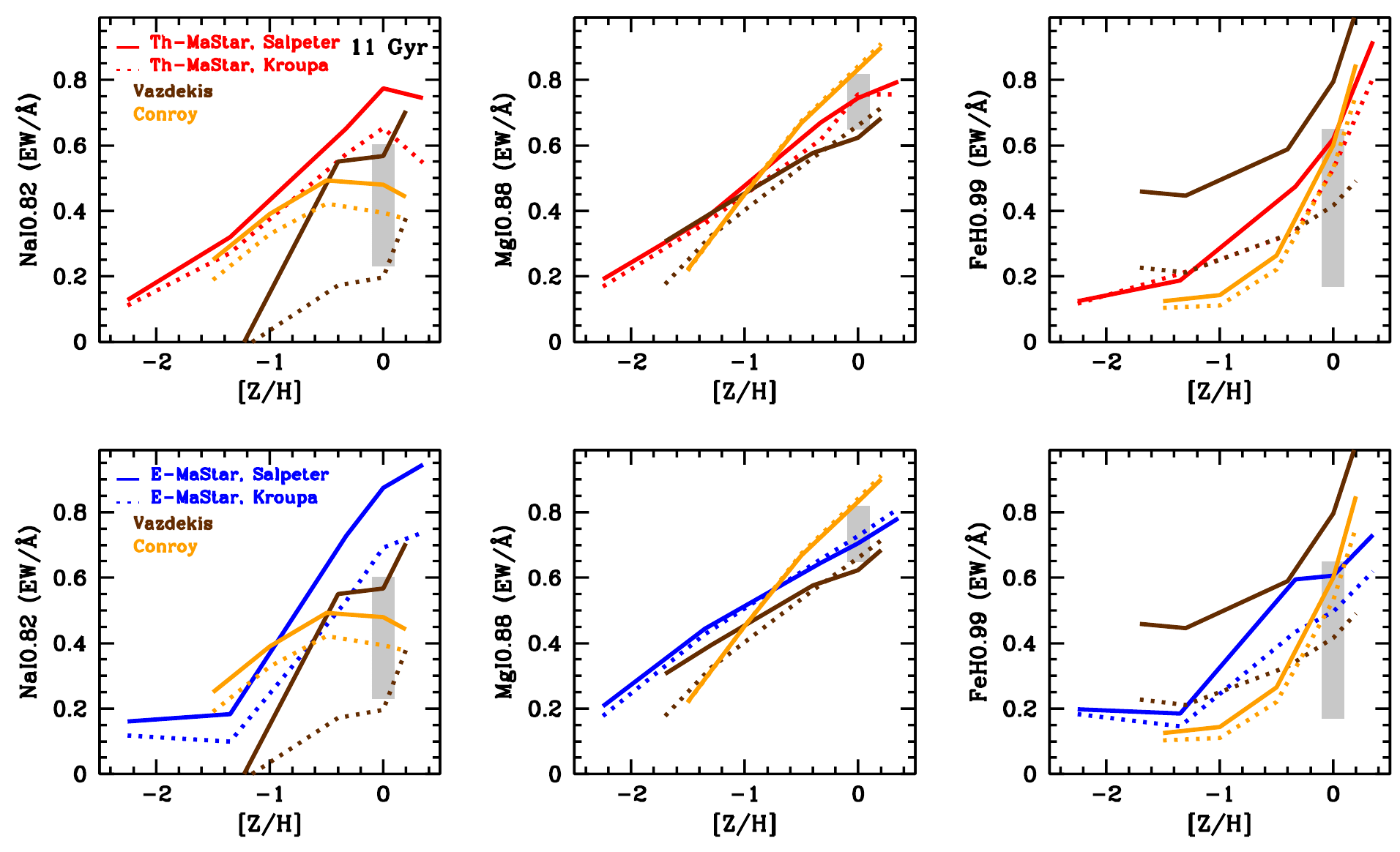}
  \caption{Model predictions for NaI0.82, MgI0.88 and FeH0.99 (from left to right), for Th- and E-MaStar models (upper and lower rows, respectively), as a function of metallicity, for population models with an age of 11 Gyr and Salpeter and Kroupa-like IMFs (solid and dotted lines, respectively). Models by \citet{vazdekis_etal_2016} and \citet{conroy_etal_2018} are shown in brown and orange for the same choice of IMFs and age, and for solar-scaled abundance ratios. Models have slightly different spectral resolutions, which however does not affect these comparisons (see text). For illustration, the grey areas show the range spanned by galaxy data from \citet{parikh_etal_2018}.}
  \label{fig:nearirindices}
\end{figure*}
\begin{figure}
  \includegraphics[width=0.49\textwidth]{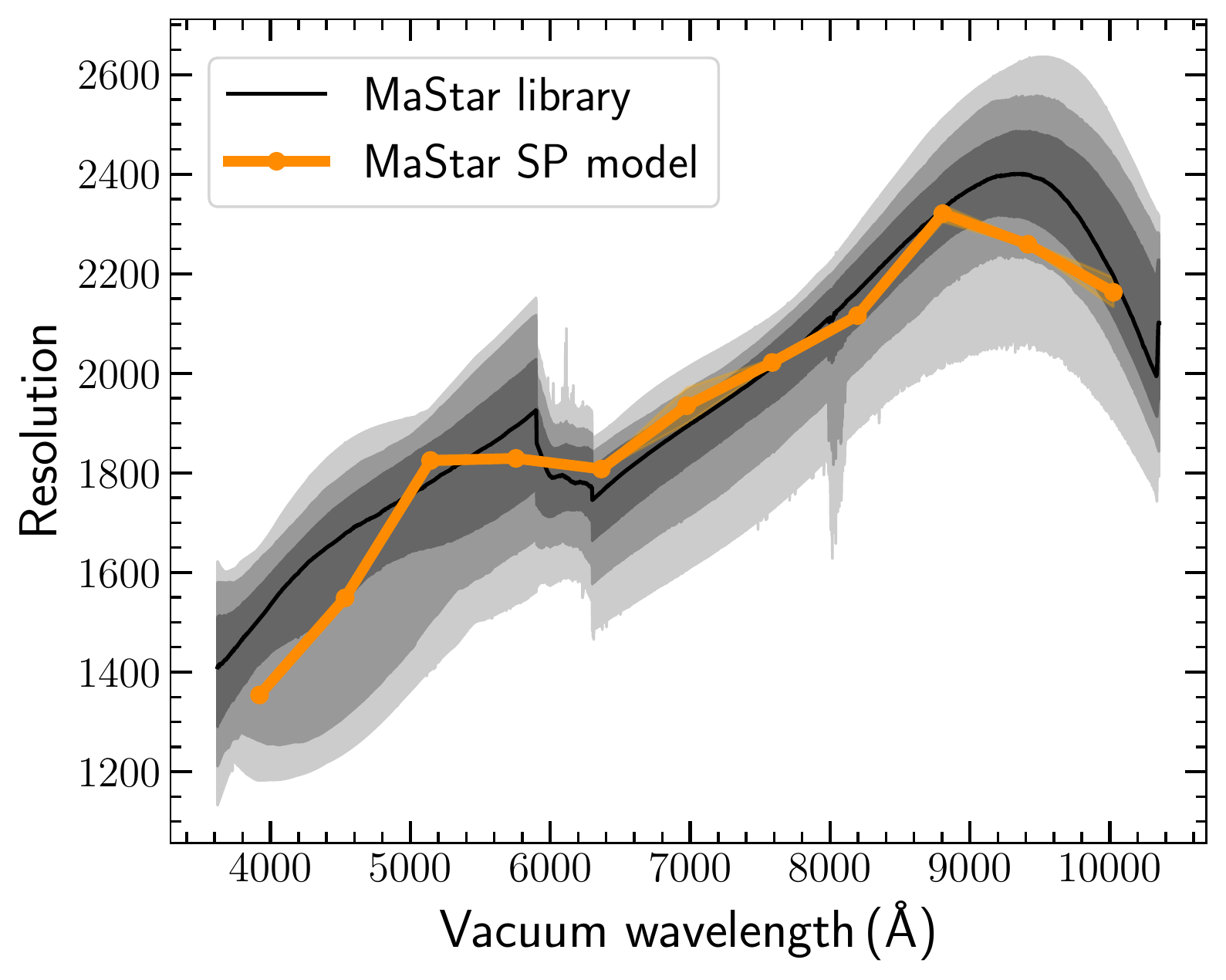}
  \caption{Median spectral resolution ($R=\lambda/\Delta\lambda$) as a function of wavelength ($\lambda$, in \AA), of MaStar stellar population models for a range of ages at solar metallicity. The grey stripes show the resolution distribution of the stars that were used to construct the models in percentiles of 68.3, 95, and 99.7 around the median, from the darkest to the lightest grey. The feature between $5900\,\text{\AA}$ and $6300\,\text{\AA}$ is due to an overlap of the blue and red camera and the feature at $\sim 8000\,\text{\AA}$ is due to slightly different pixel widths on the detector.}
    \label{fig:resolution}
\end{figure}

Line indices in the near-IR are affected by the modelling of the coldest dwarf and giant stars \citep[e.g.][]{schiavon_etal_2000,maraston_2005,conroy_2013,vazdekis_etal_2012}. It will be interesting to understand the behaviour of our new models in this particular wavelength region. In the next section we start examining a few selected features at the red edge of the MaStar wavelength range.

\begin{figure*}
  \includegraphics[width=0.49\textwidth]{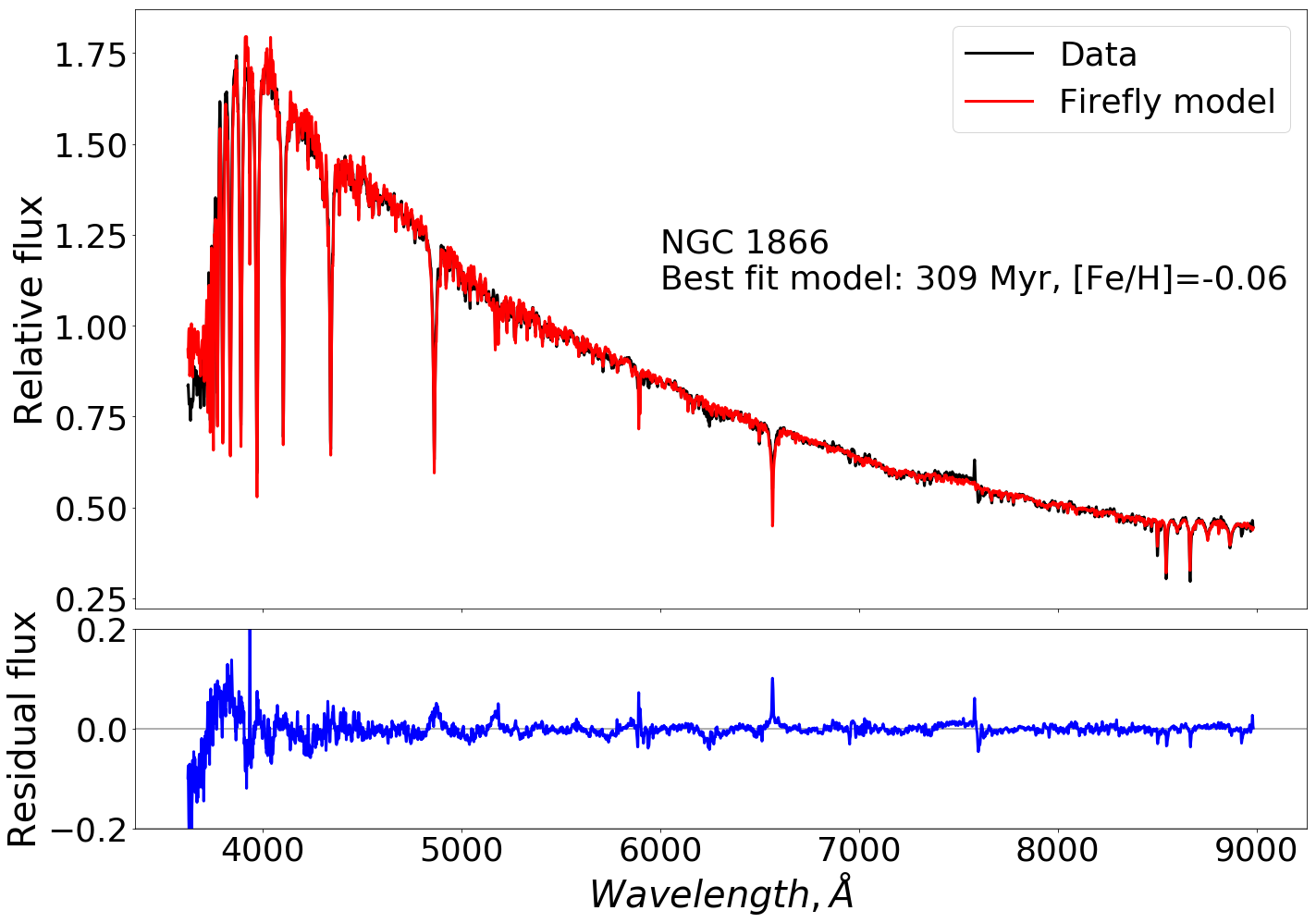}
    \includegraphics[width=0.49\textwidth]{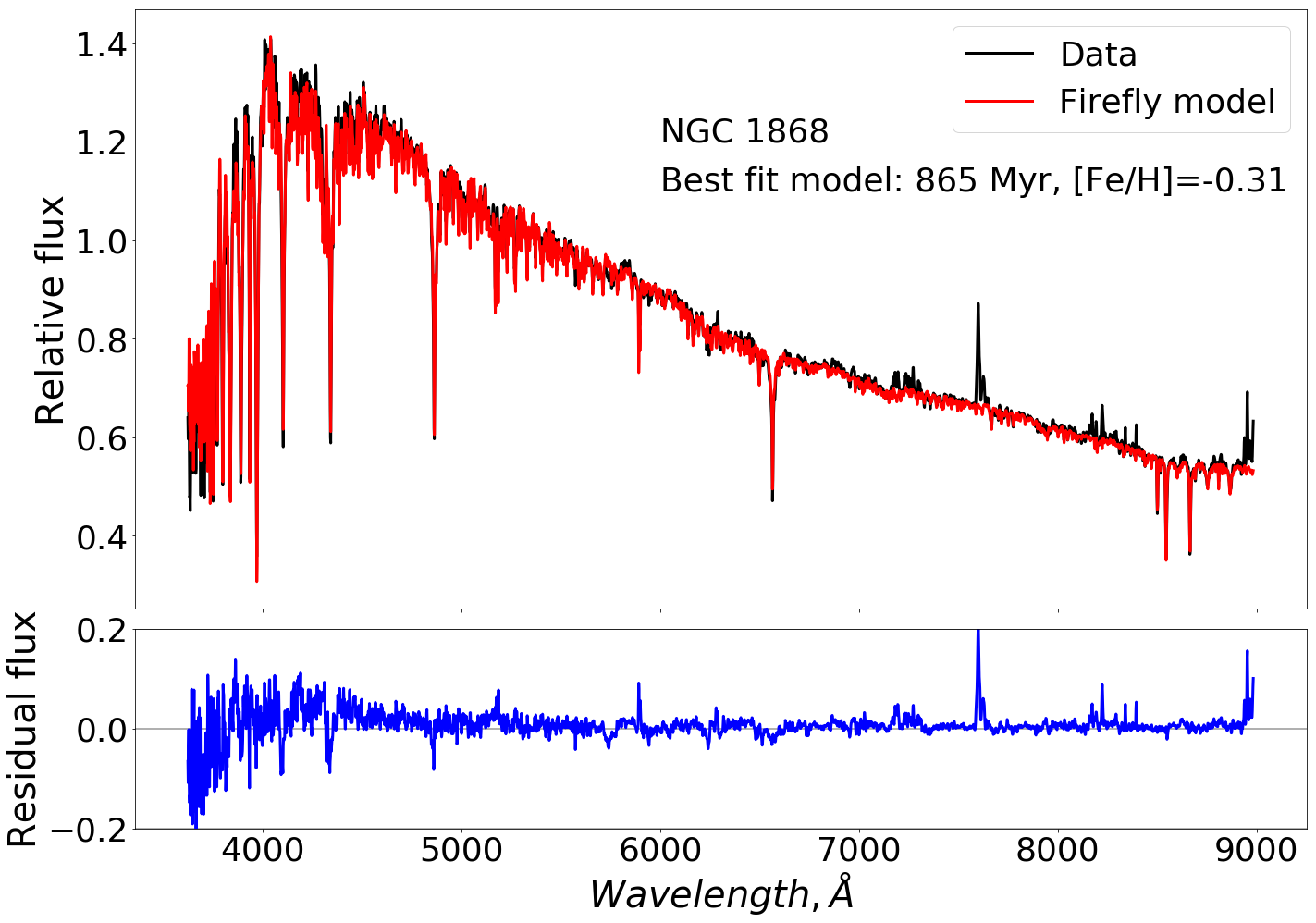}
 \includegraphics[width=0.49\textwidth]{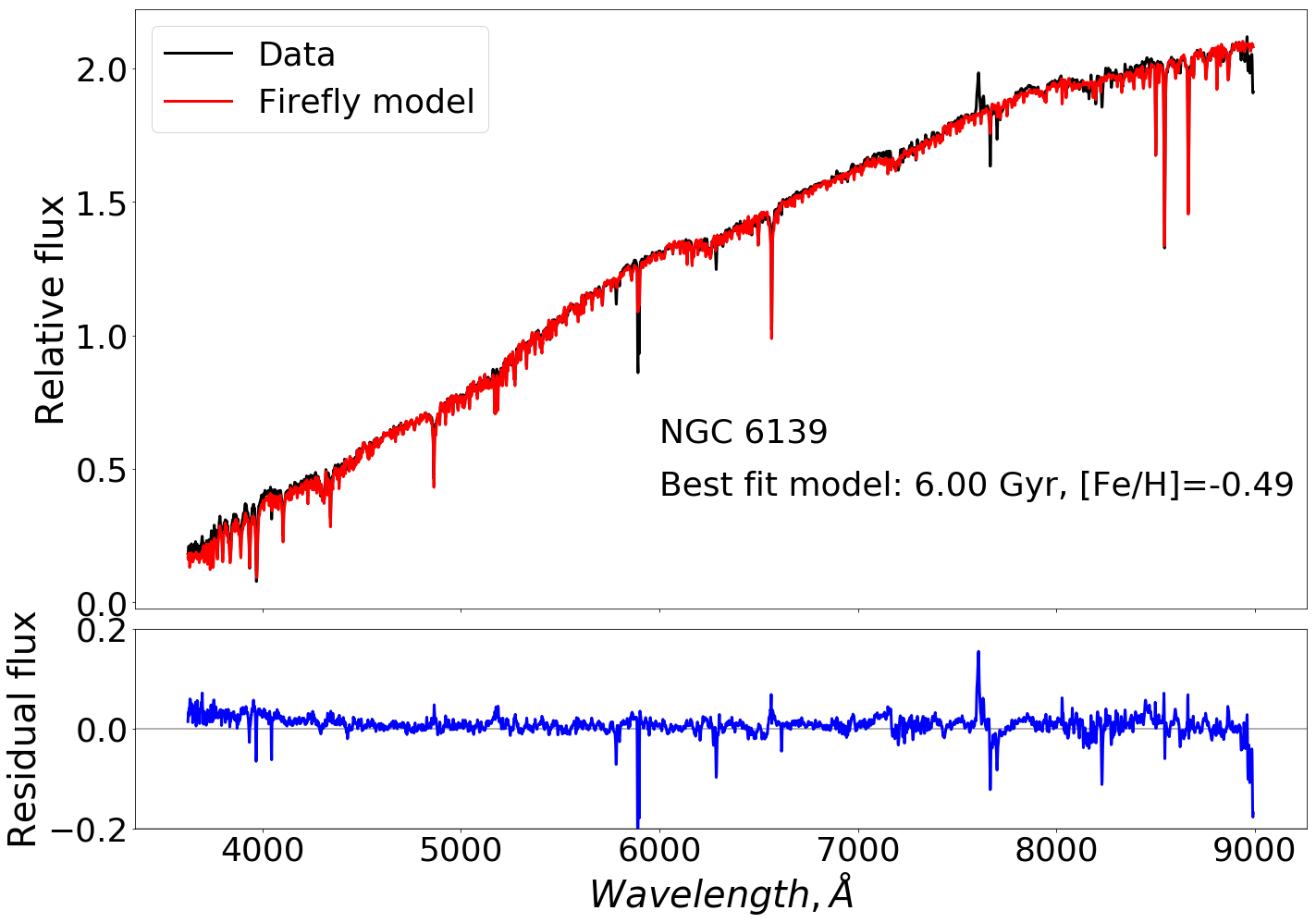}
    \includegraphics[width=0.49\textwidth]{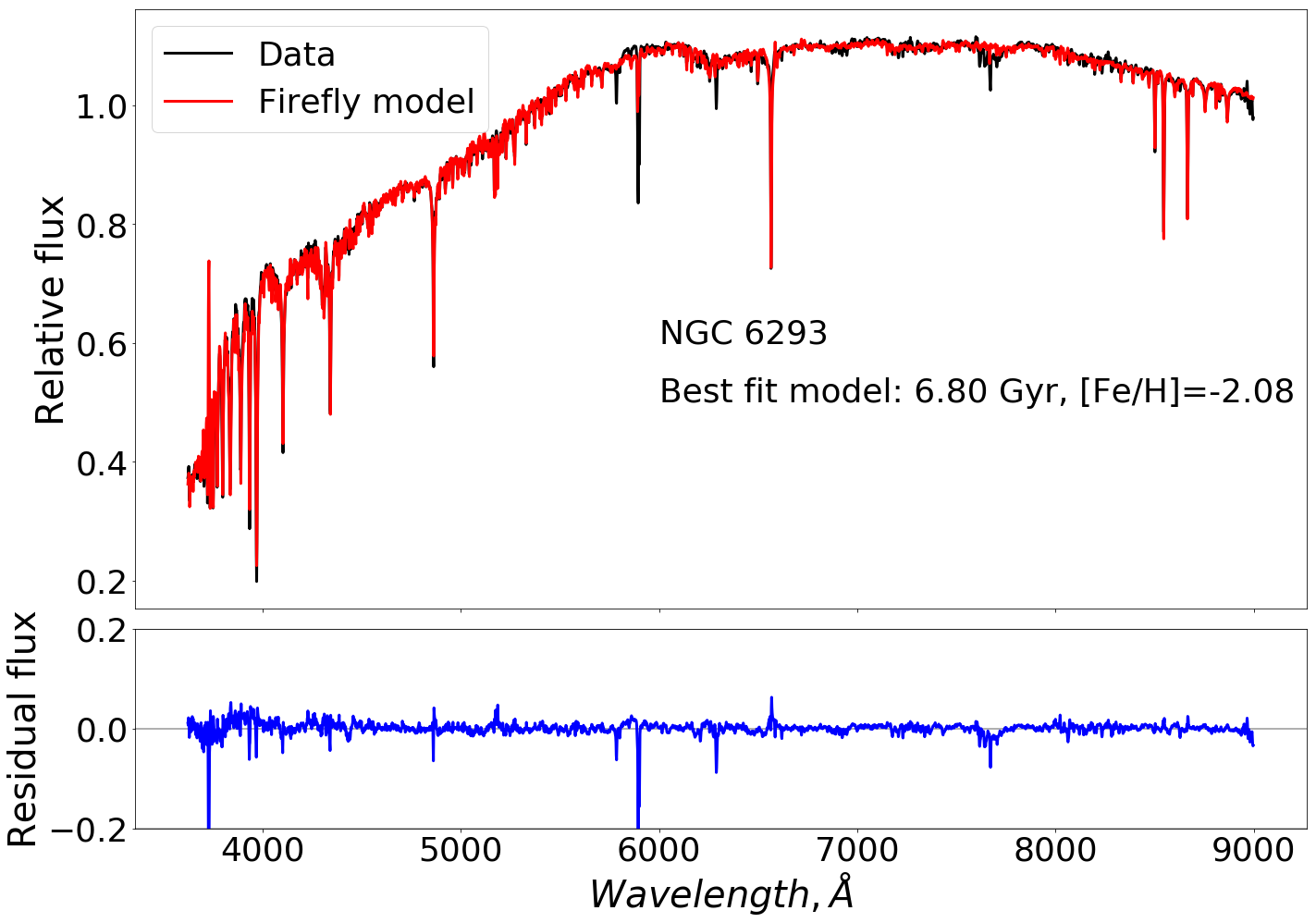}
 \includegraphics[width=0.49\textwidth]{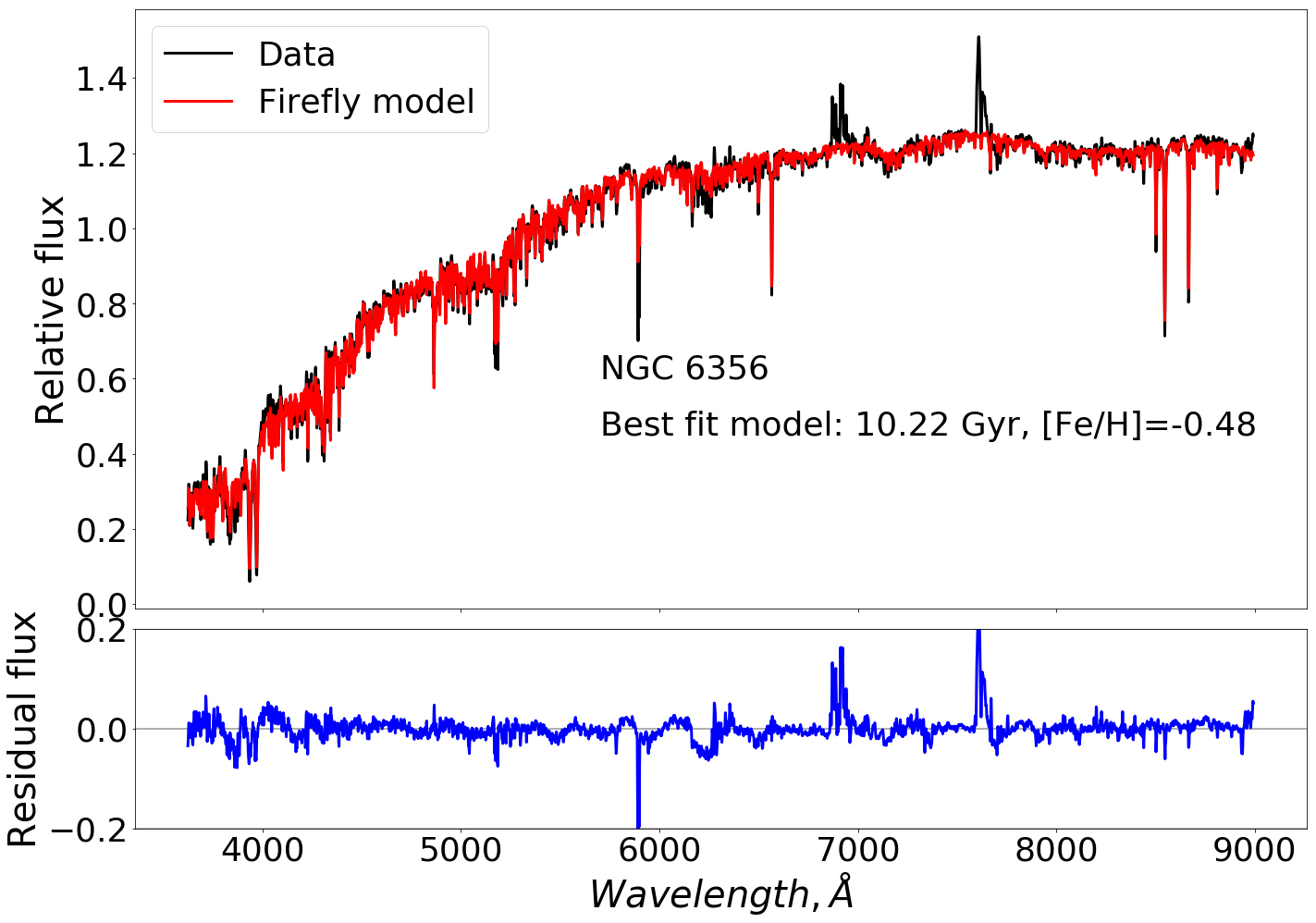}
    \includegraphics[width=0.49\textwidth]{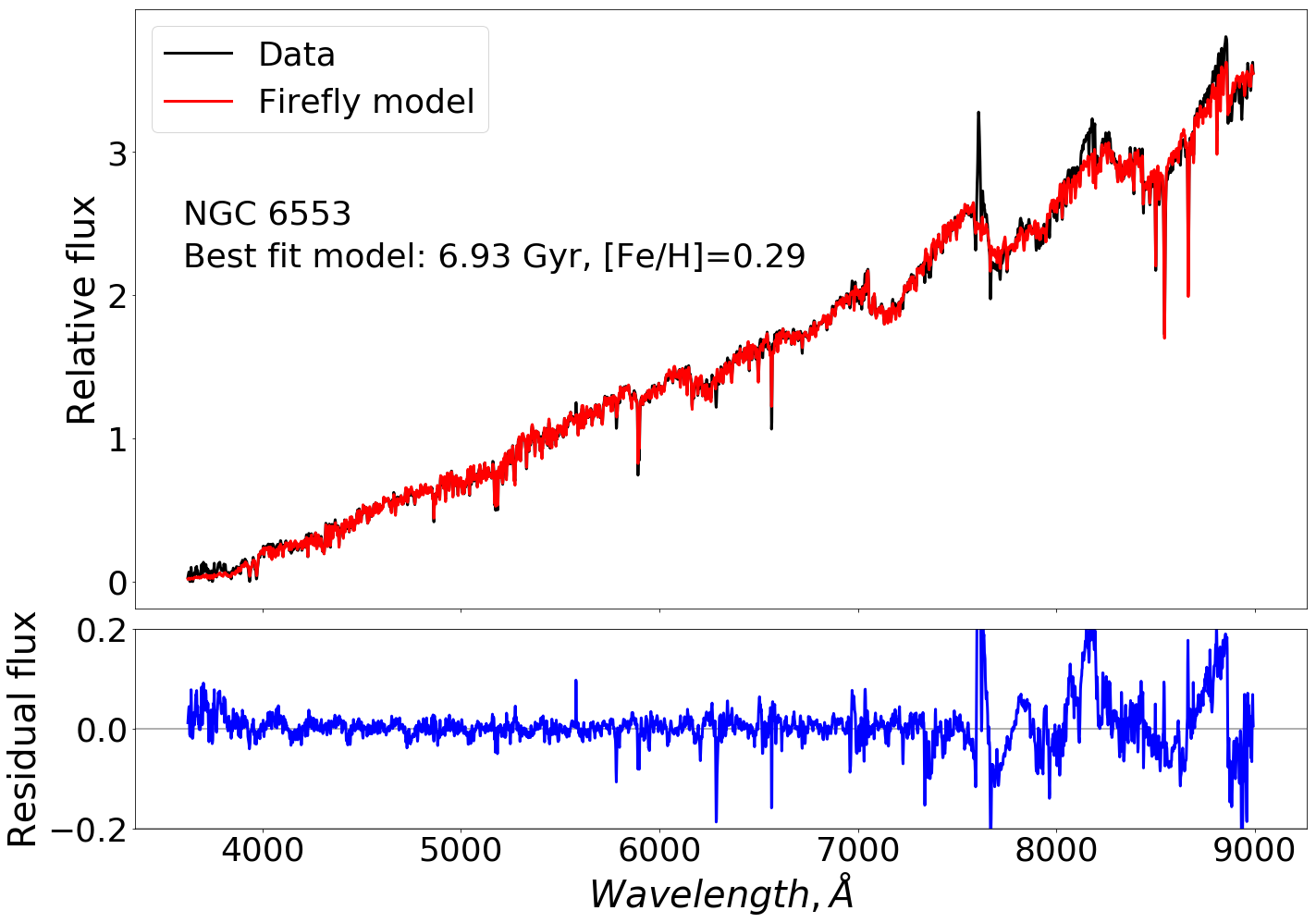}
  \caption{Examples of fittings of Th-MaStar models to GC observed spectra \citep{usher_etal_2017} for objects spanning a range in age and metallicity, from young clusters in the LMC (top row) through old metal-poor MW objects (middle row) to old metal-rich clusters in the MW Bulge (lower row). Mass-weighted best-fit values for age and metallicity are labelled on each plot. The full spectral fit was performed with the Firefly code \citep{wilkinson_etal_2017}.} 
  \label{fig:gc-fits}
\end{figure*}
\subsection{Near-IR spectral features.}
\label{sec:nearirindices}
The wavelength extension of the MaStar stellar library up to $10,000$~\AA\ allows the modelling of spectral features that have seen a renewed interest as IMF indicators for stellar populations, after \citet{conroy_and_van_dokkum_2012}  \citep[and several others afterwards, see][for a comprehensive reference list]{parikh_etal_2018}. These are for example the Sodium doublet at 8200~\AA~(hereafter NaI0.82),  the Magnesium feature at 8800~\AA~\citep[][hereafter MgI0.88]{diaz_etal_1989,conroy_and_van_dokkum_2012} and the FeH band at 9900~\AA~\citep[or Wing-Ford band,][hereafter FeH0.99]{whitford_1977}, just to name a few that have been widely used in the recent literature. 

In Figure~\ref{fig:nearirindices} we provide a first look at how the new MaStar population models behave with respect to these three features and compare them with models by \citet{vazdekis_etal_2016} and \citet{conroy_etal_2018}. For all three lines, we use the index definitions from \citet{conroy_and_van_dokkum_2012} (Table~1), which refer to wavelengths in vacuum. 

Results for Th- and E-MaStar are reported in the upper and lower rows, respectively, as red and blue lines, where solid and dotted linestyles refer to the Salpeter and Kroupa IMFs, respectively. Models by \citet{vazdekis_etal_2016} and \citet{conroy_etal_2018} are shown in brown and orange for the same choice of IMFs and age and for solar-scaled abundance ratios. In evaluating these comparisons, it should be noted that the models have a somewhat different resolution, namely the Vazdekis et al. models have a slightly higher ($R\sim2100-2500$) and the Conroy's et al. models a lower ($R=1270$) resolution as the MaStar models ($R\sim 2000$) at these wavelengths (see Figure~\ref{fig:resolution}). However, we evaluated that the correction factors are negligible compared to the difference between the models~\footnote{We evaluated the correction factors due to resolution by downgrading both the MaStar models as well as the high-resolution M11-MARCS ($R=10,000$) models with solar metallicity to the Conroy's models resolution and re-measured the indices. In both cases, we found correction factors that were negligible for NaI0.82, 6\% for MgI0.88 and 3\% for FeH0.99. For the Vazdekis models' resolution these corrections would be even smaller.} hence we are allowed to compare the indices at the original resolutions of the models.

The grey area shows the range spanned by galaxy data from \citet{parikh_etal_2018}. Data refer to spatially-resolved spectra out to the half-light radii for early-type galaxies with stellar masses in the range $10^{9.9}<\rm M^{*}/M_{\odot}<10^{10.8}$, from the MaNGA survey \citep{bundy_etal_2016}. \citet{parikh_etal_2018} perform the full stellar population analysis of the spectra, deriving total metallicities $[Z/H]$~in the range -0.1 to 0.1, which define the size of the shaded area along the x-axis. 
  
Commenting on each individual line starting from NaI0.82 (left-hand panels), a large variation among model predictions is evident. Line-strengths of Th-MaStar models (red) increase with metallicity, displaying a mild saturation around solar metallicity. Models assuming a Salpeter IMF with a larger proportion of low mass stars over those for a Kroupa IMF have larger line-strengths, in agreement with what found in other models 
\citep[see discussion in][]{conroy_and_van_dokkum_2012}. E-MaStar models (lower panels, blue lines) show a similar trend however with somewhat larger values of the index. This is due to a deficit flux from the giant component of the population in these models. As discussed earlier (see Figure~\ref{fig:ssp_2sun}) we noticed that the empirical parameters fed in the E-MaStar models lack the coldest part of the RGB and we added giant spectra in order to improve the E-MaStar spectra. However, when compared to Th-MaStar the giant contribution in E-MaStar models remains lower, as can be seen in Figure~\ref{fig:phases2sun} and this explains why the NaI0.82 of E-MaStar is larger.\footnote{Note that in the figure we only plot the indices of E-MaStar+giants models. For the standard E-MaStar models the NaI index would be even larger.} 
 
This is a point to keep in mind when using this index to draw conclusions on a galaxy IMF. Indeed, we demonstrate that an underestimation of the correct energetics for the giant population acts in boosting the index without being connected to a change in the IMF.

As for the literature models, we see that the \citet{conroy_etal_2018} models display a trend with metallicity which is similar to the MaStar models, including a saturation at high metallicity, but the index values are substantially lower. 
The \citet{vazdekis_etal_2016} models on the contrary show a sharp increase with metallicity, lying to very low values for sub-solar metallicities, do not saturate at high [Z/H] and have lower index values than the MaStar models. The Vazdekis et al. models also suggest a large effect from the two assumed IMFs, which is absent in ours and Conroy's models. 

Turning to MgI0.88 (central panels), we see that all models steadily increases with metallicity while being quite insensitive to the IMF choice \citep[as known, see][]{cenarro_etal_2009,conroy_and_van_dokkum_2012}. Th- and E-MaStar models behave very similarly for this line and cross the area covered by the galaxy data. Models by Conroy et al. display a steeper slope with metallicity and lie above the galaxy area, whereas models by Vazdekis et al. display a shallower slope and lie below the galaxy area. Again here, models by various author disagree, even when they are based on the same stellar evolutionary tracks (as is the case for the Vazdekis and Conroy's models). Clearly, the stellar spectral input plays a crucial role. 

Finally, quite a variance is also found among model predictions for the FeH0.99 band (right-hand panels), with trends similar to those seen for the other indices, namely a stronger dependence on the IMF for Vazdekis et al. models, and much larger indices for the Salpeter IMF case in these latter models, and more consistent values between MaStar and Conroy models at high metallicity, particularly in case of Th-MaStar.

In conclusion, it is clear that these indices are quite affected by the details of the modelling, as also concluded in \citet{parikh_etal_2018}. Due to the complexity of the problem and the space needed to properly address it, we move an in-depth discussion of this spectral region to a separate paper. 

\subsection{The wavelength-dependent spectral resolution of MaStar models.}
\label{sec:specres}
For a correct application of models to data, the spectral resolution of the models need to be known. In this section we derive and discuss the spectral resolution of our MaStar-based population models and demonstrate that it is the same as the input stellar spectra.

The measurement of the spectral resolution of the MaStar stellar population models follows mostly the same analysis that is described in detail in \citet{beifiori_etal_2011}. Here, we summarise the basic procedure.
To determine the resolution of the model templates, we conducted a full-spectral fitting with the Penalized Pixel-fitting method (pPXF) of \citet{cappellari_and_emsellem_2004} employing the theoretical stellar library MARCS. To be able to measure the line broadening in the spectra across the whole wavelength range, it is necessary to use an input library that covers at least the same spectral range at a higher resolution. The MARCS library complies with these criteria covering wavelengths from $1\,300\,\text{\AA}$ to $20\,000\,\text{\AA}$ at a constant high resolution of $R = 20\,000$ \citep{gustafsson_etal_2008}.

We chose 13 MaStar model templates of solar metallicity covering stellar ages from $\rm 3\,Gyr$ to $\rm 15\,Gyr$. Stellar populations younger than $\rm 3\,Gyr$ are excluded in order to avoid template mismatch with MARCS, which does not include stars hotter than $\rm 8\,000\,K$. As template input we chose a subset of stellar spectra in MARCS covering all effective temperatures between $\rm 2\,500\,K$ and $\rm 8\,000\,K$, as well as six evenly spaced surface gravities between $-1$ and $5$. With this set of templates we fitted the kinematics of each of the aforementioned models in eleven equally sized wavelength segments from $3\,633\,\text{\AA}$ to $10\,352\,\text{\AA}$.

The best fit is determined by $\chi^2$ minimisation in pixel space from a non-negative linear combination of stellar templates convolved with a Gaussian function that accounts for the broader lines of the MaStar models as compared to the MARCS templates. We did not broaden the input spectra prior to the fit to match the resolution as it would be done if we were to measure the kinematics of real observations. Instead, the sigma of the Gaussian from the best-fitting solution was used afterwards to infer the difference in resolution between MaStar and MARCS templates. For the fits, we chose to add a fourth order polynomial to account for the differences in continuum shapes. The robustness of this choice was tested and the variation of the results was found to be smaller than the spread between different models. The final resolution was obtained by adding in quadrature the resolution of the MARCS templates to the relative broadening measured from the fits.

In Fig. \ref{fig:resolution}, we show the median resolution per wavelength segment from the distribution of all fitted model templates. The orange shades give the median absolute deviation from the median. We compare our results to the original resolution of the stars that were used to construct the models. The spectral resolution can vary between each star and each observation and, thus, we show here the median resolution per wavelength from all 908 stars that entered the models\footnote{It is important to keep in mind that only a selection of stars enter each of the model templates.}. We refer to \citet{yan_etal_2016, yan_etal_2019} for details about the determination of the line spread function of stellar spectra in MaStar. The comparison in Fig. \ref{fig:resolution} between the models and the stars clearly shows that there is no significant loss of spectral resolution as a result of our population synthesis. The slight difference is likely due to template mismatch in the pPXF fitting. the median resolution curve of the stars as the resolution for the SSP models. 

When using our MaStar stellar population models, we recommend to adopt the wavelength-dependent resolution vector we make available at the model web site http://www.icg.port.ac.uk/MaStar/, which is based on the median resolution of the stars.  
\section{Testing the models with globular clusters}
\label{sec:testing}
Even at this initial stage of the MaStar library release it is instructive to examine how the new MaStar-based models perform when they are used to derive ages and metallicities of real stellar populations. The most useful test in this regard is to use Milky Way globular clusters for which ages and metallicities are determined independently of stellar population models, e.g. from Colour Magnitude Diagram (CMD) fitting for age and resolved spectroscopy for metallicity. We have been performing these tests initially suggested in \citet{fusi_alvio_1988} in all our past publications \citep[e.g.][]{maraston_1998,maraston_etal_2003,thomas_etal_2003, maraston_2005,maraston_and_stromback_2011,thomas_etal_2011}. With the new models we complete similar tests to those shown in M11 and \citet{wilkinson_etal_2017} where we used GC spectra from \citet{schiavon_etal_2005} and performed full spectral fitting on the whole wavelength range allowed by data (M11, Fig. 22 and 25-28; Wilkinson et al. 2017, Fig. 20, 21). 

In the present paper we use newly published observed spectra for GCs by \citet{usher_etal_2017}. This database offers a wider wavelength range with respect to older data, which nicely matches the one of the new MaStar models. Also the spectral resolution of these data is comparable to the MaStar one. 

Furthermore, an improvement over the tests performed in W17 is that we are now able to check the model age scale over a wider range with respect to the 'old' regime that is allowed by MW GCs, as the Usher et al. database also includes a few young/intermediate age objects from the Small and the Large Magellanic Cloud (SMC and LMC). As our current MaStar models do not extend to ages lower than $0.2-0.3~\rm Gyr$~(cfr. Table\ref{tab:grid}), we exclude those objects from the model fitting whose literature ages are younger than 0.3 Gyr. As a consequence, out of the 83 GC spectra published by Usher et al., we include 71 in our testing. For these objects, ages and metallicities have been determined using a variety of techniques/data sources. The Usher et al. work provides us with a compilation of heterogeneous literature determinations. In addition, we use homogeneous ages and metallicity determinations when available for sub-samples of GCs, as in M11 and W17. 
The model fitting to the observed spectra is performed by using our full spectral fitting code Firefly \citep{wilkinson_etal_2017} as in W17, Figure 22. In particular, no prior other than the model grid is adopted and the observed spectra are used in their full wavelength range.

In the following we first provide a qualitative comparison between the ages and metallicities derived from model fitting and those available in the literature, exploring various literature sources. We then proceed to more quantitative comparisons by showing histograms of age and metallicity differences and statistics calculated on these quantities. 
Our scope for this paper is to understand whether the models perform reasonably at this stage and whether we can single out any merit from the adopted stellar parameters. A more extensive testing with GC spectra, including assessing the effect of wavelength range, stellar library and literature models is the matter of another paper. 
\begin{figure*}
  \includegraphics[width=0.49\textwidth]{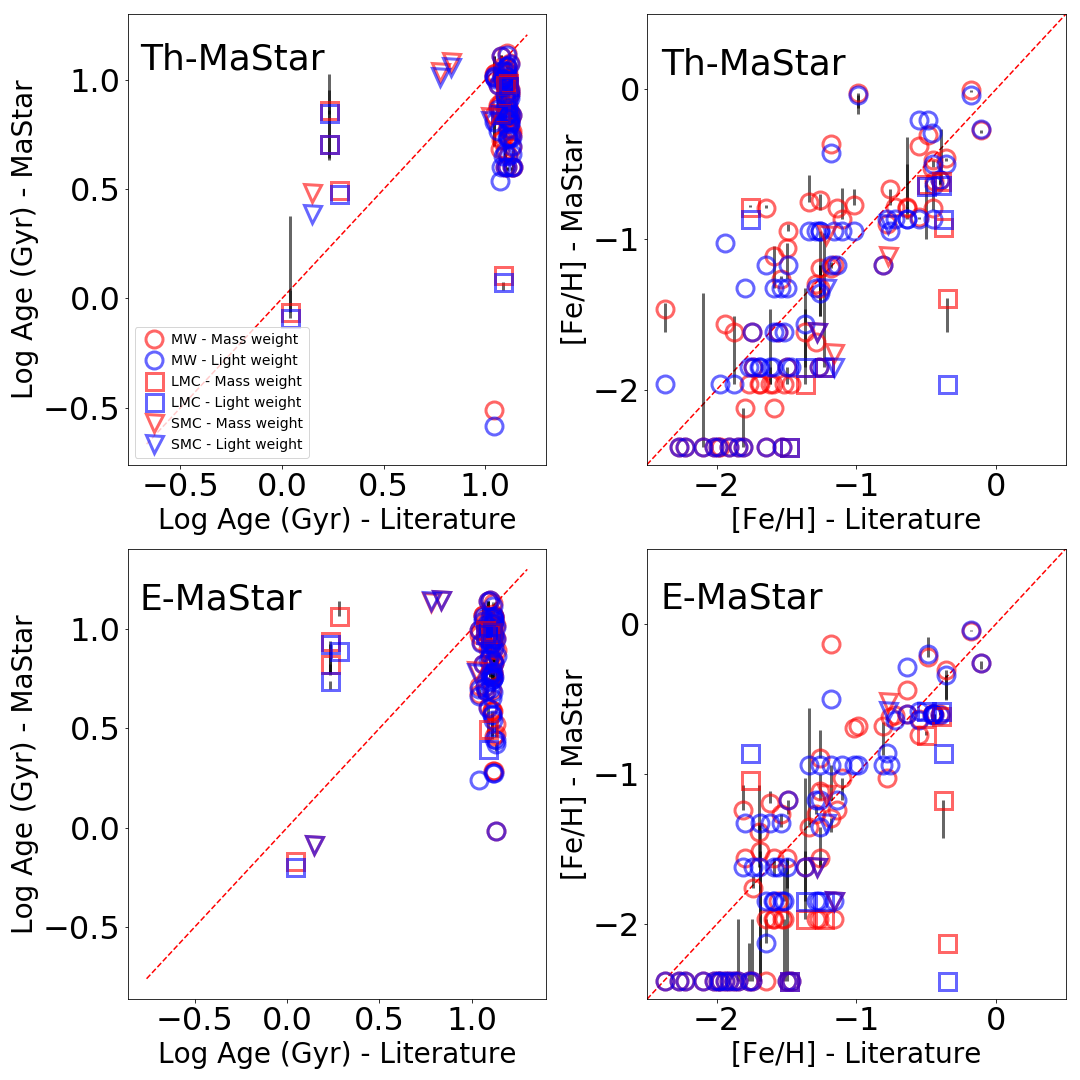}
  \includegraphics[width=0.49\textwidth]{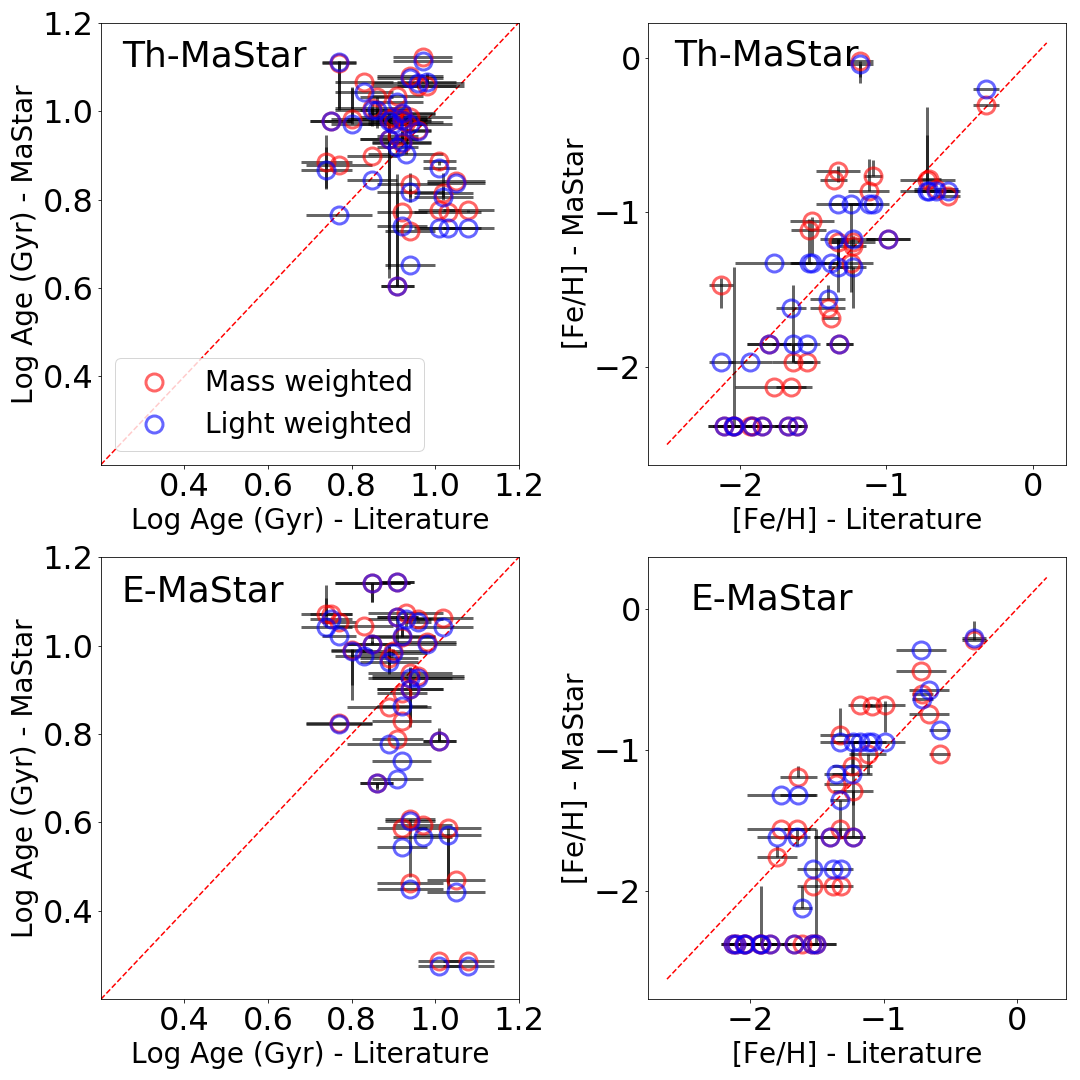}
  \caption{Comparison of ages and metallicities obtained by fitting Th-MaStar and E-MaStar models (upper and lower rows, respectively) to observed GC spectra from \citet{usher_etal_2017} with the Firefly spectral fitting code \citep{wilkinson_etal_2017}, with literature values. Different symbols denote MW, LMC or SMC GCs. Mass-weighted and light-weighted quantities are distinguished through red and blue colours, their errors shown as black lines. Literature values are from the compilation by Usher et al. for the left-hand set of plots, and from \citet{deangeli_etal_2005} for the right-hand plots. The cross-match between the Usher et al. and the De Angeli et al. databases gives us 34 GCs. This database only includes MW GCs.}
  \label{fig:gc}
\end{figure*}
Figure~\ref{fig:gc} compares the ages and metallicities we obtain for the Usher et al. sample of GCs by fitting Th-MaStar and E-MaStar models (for a Kroupa IMF) to the observed spectra with the Firefly full spectral fitting code, to available literature values. Different symbols distinguish whether the objects are from the MW, LMC or SMC. Light-weighted and mass-weighted fitting results are shown in blue and red, respectively. 
\begin{figure}
  \includegraphics[width=0.5\textwidth]{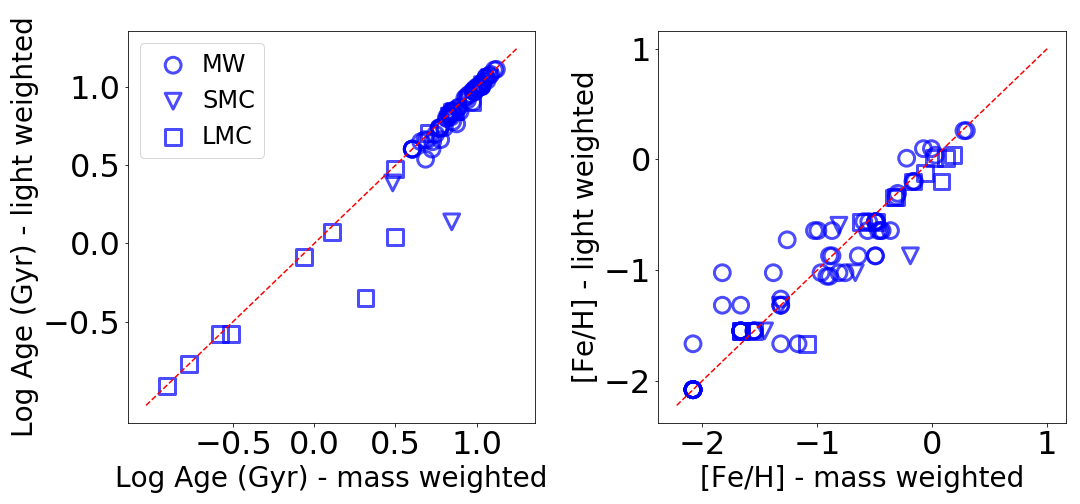}
  \caption{Comparison between mass-weighted and light-weighted ages and metallicities (left-hand and right-hand panel, respectively) for the fitting results of the Usher et al. GC observed spectra shown in Figure~\ref{fig:gc}.}
  \label{fig:gc-lightmass}
\end{figure}

Metallicities derived in the literature are generally tied to scales representing the Fe abundance [Fe/H] rather than the total metallicity [Z/H] the models would like to trace. In addition, several MW GCs have $\alpha$-enhanced chemical compositions, implying the [Fe/H] is offset from the actual metallicity. In order to make a more meaningful
comparison, as in W17 we shifted the model fitting values by $-0.3$~dex, which corrects [Z/H] to [Fe/H] for a $\alpha$-enhancement value around $[\alpha/Fe]=0.3$,
using the scaling by Thomas et al. (2003). In the plots we shall use the notation [Fe/H] when we compare model fittings to GC data just in order to maintain the notations that are adopted in those paper.

As literature values, we use the compilation from Usher et al. for the left-hand set of panels, and the homogenous compilation of ages from \citet{deangeli_etal_2005} for MW GCs (i.e., LMC and SMC objects are not included in this plot) for the right-hand panels. These were obtained by fitting isochrones to HST CMD data from \citet{piotto_etal_2002} adopting the Zinn \& West \citep[][]{zinn_west_1984} metallicity scale. We chose this combination because it gives us the highest number of GCs (34) in common with Usher et al.\footnote{Other possible combinations, i.e. using ground-based data and the Carretta \& Gratton \citep{carretta_and_gratton_1997} metallicity scale, will be discussed in a publication devoted to matching our models to GC data.}  

With reference to the comparison with the Usher et al. compilation of parameters (left-hand panels), it can be appreciated that - in spite of a general acceptable agreement - the model fitting stretches ages towards lower values. This trend is somewhat more pronounced for the E-MaStar models. The intermediate ages of LMC/SMC GCs are well recovered especially by Th-MaStar models.

When we use the De Angeli et al. work as literature values (right-hand panels), It can be seen that the old ages of MW GCs are well recovered by the Th-MaStar models, albeit with some scatter. The E-MaStar models on the other hand lead to lower ages for a notable number of GCs.

Before proceeding with a quantitative assessment of the comparisons shown in Figure~\ref{fig:gc}, we comment on the relation between mass and light-weighted quantities. The option of having two types of population model derivations is useful for the analysis of galaxies, where the existence of multiple stellar generations makes the youngest one dominate the light and produces an offset between light-weighted and mass-weighted quantities, a phenomenon which \citet{maraston_etal_2010} dubbed 'overshining'. 
Figure~\ref{fig:gc-lightmass} directly compares light- and mass-weighted ages and metallicities for the Usher et al. sample. We see that light and mass-weighted ages are in excellent agreement as expected for simple populations such as GCs\footnote{The case of multiple populations in GCs as revealed by CMD studies \citep[e.g.][]{piotto_etal_2007} either refers to metallicity spread at a constant age, or to a too small age spread to be revealed in integrated spectra.}. Interestingly, there is a slightly larger spread between light- and mass weighted metallicities, especially around $[Z/H]\sim-1$~where there is also a spread in the HB morphology \citep[e.g.][]{harris_1996}. 
\begin{figure*}
  \includegraphics[width=0.49\textwidth]{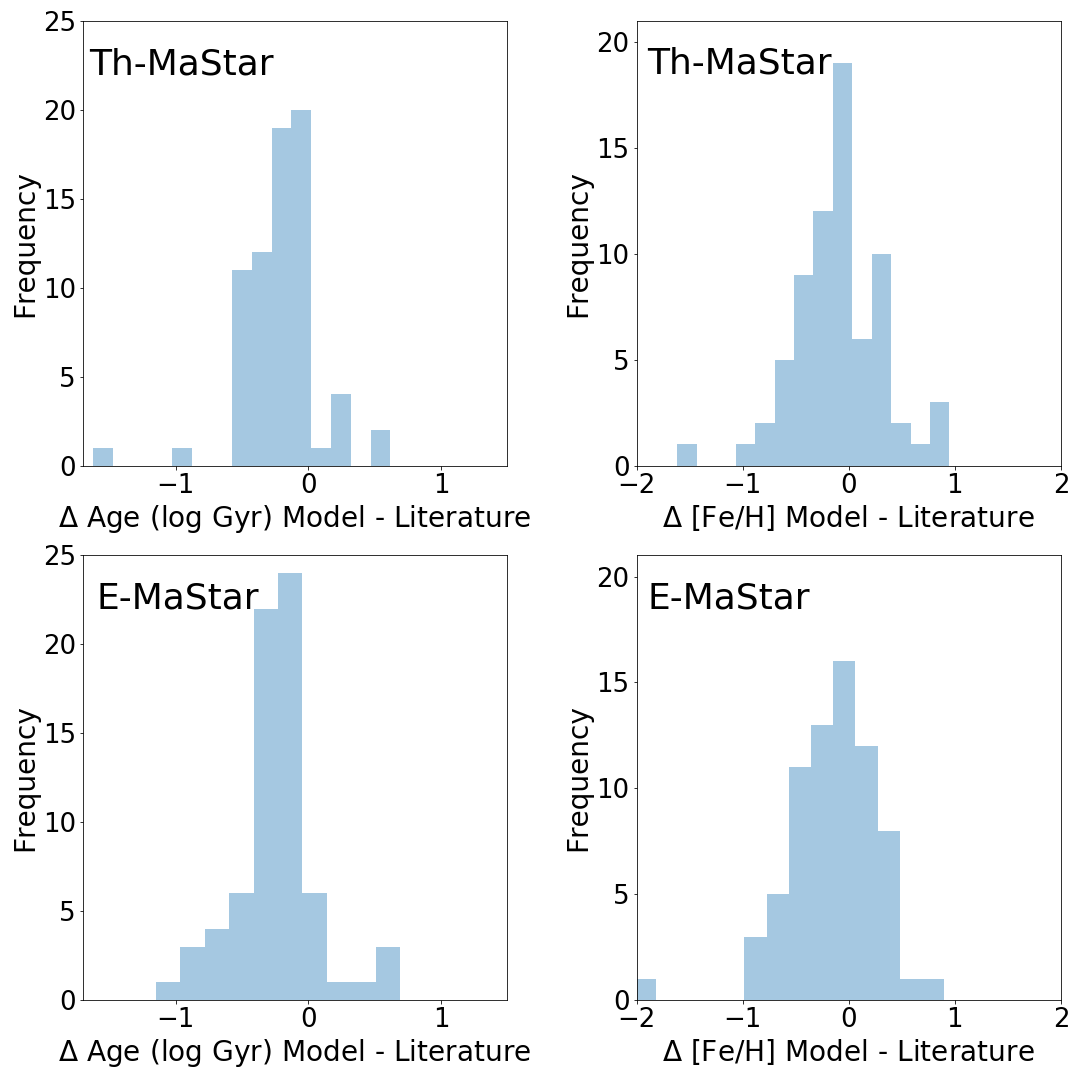}
   \includegraphics[width=0.49\textwidth]{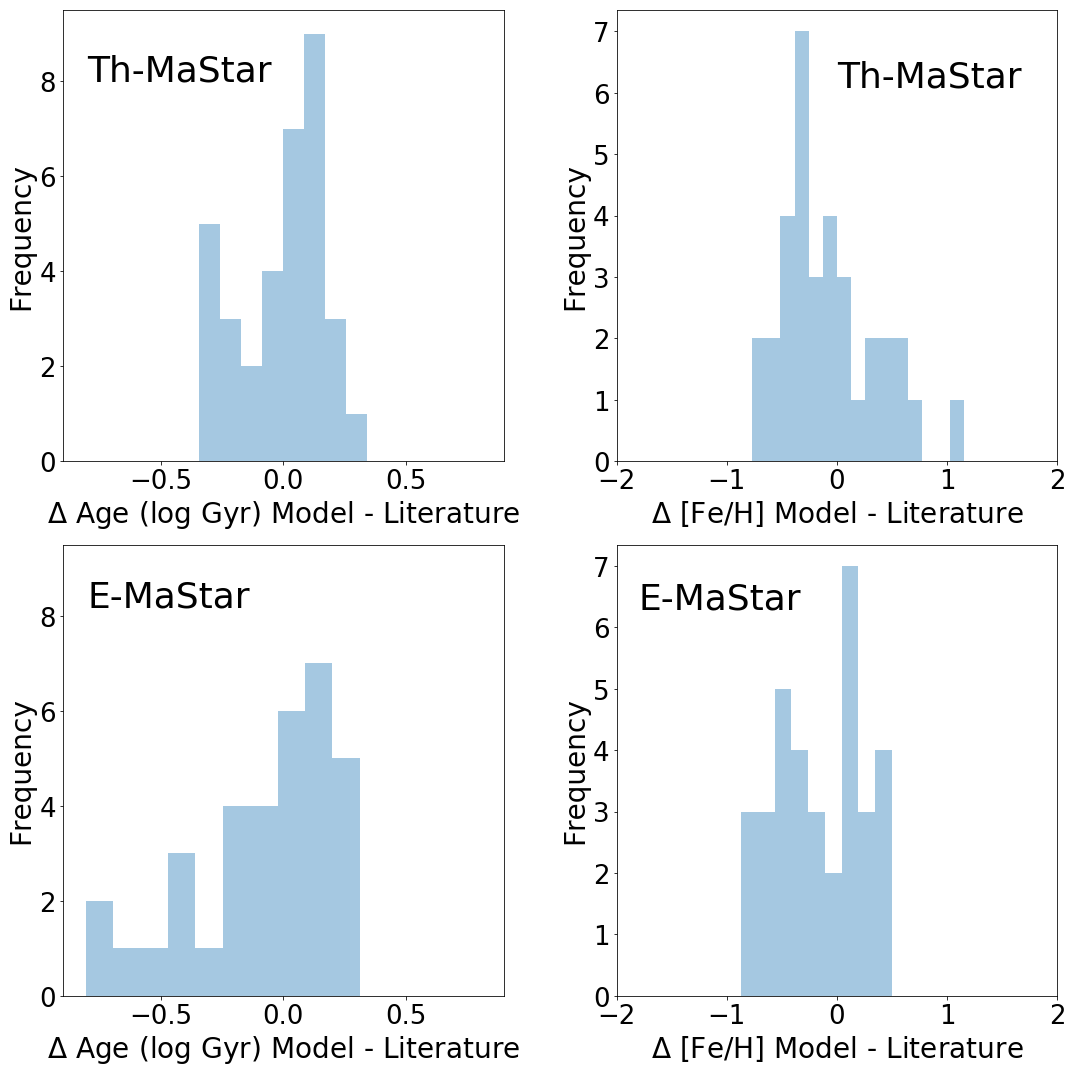}
  \caption{Histograms of age (in log Gyr) and metallicity (in dex) differences between quantities obtained by fitting Th-MaStar and E-MaStar models (upper and lower rows, respectively) to GC spectra and literature values from Usher et al. (2017) (left-hand plots) and De Angeli et al. (2015, right-hand plots). These histograms refer to the values plotted in Figures~\ref{fig:gc}.}
  \label{fig:gchist}
\end{figure*}

For the quantitative comparisons we decided to use the mass-weighted quantities, in the assumption that CMD ages and metallicities from resolved spectroscopy are closer to be representative of the whole population rather than any sub-population which may be bright in a light-weighted sense. The effect of adopting this option has negligible effects on the final results.
Figure~\ref{fig:gchist} shows histograms of differences between model-derived mass-weighted ages and metallicities and the equivalent from the literature (from the compilation of Usher et al. or as derived by De Angeli et al., respectively, as plotted in Figure~\ref{fig:gc}) for Th- and E-Mastar models (upper and lower rows, respectively). The histograms are used to calculate the statistics given in Table~\ref{tab:gcstats}, namely the median offset between estimated and literature parameters, the width of the distributions and the median error. The latter is the combined median error of our parameter estimation and the literature\footnote{As Usher et al. do not provide errors, we adopt those from de Angeli et al.}. 

In the case of the homogeneous age derivation by De Angeli et al., we find that Th-MaStar models allow the determination of ages with a systematic offset as low as 0.04 dex (corresponding to $9\%$), which is smaller than the median error. The offset in the age determinations using E-MaStar models is just zero, but the scatter is larger, as is also visible in the plots. This case proves that we need to consider both figures in these comparisons.

In case of metallicity, the offsets are 0.09 and 0.12 for Th- and E-MaStar models (corresponding to $23\%$ and $32\%$) when referring to the De Angeli et al. dataset. 

Deviations are somewhat larger when we use the compilation by Usher et al., which is however an heterogeneous assembly of literature determination hence it provides a weaker constraint to the models. In this case, the performance of the models is identical. 
\subsection{A glimpse of model galaxy fitting performance}
\label{sec:galfit}
\begin{figure*}
   \includegraphics[width=0.9\textwidth]{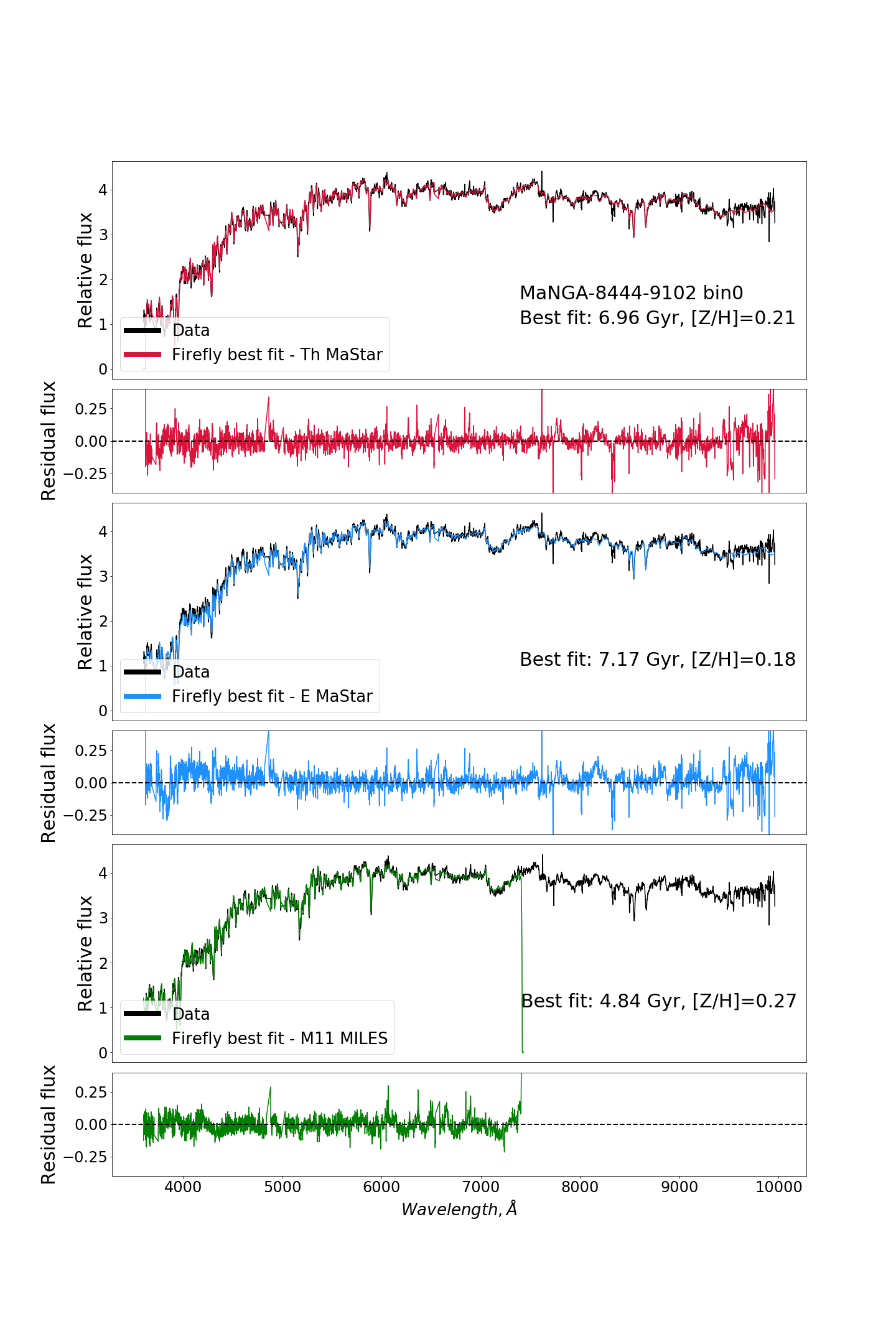}
  \caption{Example full spectral fitting of a MaNGA galaxy spectrum with Th-MaStar, E-MaStar and M11-MILES models (from top to bottom, respectively). The derived galaxy parameters (light-weighted) are labelled and model residuals are shown underneath each fit.}
  \label{fig:galfit}
\end{figure*}

The results obtained with the globular cluster testing described previously should have positive implications for galaxy evolution studies, when we shall use the models to obtain the physical properties of galaxies. We give a taste of the fit quality achieved with the new models by showing in Figure~\ref{fig:galfit} the fits to a SDSS/MaNGA galaxy spectrum, using Th-MaStar, E-MaStar and M11-MILES models. It should be stressed that the last type of model offers a shorter wavelength extension with respect to the new MaStar models. The galaxy spectrum was selected for a galaxy displaying a prominent absorption-line spectrum in order to offer a stronger challenge to the models. As before the fits are performed with Firefly. 
\begin{table*}
\centering
{\footnotesize 
\begin{tabular}{ l | cccccc |}
\hline\hline
Model & Literature data & Parameter & Median difference & $\sigma$ difference & Median error\\
 & & & (model - literature) & & \\ \hline \hline
 Th-MaStar & Usher et al. (2017) & Age (log Gyr) & -0.18 & 0.29 & 0.01 \\ 
  & & [Fe/H] & -0.15 & 0.43 & 0.0 \\
 E-MaStar & & Age (log Gyr) & -0.19 & 0.31 & 0.01 \\
  & & [Fe/H] & -0.15 & 0.44 & 0.0 \\ \hline 
 Th-MaStar &  De Angeli et al. (2005) HST & Age (log Gyr) & 0.04 & 0.17 & 0.17 \\ 
  & & [Fe/H] & -0.09 & 0.55 & 0.08 \\ 
 E-MaStar & & Age (log Gyr) & 0.0 & 0.29 & 0.07 \\ 
  & & [Fe/H] & -0.12 & 0.58 & 0.08 \\ \hline
 \end{tabular}
\caption[Summary of statistics for globular cluster fitting.]{Quantitative comparison of ages and metallicities obtained by fitting GC spectra from Usher et al. (2017) with Th- and E-MaStar models, with various age and metallicity determinations from the literature. We calculate the median of the differences between parameters from full spectral fitting and from the literature, in addition to the standard deviation and median error.
\label{tab:gcstats}}
} 
\end{table*}

First of all, the galaxy parameters - age and metallicity - obtained with both MaStar models are well consistent. 
Those derived using M11-MILES are somewhat different, the age is 2 Gyr younger and the metallicity is higher. The derivation of population parameters is affected by the fitted wavelength range \citep[e.g.][]{pforr_etal_2012}, in dependence of the galaxy star formation history, reddening, etc. We shall explore these effects when fitting the whole MaNGA galaxy sample. 

From Figure~\ref{fig:galfit} is evident that the fits obtained using the new MaStar models are good, in particular the Th-flavour give small residuals over the majority of the observed wavelength range. There are noisy spells at the edges of the wavelength range, which is reported in \citet{yan_etal_2019} for a fraction of MaStar spectra. The fit with M11-MILES looks similarly good therefore we performed a quantitative evaluation of residuals. 
\begin{figure*}
    \includegraphics[width=0.7\textwidth]{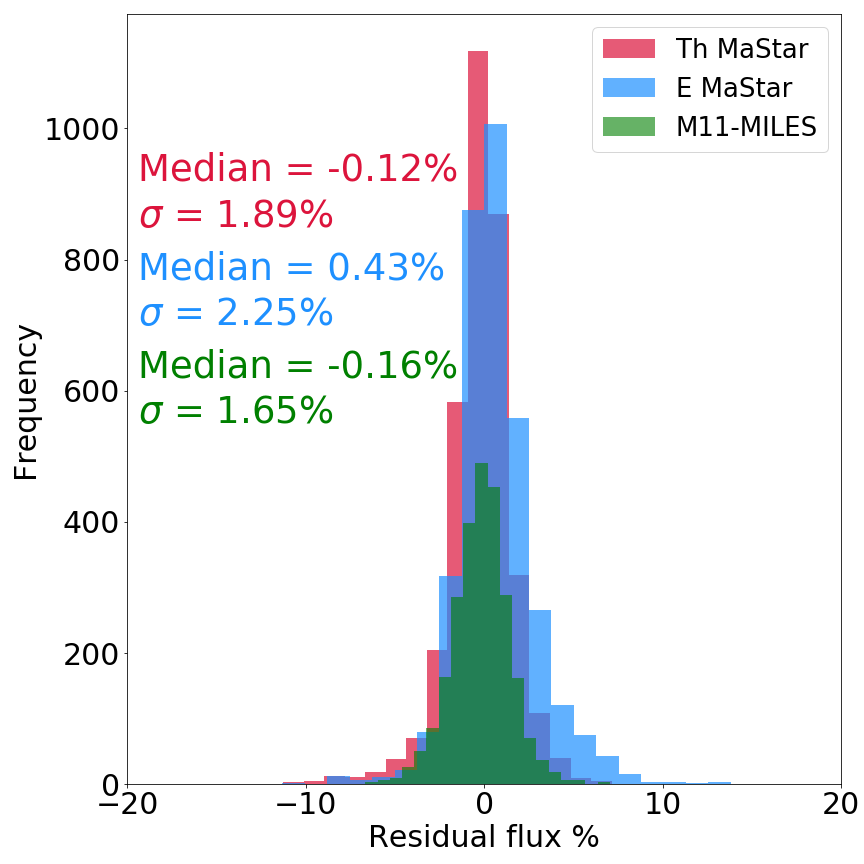} 
   \includegraphics[width=0.8\textwidth]{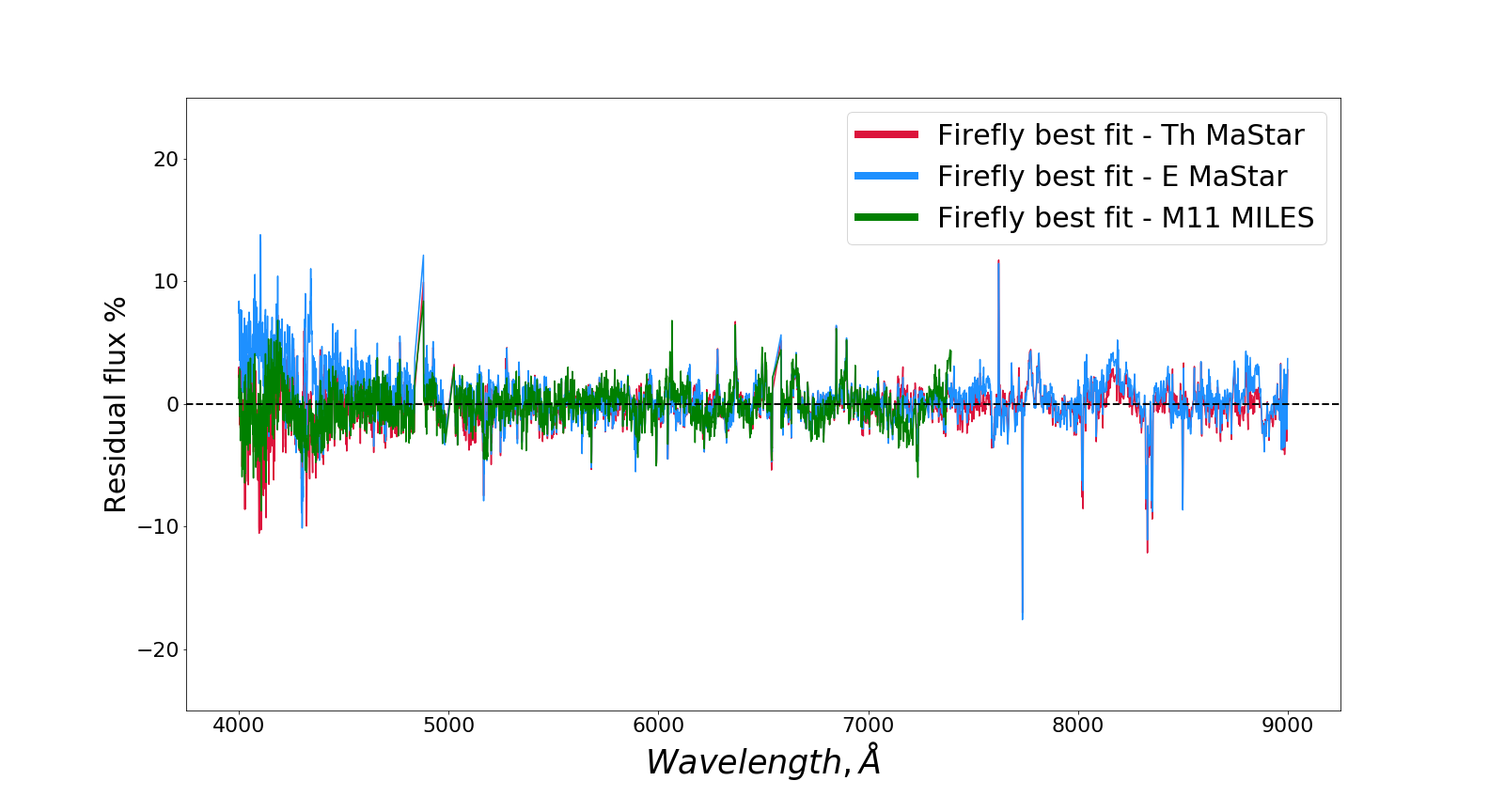}
  \caption{{\it Upper panel}: Distributions of percentage residuals per wavelength for the three models. Median and $\sigma$'s refer to residuals in the wavelength range $4000-9000$~\AA\ for E- and Th-Mastar models, to $4000-7200$~\AA\ for M11, as shown in the lower panel.}
  \label{fig:galres}
\end{figure*}
Figure~\ref{fig:galres} shows the distributions of percentage residuals per wavelength (over $4000-9000$ \AA\ for Mastar models, over $\lambda >4000-7200$~\AA\ for M11-MILES) for the three model fits of Figure~\ref{fig:galfit}. Median residuals and their dispersion are small (a few percents) in all cases, but particularly for Th-MaStar and compare favourably with the results obtained with M11-MILES which represents the state-of-art we confront to. Noteworthy, the residuals obtained with the new MaStar models are not substantially worse in spite of being evaluated over a larger wavelength range, in other words
the fit quality does not degrade over the new wavelength range.
This is very reassuring.  Naturally, these comments refer to this one galaxy case and should not be generalised at this stage, but nonetheless give us a positive signal.
\section{Summary and Conclusive remarks}
\label{sec:summary}
Stellar population synthesis models describing the energetic emission and stellar mass distribution of stellar systems (galaxies and star clusters) find ubiquitous applications in astrophysics and cosmology. They allow the determination of galaxy physical properties - such as formation epoch, star and chemical enrichment history, luminous versus dark matter - from data. They are also a key component of galaxy formation and evolution models, such as semi-analytic models and hydro-dynamical simulations. Population models are also used to tailor cosmological surveys as they determine the luminosity and colours of target galaxies, the signal-to-noise ratio required to resolve absorption features, hence the redshift accuracy. Stellar population models have reached a high standard \citep[e.g.][]{conroy_2013}, but critical improvements are still required to attain precision astrophysics in the epoch of precision cosmology. 

In this paper we take a step forward by calculating stellar population models based on MaStar \citep{yan_etal_2019}, the empirical stellar library being currently acquired with the Sloan Telescope \citep{gunn_etal_2006}. The first release which we use here comprises over 8000 spectra referred to $\sim 3000$ individual Milky Way stars. The models cover a wide contiguous wavelength range (0.36-1$\mu$m) at the spectral resolution of $R\sim1800$, the same resolution and flux calibration of the MaNGA galaxy data. The entry spectra have typically a high signal-to-noise ratio ($S/N>50$), which helps achieving small statistical errors.

Originally in this paper, we explore the effect on population models of two independent choices of the stellar parameters - surface gravity, effective temperature and metallicity $logg$, $T_{\rm eff}$, $[Z/H]$. One set is tied to empirical spectra from the MILES stellar library, where parameters are obtained as combinations of those of the MILES spectra that contribute to best-fit each MaStar spectrum \citep[][Chen et al. {\it in prep.}]{yan_etal_2019}. \citet{yan_etal_2019} experimented with these MILES-based parameters and concluded that the parameter 'chemical composition' is more robust when taken from spectroscopic surveys which the target selection was based on, namely APOGEE, SEGUE and LAMOST \citep[called by]['input parameters']{yan_etal_2019}. We retain their conclusion here and utilise as our 'E' (empirical) parameters a combination of MILES-derived gravities and effective temperatures and chemical composition from these other sources. 

The other parameter set is obtained by fitting theoretical spectra from model atmospheres to each MaStar spectrum individually (Hill et al. {\it in prep.}). One advantage of this 'Th' (theoretical) approach are that it allows the widest possible parameter range to be included without being tied to any particular previous library. This is crucial as one of the aims of MaStar is to achieve a wider coverage of stellar parameters than previous libraries. The other advantage is to obtain a single fitting result for each stellar spectrum rather than a combination as indeed stars are single objects.

We then calculated stellar population models based on each parameter set  independently, for the ages and metallicities allowed by the stellar parameter coverage. The derived stellar population model sets - E-MaStar and Th-MaStar - share the same energetics (from M05) - hence their comparison allows us to pinpoint the effect of the new MaStar stellar library and its parameters. 

First of all, both sets allow us to model the five metallicity grid points of M05 as we could do with M11-MILES models, namely -2.25; -1.35; -0.33; 0.0; 0.35. The parameter coverage varies within these metallicity pillars due to stellar distributions in the Milky Way, as is well known, however MaStar gives us a number of spectra/stellar phases which is usually larger than ten and in some cases including hundreds of spectra.
The model ages we are able to describe with the first MaStar release encompass the intermediate-age/old regime of stellar populations and are generally larger than 0.2 Gyr (cfr. Table 1).   

The lower age limit depends on the assigned parameters. The Th parameter set permits the reach of lower ages (at given metallicity) as a combined effect of no priors in the fitting grid and of not combining parameters from different fits. 
This effect is remarkable at low metallicity. At $[Z/H]=-1.35$~we could push the calculation of models down to 0.5 Gyr, whereas M11-MILES could only cover ages larger than 2 Gyr. In the lowest metallicity bin ($[Z/H]=-2.25$) we now have models as young as 1 Gyr whereas M11-MILES was limited to 6 Gyr. Also, we are now able to include Blue Horizontal Branch (BHB) stars with a good coverage of effective temperature such that we are able to model the full range of HB morphology of the M05 models. These features at low metallicity should help the fitting of dwarf galaxies and star clusters in the local Universe and of high-redshift, metal-poor galaxies. 

Notably, and especially with Th-MaStar, we are able to include spectra for dwarf MS stars close to the H-burning limit of $\sim~0.1~M_{\odot}$. This remarkable achievement improves on the deficiency of dwarf MS spectra that affected all previous empirical libraries for population synthesis modelling (cfr. discussion in M11).

Another interesting result pertains to the slope of the RGB, which is dependent on parameters such as the mixing-length, which are hard to calibrate as a function of metallicity \citep[see discussion and references in][]{maraston_2005}. We find that the RGB stellar parameters for MaStar spectra, especially those obtained with the Th-set, nicely line up along the predicted stellar evolution parameters used in the M05 model virtually at all metallicities. 
Comparisons of these kind with other stellar evolution computations, which we shall pursue as a parallel project, may be able to shed light on the still-elusive mixing-length and help assessing galaxy evolution results, as the RGB slope affects the colours and absorption features of model galaxies (cfr. Maraston~2005, Fig. 9).   
A striking feature of the new models based on theoretical parameters (Th-MaStar) is the description of near-IR bands in old and metal-rich populations, which result from the RGB empirical spectra. We further show that the treatment of the giant phases has a direct effect on the predicted spectral absorption features at wavelengths larger than $\sim0.7 \mu$, e.g. NaI0.82, MgI0.88 and FeH0.99.  We also find sizable difference between ours and other models from the literature. We shall compare these predictions to data in future work.

We then tested how well the new models are able to recover independently determined ages and metallicities of GCs in the Milky Way and the Magellanic Clouds, by performing quantitative testing on various databases from the literature. In particular, we derive ages and metallicities for a sample of 71 GCs drawn from the recent Usher et al. (2017) database of spectra, which very conveniently cover the same wavelength range as MaStar spectra and include young/intermediate age objects from the Magellanic Clouds. Age and metallicity were derived through full spectral fitting of the new MaStar models we present here using the Firefly code \citep{wilkinson_etal_2017}. 

The new MaStar models are able to recover ages and metallicities of GCs with systematic offsets that can be as low as $9\%$~when homogeneously-derived GC ages are considered for the comparison. Metallicity is determined with an offset of $20\%$~in the best case. The performance of the two model flavours is similar, with Th-MaStar gaining a somewhat smaller scatter and systematic offset.

We then performed the full spectral fitting of a MaNGA galaxy spectrum, to provide one example of model performance with galaxies. We selected a red-looking galaxy with a marked absorption-line spectrum to offer a harder challenge to the models. We found that consistent ages and metallicities are derived with both MaStar population models, which also do not differ much to those obtained with M11-MILES. In addition, we find that the percentage residual between models and data is just a few percent for all models. In particular, the new models are reaching the high standard set by e.g. M11-MILES which is based on the state-of art library. Noteworthy, the residuals do not increase over the newly explored wavelength range. 

With this first paper, we have tried to push the stellar population model calculations using the first MaStar release. 
We have also shown that in spite of differences in stellar parameters derivation, the resulting models are overall quite consistent, with exceptions possibly located in the near-IR. 
We have also improved our understanding of how to best calculate stellar parameters for MaStar spectra, probing that parameters derived via spectral fitting of theoretical spectra from model atmosphere exploiting the full wavelength range of the data yields physically sound parameters. We are currently working on improving the stellar parameter determination for the whole of MaStar.

Our models will be updated following progress in MaStar spectra acquisition. We are going to significantly expand the parameter coverage in our final release allowing us to probe younger ages and get models for different alpha-enhancements.

We expect these models to leverage galaxy evolution studies, in particular the analysis of the SDSS-legacy and SDSS-IV database as they are based on either a similar or the exact same instrument. Models are available at http://www.icg.port.ac.uk/MaStar/.

\section*{Acknowledgments}
We are grateful to Michele Cappellari for support in the use of pPXF. We acknowledge the Referee for an encouraging report. Numerical computations were done on the Sciama High Performance Compute (HPC) cluster which is supported by the ICG, SEPnet and the University of Portsmouth.
The Science, Technology and Facilities Council (STFC) is acknowledged for support through the Consolidated Grant Cosmology and Astrophysics at Portsmouth, ST/S000550/1.
Funding for the Sloan Digital Sky Survey IV has been provided by the Alfred P. Sloan Foundation, the U.S. Department of Energy Office of Science, and the Participating Institutions. The MaStar project is partly supported by NSF grant AST-1715898.
SDSS acknowledges support and resources from the Center for High-Performance Computing at the University of Utah. The SDSS web site is \url{www.sdss.org}.
SDSS is managed by the Astrophysical Research Consortium for the Participating Institutions of the SDSS Collaboration including the Brazilian Participation Group, the Carnegie Institution for Science, Carnegie Mellon University, the Chilean Participation Group, the French Participation Group, Harvard-Smithsonian Center for Astrophysics, Instituto de Astrof\'{i}sica de Canarias, The Johns Hopkins University, Kavli Institute for the Physics and Mathematics of the Universe (IPMU) / University of Tokyo, Lawrence Berkeley National Laboratory, Leibniz Institut f\"ur Astrophysik Potsdam (AIP), Max-Planck-Institut f\"ur Astronomie (MPIA Heidelberg), Max-Planck-Institut f\"ur Astrophysik (MPA Garching), Max-Planck-Institut f\"ur Extraterrestrische Physik (MPE), National Astronomical Observatories of China, New Mexico State University, New York University, University of Notre Dame, Observat\'{o}rio Nacional / MCTI, The Ohio State University, Pennsylvania State University, Shanghai Astronomical Observatory, United Kingdom Participation Group, Universidad Nacional Aut\'{o}noma de M\'{e}xico, University of Arizona, University of Colorado Boulder, University of Oxford, University of Portsmouth, University of Utah, University of Virginia, University of Washington, University of Wisconsin, Vanderbilt University, and Yale University.




\bibliographystyle{mnras}
\bibliography{MaStar} 

\begin{thebibliography}{}
\makeatletter
\relax
\def\mn@urlcharsother{\let\do\@makeother \do\$\do\&\do\#\do\^\do\_\do\%\do\~}
\def\mn@doi{\begingroup\mn@urlcharsother \@ifnextchar [ {\mn@doi@}
  {\mn@doi@[]}}
\def\mn@doi@[#1]#2{\def\@tempa{#1}\ifx\@tempa\@empty \href
  {http://dx.doi.org/#2} {doi:#2}\else \href {http://dx.doi.org/#2} {#1}\fi
  \endgroup}
\def\mn@eprint#1#2{\mn@eprint@#1:#2::\@nil}
\def\mn@eprint@arXiv#1{\href {http://arxiv.org/abs/#1} {{\tt arXiv:#1}}}
\def\mn@eprint@dblp#1{\href {http://dblp.uni-trier.de/rec/bibtex/#1.xml}
  {dblp:#1}}
\def\mn@eprint@#1:#2:#3:#4\@nil{\def\@tempa {#1}\def\@tempb {#2}\def\@tempc
  {#3}\ifx \@tempc \@empty \let \@tempc \@tempb \let \@tempb \@tempa \fi \ifx
  \@tempb \@empty \def\@tempb {arXiv}\fi \@ifundefined
  {mn@eprint@\@tempb}{\@tempb:\@tempc}{\expandafter \expandafter \csname
  mn@eprint@\@tempb\endcsname \expandafter{\@tempc}}}

\bibitem[\protect\citeauthoryear{{Allende Prieto} et~al.,}{{Allende Prieto}
  et~al.}{2008}]{allendeprieto_etal_2008}
{Allende Prieto} C.,  et~al., 2008, \mn@doi [\aj]
  {10.1088/0004-6256/136/5/2070}, \href
  {https://ui.adsabs.harvard.edu/abs/2008AJ....136.2070A} {136, 2070}

\bibitem[\protect\citeauthoryear{{Allende Prieto} et~al.,}{{Allende Prieto}
  et~al.}{2014}]{allendeprieto_etal_2014}
{Allende Prieto} C.,  et~al., 2014, \mn@doi [\aap]
  {10.1051/0004-6361/201424053}, \href
  {https://ui.adsabs.harvard.edu/abs/2014A&A...568A...7A} {568, A7}

\bibitem[\protect\citeauthoryear{{Baugh}}{{Baugh}}{2006}]{baugh_2006}
{Baugh} C.~M.,  2006, \mn@doi [Reports on Progress in Physics]
  {10.1088/0034-4885/69/12/R02}, \href
  {https://ui.adsabs.harvard.edu/abs/2006RPPh...69.3101B} {69, 3101}

\bibitem[\protect\citeauthoryear{{Beifiori}, {Maraston}, {Thomas}  \&
  {Johansson}}{{Beifiori} et~al.}{2011}]{beifiori_etal_2011}
{Beifiori} A.,  {Maraston} C.,  {Thomas} D.,   {Johansson} J.,  2011, \mn@doi
  [\aap] {10.1051/0004-6361/201016323}, \href
  {https://ui.adsabs.harvard.edu/abs/2011A&A...531A.109B} {531, A109}

\bibitem[\protect\citeauthoryear{{Blanton} et~al.,}{{Blanton}
  et~al.}{2017}]{blanton_etal_2017}
{Blanton} M.~R.,  et~al., 2017, \mn@doi [\aj] {10.3847/1538-3881/aa7567}, \href
  {https://ui.adsabs.harvard.edu/abs/2017AJ....154...28B} {154, 28}

\bibitem[\protect\citeauthoryear{{Bohlin}, {M{\'e}sz{\'a}ros}, {Fleming},
  {Gordon}, {Koekemoer}  \& {Kov{\'a}cs}}{{Bohlin}
  et~al.}{2017}]{bohlin_etal_2017}
{Bohlin} R.~C.,  {M{\'e}sz{\'a}ros} S.,  {Fleming} S.~W.,  {Gordon} K.~D.,
  {Koekemoer} A.~M.,   {Kov{\'a}cs} J.,  2017, \mn@doi [\aj]
  {10.3847/1538-3881/aa6ba9}, \href
  {http://adsabs.harvard.edu/abs/2017AJ....153..234B} {153, 234}

\bibitem[\protect\citeauthoryear{{Bruzual} \& {Charlot}}{{Bruzual} \&
  {Charlot}}{2003}]{bruzual_and_charlot_2003}
{Bruzual} G.,  {Charlot} S.,  2003, \mn@doi [\mnras]
  {10.1046/j.1365-8711.2003.06897.x}, \href
  {http://adsabs.harvard.edu/abs/2003MNRAS.344.1000B} {344, 1000}

\bibitem[\protect\citeauthoryear{{Bruzual A.}}{{Bruzual
  A.}}{1983}]{bruzual_1983}
{Bruzual A.} G.,  1983, \mn@doi [\apj] {10.1086/161352}, \href
  {https://ui.adsabs.harvard.edu/abs/1983ApJ...273..105B} {273, 105}

\bibitem[\protect\citeauthoryear{{Bundy} et~al.,}{{Bundy}
  et~al.}{2015}]{bundy_etal_2016}
{Bundy} K.,  et~al., 2015, \mn@doi [\apj] {10.1088/0004-637X/798/1/7}, \href
  {https://ui.adsabs.harvard.edu/abs/2015ApJ...798....7B} {798, 7}

\bibitem[\protect\citeauthoryear{{Cappellari}}{{Cappellari}}{2017}]{ppxf}
{Cappellari} M.,  2017, \mn@doi [\mnras] {10.1093/mnras/stw3020}, \href
  {https://ui.adsabs.harvard.edu/abs/2017MNRAS.466..798C} {466, 798}

\bibitem[\protect\citeauthoryear{{Cappellari} \& {Emsellem}}{{Cappellari} \&
  {Emsellem}}{2004}]{cappellari_and_emsellem_2004}
{Cappellari} M.,  {Emsellem} E.,  2004, \mn@doi [\pasp] {10.1086/381875}, \href
  {http://adsabs.harvard.edu/abs/2004PASP..116..138C} {116, 138}

\bibitem[\protect\citeauthoryear{{Carretta} \& {Gratton}}{{Carretta} \&
  {Gratton}}{1997}]{carretta_and_gratton_1997}
{Carretta} E.,  {Gratton} R.~G.,  1997, \mn@doi [\aaps] {10.1051/aas:1997116},
  \href {https://ui.adsabs.harvard.edu/abs/1997A&AS..121...95C} {121, 95}

\bibitem[\protect\citeauthoryear{{Cassisi}, {Castellani}  \&
  {Castellani}}{{Cassisi} et~al.}{1997}]{cassisi_etal_1997}
{Cassisi} S.,  {Castellani} M.,   {Castellani} V.,  1997, \aap, \href
  {https://ui.adsabs.harvard.edu/abs/1997A&A...317..108C} {317, 108}

\bibitem[\protect\citeauthoryear{{Cenarro}, {Cardiel}, {Vazdekis}  \&
  {Gorgas}}{{Cenarro} et~al.}{2009}]{cenarro_etal_2009}
{Cenarro} A.~J.,  {Cardiel} N.,  {Vazdekis} A.,   {Gorgas} J.,  2009, \mn@doi
  [\mnras] {10.1111/j.1365-2966.2009.14839.x}, \href
  {https://ui.adsabs.harvard.edu/abs/2009MNRAS.396.1895C} {396, 1895}

\bibitem[\protect\citeauthoryear{{Conroy}}{{Conroy}}{2013}]{conroy_2013}
{Conroy} C.,  2013, \mn@doi [\araa] {10.1146/annurev-astro-082812-141017},
  \href {http://adsabs.harvard.edu/abs/2013ARA%26A..51..393C} {51, 393}

\bibitem[\protect\citeauthoryear{{Conroy} \& {van Dokkum}}{{Conroy} \& {van
  Dokkum}}{2012}]{conroy_and_van_dokkum_2012}
{Conroy} C.,  {van Dokkum} P.,  2012, \mn@doi [\apj]
  {10.1088/0004-637X/747/1/69}, \href
  {http://adsabs.harvard.edu/abs/2012ApJ...747...69C} {747, 69}

\bibitem[\protect\citeauthoryear{{Conroy}, {Gunn}  \& {White}}{{Conroy}
  et~al.}{2009}]{conroy_etal_2009}
{Conroy} C.,  {Gunn} J.~E.,   {White} M.,  2009, \mn@doi [\apj]
  {10.1088/0004-637X/699/1/486}, \href
  {https://ui.adsabs.harvard.edu/abs/2009ApJ...699..486C} {699, 486}

\bibitem[\protect\citeauthoryear{{Conroy}, {Villaume}, {van Dokkum}  \&
  {Lind}}{{Conroy} et~al.}{2018}]{conroy_etal_2018}
{Conroy} C.,  {Villaume} A.,  {van Dokkum} P.~G.,   {Lind} K.,  2018, \mn@doi
  [\apj] {10.3847/1538-4357/aaab49}, \href
  {https://ui.adsabs.harvard.edu/abs/2018ApJ...854..139C} {854, 139}

\bibitem[\protect\citeauthoryear{{Cui} et~al.,}{{Cui} et~al.}{2012}]{lamost1}
{Cui} X.-Q.,  et~al., 2012, \mn@doi [Research in Astronomy and Astrophysics]
  {10.1088/1674-4527/12/9/003}, \href
  {https://ui.adsabs.harvard.edu/abs/2012RAA....12.1197C} {12, 1197}

\bibitem[\protect\citeauthoryear{{De Angeli}, {Piotto}, {Cassisi}, {Busso},
  {Recio-Blanco}, {Salaris}, {Aparicio}  \& {Rosenberg}}{{De Angeli}
  et~al.}{2005}]{deangeli_etal_2005}
{De Angeli} F.,  {Piotto} G.,  {Cassisi} S.,  {Busso} G.,  {Recio-Blanco} A.,
  {Salaris} M.,  {Aparicio} A.,   {Rosenberg} A.,  2005, \mn@doi [\aj]
  {10.1086/430723}, \href
  {https://ui.adsabs.harvard.edu/abs/2005AJ....130..116D} {130, 116}

\bibitem[\protect\citeauthoryear{{Deng} et~al.,}{{Deng} et~al.}{2012}]{lamost3}
{Deng} L.-C.,  et~al., 2012, \mn@doi [Research in Astronomy and Astrophysics]
  {10.1088/1674-4527/12/7/003}, \href
  {https://ui.adsabs.harvard.edu/abs/2012RAA....12..735D} {12, 735}

\bibitem[\protect\citeauthoryear{{Diaz}, {Terlevich}  \& {Terlevich}}{{Diaz}
  et~al.}{1989}]{diaz_etal_1989}
{Diaz} A.~I.,  {Terlevich} E.,   {Terlevich} R.,  1989, \mn@doi [\mnras]
  {10.1093/mnras/239.2.325}, \href
  {https://ui.adsabs.harvard.edu/abs/1989MNRAS.239..325D} {239, 325}

\bibitem[\protect\citeauthoryear{{Drory} et~al.,}{{Drory}
  et~al.}{2015}]{drory_etal_2015}
{Drory} N.,  et~al., 2015, \mn@doi [\aj] {10.1088/0004-6256/149/2/77}, \href
  {https://ui.adsabs.harvard.edu/abs/2015AJ....149...77D} {149, 77}

\bibitem[\protect\citeauthoryear{{Fioc} \& {Rocca-Volmerange}}{{Fioc} \&
  {Rocca-Volmerange}}{1997}]{fioc_and_rocca_1997}
{Fioc} M.,  {Rocca-Volmerange} B.,  1997, \aap, \href
  {https://ui.adsabs.harvard.edu/abs/1997A&A...326..950F} {500, 507}

\bibitem[\protect\citeauthoryear{{Garc{\'\i}a P{\'e}rez} et~al.,}{{Garc{\'\i}a
  P{\'e}rez} et~al.}{2016}]{garcia_etal_2016}
{Garc{\'\i}a P{\'e}rez} A.~E.,  et~al., 2016, \mn@doi [\aj]
  {10.3847/0004-6256/151/6/144}, \href
  {https://ui.adsabs.harvard.edu/abs/2016AJ....151..144G} {151, 144}

\bibitem[\protect\citeauthoryear{{Girardi}, {Bressan}, {Bertelli}  \&
  {Chiosi}}{{Girardi} et~al.}{2000}]{girardi_etal_2000}
{Girardi} L.,  {Bressan} A.,  {Bertelli} G.,   {Chiosi} C.,  2000, \mn@doi
  [\aaps] {10.1051/aas:2000126}, \href
  {http://adsabs.harvard.edu/abs/2000A%26AS..141..371G} {141, 371}

\bibitem[\protect\citeauthoryear{{Gunn} et~al.,}{{Gunn}
  et~al.}{2006}]{gunn_etal_2006}
{Gunn} J.~E.,  et~al., 2006, \mn@doi [\aj] {10.1086/500975}, \href
  {https://ui.adsabs.harvard.edu/abs/2006AJ....131.2332G} {131, 2332}

\bibitem[\protect\citeauthoryear{{Gustafsson}, {Edvardsson}, {Eriksson},
  {J{\o}rgensen}, {Nordlund}  \& {Plez}}{{Gustafsson}
  et~al.}{2008}]{gustafsson_etal_2008}
{Gustafsson} B.,  {Edvardsson} B.,  {Eriksson} K.,  {J{\o}rgensen} U.~G.,
  {Nordlund} {\AA}.,   {Plez} B.,  2008, \mn@doi [\aap]
  {10.1051/0004-6361:200809724}, \href
  {http://adsabs.harvard.edu/abs/2008A%26A...486..951G} {486, 951}

\bibitem[\protect\citeauthoryear{{Harris}}{{Harris}}{1996}]{harris_1996}
{Harris} W.~E.,  1996, \mn@doi [\aj] {10.1086/118116}, \href
  {https://ui.adsabs.harvard.edu/abs/1996AJ....112.1487H} {112, 1487}

\bibitem[\protect\citeauthoryear{{Holtzman} et~al.,}{{Holtzman}
  et~al.}{2015}]{holtzman_etal_2015}
{Holtzman} J.~A.,  et~al., 2015, \mn@doi [\aj] {10.1088/0004-6256/150/5/148},
  \href {https://ui.adsabs.harvard.edu/abs/2015AJ....150..148H} {150, 148}

\bibitem[\protect\citeauthoryear{{Holtzman} et~al.,}{{Holtzman}
  et~al.}{2018}]{holtzman_etal_2018}
{Holtzman} J.~A.,  et~al., 2018, \mn@doi [\aj] {10.3847/1538-3881/aad4f9},
  \href {https://ui.adsabs.harvard.edu/abs/2018AJ....156..125H} {156, 125}

\bibitem[\protect\citeauthoryear{{Kauffmann}, {White}  \&
  {Guiderdoni}}{{Kauffmann} et~al.}{1993}]{kauffmann_etal_1993}
{Kauffmann} G.,  {White} S.~D.~M.,   {Guiderdoni} B.,  1993, \mn@doi [\mnras]
  {10.1093/mnras/264.1.201}, \href
  {https://ui.adsabs.harvard.edu/abs/1993MNRAS.264..201K} {264, 201}

\bibitem[\protect\citeauthoryear{{Koleva}, {Prugniel}, {De Rijcke}, {Zeilinger}
   \& {Michielsen}}{{Koleva} et~al.}{2009}]{koleva_etal_2009}
{Koleva} M.,  {Prugniel} P.,  {De Rijcke} S.,  {Zeilinger} W.~W.,
  {Michielsen} D.,  2009, \mn@doi [Astronomische Nachrichten]
  {10.1002/asna.200911272}, \href
  {http://adsabs.harvard.edu/abs/2009AN....330..960K} {330, 960}

\bibitem[\protect\citeauthoryear{{Korn}, {Maraston}  \& {Thomas}}{{Korn}
  et~al.}{2005}]{korn_etal_2005}
{Korn} A.~J.,  {Maraston} C.,   {Thomas} D.,  2005, \mn@doi [\aap]
  {10.1051/0004-6361:20042126}, \href
  {https://ui.adsabs.harvard.edu/abs/2005A&A...438..685K} {438, 685}

\bibitem[\protect\citeauthoryear{{Kurucz}}{{Kurucz}}{1979}]{kurucz_1979}
{Kurucz} R.~L.,  1979, \mn@doi [\apjs] {10.1086/190589}, \href
  {https://ui.adsabs.harvard.edu/abs/1979ApJS...40....1K} {40, 1}

\bibitem[\protect\citeauthoryear{{Lan{\c c}on} \& {Mouhcine}}{{Lan{\c c}on} \&
  {Mouhcine}}{2002}]{lancon_and_mouhcine_2002}
{Lan{\c c}on} A.,  {Mouhcine} M.,  2002, \mn@doi [\aap]
  {10.1051/0004-6361:20020585}, \href
  {http://adsabs.harvard.edu/abs/2002A%26A...393..167L} {393, 167}

\bibitem[\protect\citeauthoryear{{Lan{\c c}on} \& {Wood}}{{Lan{\c c}on} \&
  {Wood}}{2000}]{lancon_and_wood_2000}
{Lan{\c c}on} A.,  {Wood} P.~R.,  2000, \mn@doi [\aaps] {10.1051/aas:2000269},
  \href {http://adsabs.harvard.edu/abs/2000A%26AS..146..217L} {146, 217}

\bibitem[\protect\citeauthoryear{{Le Borgne} et~al.,}{{Le Borgne}
  et~al.}{2003}]{stelib}
{Le Borgne} J.~F.,  et~al., 2003, \mn@doi [\aap] {10.1051/0004-6361:20030243},
  \href {https://ui.adsabs.harvard.edu/abs/2003A&A...402..433L} {402, 433}

\bibitem[\protect\citeauthoryear{{Lee} et~al.,}{{Lee}
  et~al.}{2008a}]{lee_etal_2008a}
{Lee} Y.~S.,  et~al., 2008a, \mn@doi [\aj] {10.1088/0004-6256/136/5/2022},
  \href {https://ui.adsabs.harvard.edu/abs/2008AJ....136.2022L} {136, 2022}

\bibitem[\protect\citeauthoryear{{Lee} et~al.,}{{Lee}
  et~al.}{2008b}]{lee_etal_2008b}
{Lee} Y.~S.,  et~al., 2008b, \mn@doi [\aj] {10.1088/0004-6256/136/5/2050},
  \href {https://ui.adsabs.harvard.edu/abs/2008AJ....136.2050L} {136, 2050}

\bibitem[\protect\citeauthoryear{{Leitherer} et~al.,}{{Leitherer}
  et~al.}{1999}]{leitherer_et_al_1999}
{Leitherer} C.,  et~al., 1999, \mn@doi [\apjs] {10.1086/313233}, \href
  {https://ui.adsabs.harvard.edu/abs/1999ApJS..123....3L} {123, 3}

\bibitem[\protect\citeauthoryear{{Lejeune}, {Cuisinier}  \& {Buser}}{{Lejeune}
  et~al.}{1997}]{lejeune_etal_1997}
{Lejeune} T.,  {Cuisinier} F.,   {Buser} R.,  1997, \mn@doi [\aaps]
  {10.1051/aas:1997373}, \href
  {http://adsabs.harvard.edu/abs/1997A%26AS..125..229L} {125, 229}

\bibitem[\protect\citeauthoryear{{Luo} et~al.,}{{Luo} et~al.}{2015}]{lamost4}
{Luo} A.~L.,  et~al., 2015, \mn@doi [Research in Astronomy and Astrophysics]
  {10.1088/1674-4527/15/8/002}, \href
  {https://ui.adsabs.harvard.edu/abs/2015RAA....15.1095L} {15, 1095}

\bibitem[\protect\citeauthoryear{{Majewski} et~al.,}{{Majewski}
  et~al.}{2017}]{apogee}
{Majewski} S.~R.,  et~al., 2017, \mn@doi [\aj] {10.3847/1538-3881/aa784d},
  \href {https://ui.adsabs.harvard.edu/abs/2017AJ....154...94M} {154, 94}

\bibitem[\protect\citeauthoryear{{Maraston}}{{Maraston}}{1998}]{maraston_1998}
{Maraston} C.,  1998, \mn@doi [\mnras] {10.1046/j.1365-8711.1998.01947.x},
  \href {https://ui.adsabs.harvard.edu/abs/1998MNRAS.300..872M} {300, 872}

\bibitem[\protect\citeauthoryear{{Maraston}}{{Maraston}}{2005}]{maraston_2005}
{Maraston} C.,  2005, \mn@doi [\mnras] {10.1111/j.1365-2966.2005.09270.x},
  \href {http://adsabs.harvard.edu/abs/2005MNRAS.362..799M} {362, 799}

\bibitem[\protect\citeauthoryear{{Maraston} \& {Str{\"o}mb{\"a}ck}}{{Maraston}
  \& {Str{\"o}mb{\"a}ck}}{2011}]{maraston_and_stromback_2011}
{Maraston} C.,  {Str{\"o}mb{\"a}ck} G.,  2011, \mn@doi [\mnras]
  {10.1111/j.1365-2966.2011.19738.x}, \href
  {http://adsabs.harvard.edu/abs/2011MNRAS.418.2785M} {418, 2785}

\bibitem[\protect\citeauthoryear{{Maraston}, {Greggio}, {Renzini}, {Ortolani},
  {Saglia}, {Puzia}  \& {Kissler-Patig}}{{Maraston}
  et~al.}{2003}]{maraston_etal_2003}
{Maraston} C.,  {Greggio} L.,  {Renzini} A.,  {Ortolani} S.,  {Saglia} R.~P.,
  {Puzia} T.~H.,   {Kissler-Patig} M.,  2003, \mn@doi [\aap]
  {10.1051/0004-6361:20021723}, \href
  {https://ui.adsabs.harvard.edu/abs/2003A&A...400..823M} {400, 823}

\bibitem[\protect\citeauthoryear{{Maraston}, {Pforr}, {Renzini}, {Daddi},
  {Dickinson}, {Cimatti}  \& {Tonini}}{{Maraston}
  et~al.}{2010}]{maraston_etal_2010}
{Maraston} C.,  {Pforr} J.,  {Renzini} A.,  {Daddi} E.,  {Dickinson} M.,
  {Cimatti} A.,   {Tonini} C.,  2010, \mn@doi [\mnras]
  {10.1111/j.1365-2966.2010.16973.x}, \href
  {https://ui.adsabs.harvard.edu/abs/2010MNRAS.407..830M} {407, 830}

\bibitem[\protect\citeauthoryear{{M{\'e}sz{\'a}ros} et~al.,}{{M{\'e}sz{\'a}ros}
  et~al.}{2012}]{meszaros_etal_2012}
{M{\'e}sz{\'a}ros} S.,  et~al., 2012, \mn@doi [\aj]
  {10.1088/0004-6256/144/4/120}, \href
  {http://adsabs.harvard.edu/abs/2012AJ....144..120M} {144, 120}

\bibitem[\protect\citeauthoryear{{Nidever} et~al.,}{{Nidever}
  et~al.}{2015}]{nidever_etal_2015}
{Nidever} D.~L.,  et~al., 2015, \mn@doi [\aj] {10.1088/0004-6256/150/6/173},
  \href {https://ui.adsabs.harvard.edu/abs/2015AJ....150..173N} {150, 173}

\bibitem[\protect\citeauthoryear{{Parikh} et~al.,}{{Parikh}
  et~al.}{2018}]{parikh_etal_2018}
{Parikh} T.,  et~al., 2018, \mn@doi [\mnras] {10.1093/mnras/sty785}, \href
  {https://ui.adsabs.harvard.edu/abs/2018MNRAS.477.3954P} {477, 3954}

\bibitem[\protect\citeauthoryear{{Pforr}, {Maraston}  \& {Tonini}}{{Pforr}
  et~al.}{2012}]{pforr_etal_2012}
{Pforr} J.,  {Maraston} C.,   {Tonini} C.,  2012, \mn@doi [\mnras]
  {10.1111/j.1365-2966.2012.20848.x}, \href
  {https://ui.adsabs.harvard.edu/abs/2012MNRAS.422.3285P} {422, 3285}

\bibitem[\protect\citeauthoryear{{Pickles}}{{Pickles}}{1998}]{pickles_1998}
{Pickles} A.~J.,  1998, \mn@doi [\pasp] {10.1086/316197}, \href
  {http://adsabs.harvard.edu/abs/1998PASP..110..863P} {110, 863}

\bibitem[\protect\citeauthoryear{{Piotto} et~al.,}{{Piotto}
  et~al.}{2002}]{piotto_etal_2002}
{Piotto} G.,  et~al., 2002, \mn@doi [\aap] {10.1051/0004-6361:20020820}, \href
  {https://ui.adsabs.harvard.edu/abs/2002A&A...391..945P} {391, 945}

\bibitem[\protect\citeauthoryear{{Piotto} et~al.,}{{Piotto}
  et~al.}{2007}]{piotto_etal_2007}
{Piotto} G.,  et~al., 2007, \mn@doi [\apjl] {10.1086/518503}, \href
  {https://ui.adsabs.harvard.edu/abs/2007ApJ...661L..53P} {661, L53}

\bibitem[\protect\citeauthoryear{{Prugniel} \& {Soubiran}}{{Prugniel} \&
  {Soubiran}}{2001}]{elodie}
{Prugniel} P.,  {Soubiran} C.,  2001, \mn@doi [\aap]
  {10.1051/0004-6361:20010163}, \href
  {https://ui.adsabs.harvard.edu/abs/2001A&A...369.1048P} {369, 1048}

\bibitem[\protect\citeauthoryear{{Prugniel}, {Vauglin}  \& {Koleva}}{{Prugniel}
  et~al.}{2011}]{prugniel_etal_2011}
{Prugniel} P.,  {Vauglin} I.,   {Koleva} M.,  2011, \mn@doi [\aap]
  {10.1051/0004-6361/201116769}, \href
  {http://adsabs.harvard.edu/abs/2011A%26A...531A.165P} {531, A165}

\bibitem[\protect\citeauthoryear{{Renzini} \& {Fusi Pecci}}{{Renzini} \& {Fusi
  Pecci}}{1988}]{fusi_alvio_1988}
{Renzini} A.,  {Fusi Pecci} F.,  1988, \mn@doi [\araa]
  {10.1146/annurev.aa.26.090188.001215}, \href
  {https://ui.adsabs.harvard.edu/abs/1988ARA&A..26..199R} {26, 199}

\bibitem[\protect\citeauthoryear{{S{\'a}nchez-Bl{\'a}zquez}
  et~al.,}{{S{\'a}nchez-Bl{\'a}zquez} et~al.}{2006}]{miles}
{S{\'a}nchez-Bl{\'a}zquez} P.,  et~al., 2006, \mn@doi [\mnras]
  {10.1111/j.1365-2966.2006.10699.x}, \href
  {https://ui.adsabs.harvard.edu/abs/2006MNRAS.371..703S} {371, 703}

\bibitem[\protect\citeauthoryear{{Schaller}, {Schaerer}, {Meynet}  \&
  {Maeder}}{{Schaller} et~al.}{1992}]{genevatracks}
{Schaller} G.,  {Schaerer} D.,  {Meynet} G.,   {Maeder} A.,  1992, \aaps, \href
  {https://ui.adsabs.harvard.edu/abs/1992A&AS...96..269S} {96, 269}

\bibitem[\protect\citeauthoryear{{Schiavon}, {Barbuy}  \& {Bruzual
  A.}}{{Schiavon} et~al.}{2000}]{schiavon_etal_2000}
{Schiavon} R.~P.,  {Barbuy} B.,   {Bruzual A.} G.,  2000, \mn@doi [\apj]
  {10.1086/308533}, \href
  {https://ui.adsabs.harvard.edu/abs/2000ApJ...532..453S} {532, 453}

\bibitem[\protect\citeauthoryear{{Schiavon}, {Rose}, {Courteau}  \&
  {MacArthur}}{{Schiavon} et~al.}{2005}]{schiavon_etal_2005}
{Schiavon} R.~P.,  {Rose} J.~A.,  {Courteau} S.,   {MacArthur} L.~A.,  2005,
  \mn@doi [\apjs] {10.1086/431148}, \href
  {https://ui.adsabs.harvard.edu/abs/2005ApJS..160..163S} {160, 163}

\bibitem[\protect\citeauthoryear{{Schlegel}, {Finkbeiner}  \&
  {Davis}}{{Schlegel} et~al.}{1998}]{schlegel_etal_1998}
{Schlegel} D.~J.,  {Finkbeiner} D.~P.,   {Davis} M.,  1998, \mn@doi [\apj]
  {10.1086/305772}, \href
  {https://ui.adsabs.harvard.edu/abs/1998ApJ...500..525S} {500, 525}

\bibitem[\protect\citeauthoryear{{Smee} et~al.,}{{Smee}
  et~al.}{2013}]{smee_etal_2013}
{Smee} S.~A.,  et~al., 2013, \mn@doi [\aj] {10.1088/0004-6256/146/2/32}, \href
  {https://ui.adsabs.harvard.edu/abs/2013AJ....146...32S} {146, 32}

\bibitem[\protect\citeauthoryear{{Tayar} et~al.,}{{Tayar}
  et~al.}{2017}]{tayar_etal_2017}
{Tayar} J.,  et~al., 2017, \mn@doi [\apj] {10.3847/1538-4357/aa6a1e}, \href
  {https://ui.adsabs.harvard.edu/abs/2017ApJ...840...17T} {840, 17}

\bibitem[\protect\citeauthoryear{{Thomas}, {Maraston}  \& {Bender}}{{Thomas}
  et~al.}{2003}]{thomas_etal_2003}
{Thomas} D.,  {Maraston} C.,   {Bender} R.,  2003, \mn@doi [\mnras]
  {10.1046/j.1365-8711.2003.06248.x}, \href
  {http://adsabs.harvard.edu/abs/2003MNRAS.339..897T} {339, 897}

\bibitem[\protect\citeauthoryear{{Thomas}, {Maraston}, {Bender}  \& {Mendes de
  Oliveira}}{{Thomas} et~al.}{2005}]{thomas_etal_2005}
{Thomas} D.,  {Maraston} C.,  {Bender} R.,   {Mendes de Oliveira} C.,  2005,
  \mn@doi [\apj] {10.1086/426932}, \href
  {http://adsabs.harvard.edu/abs/2005ApJ...621..673T} {621, 673}

\bibitem[\protect\citeauthoryear{{Thomas}, {Maraston}  \& {Johansson}}{{Thomas}
  et~al.}{2011}]{thomas_etal_2011}
{Thomas} D.,  {Maraston} C.,   {Johansson} J.,  2011, \mn@doi [\mnras]
  {10.1111/j.1365-2966.2010.18049.x}, \href
  {https://ui.adsabs.harvard.edu/abs/2011MNRAS.412.2183T} {412, 2183}

\bibitem[\protect\citeauthoryear{{Tinsley}}{{Tinsley}}{1972}]{tinsley_1972}
{Tinsley} B.~M.,  1972, \mn@doi [\apj] {10.1086/151793}, \href
  {https://ui.adsabs.harvard.edu/abs/1972ApJ...178..319T} {178, 319}

\bibitem[\protect\citeauthoryear{{Trager}, {Worthey}, {Faber}, {Burstein}  \&
  {Gonz{\'a}lez}}{{Trager} et~al.}{1998}]{trager_etal_1998}
{Trager} S.~C.,  {Worthey} G.,  {Faber} S.~M.,  {Burstein} D.,   {Gonz{\'a}lez}
  J.~J.,  1998, \mn@doi [\apjs] {10.1086/313099}, \href
  {http://adsabs.harvard.edu/abs/1998ApJS..116....1T} {116, 1}

\bibitem[\protect\citeauthoryear{{Usher} et~al.,}{{Usher}
  et~al.}{2017}]{usher_etal_2017}
{Usher} C.,  et~al., 2017, \mn@doi [\mnras] {10.1093/mnras/stx713}, \href
  {http://adsabs.harvard.edu/abs/2017MNRAS.468.3828U} {468, 3828}

\bibitem[\protect\citeauthoryear{{Vazdekis}, {Casuso}, {Peletier}  \&
  {Beckman}}{{Vazdekis} et~al.}{1996}]{vazdekis_etal_1996}
{Vazdekis} A.,  {Casuso} E.,  {Peletier} R.~F.,   {Beckman} J.~E.,  1996,
  \mn@doi [\apjs] {10.1086/192340}, \href
  {http://adsabs.harvard.edu/abs/1996ApJS..106..307V} {106, 307}

\bibitem[\protect\citeauthoryear{{Vazdekis}, {S{\'a}nchez-Bl{\'a}zquez},
  {Falc{\'o}n-Barroso}, {Cenarro}, {Beasley}, {Cardiel}, {Gorgas}  \&
  {Peletier}}{{Vazdekis} et~al.}{2010}]{vazdekis_etal_2010}
{Vazdekis} A.,  {S{\'a}nchez-Bl{\'a}zquez} P.,  {Falc{\'o}n-Barroso} J.,
  {Cenarro} A.~J.,  {Beasley} M.~A.,  {Cardiel} N.,  {Gorgas} J.,   {Peletier}
  R.~F.,  2010, \mn@doi [\mnras] {10.1111/j.1365-2966.2010.16407.x}, \href
  {http://adsabs.harvard.edu/abs/2010MNRAS.404.1639V} {404, 1639}

\bibitem[\protect\citeauthoryear{{Vazdekis}, {Ricciardelli}, {Cenarro},
  {Rivero-Gonz{\'a}lez}, {D{\'{\i}}az-Garc{\'{\i}}a}  \&
  {Falc{\'o}n-Barroso}}{{Vazdekis} et~al.}{2012}]{vazdekis_etal_2012}
{Vazdekis} A.,  {Ricciardelli} E.,  {Cenarro} A.~J.,  {Rivero-Gonz{\'a}lez}
  J.~G.,  {D{\'{\i}}az-Garc{\'{\i}}a} L.~A.,   {Falc{\'o}n-Barroso} J.,  2012,
  preprint, \href {http://adsabs.harvard.edu/abs/2012arXiv1205.5496V} {}
  (\mn@eprint {arXiv} {1205.5496})

\bibitem[\protect\citeauthoryear{{Vazdekis}, {Koleva}, {Ricciardelli},
  {R{\"o}ck}  \& {Falc{\'o}n-Barroso}}{{Vazdekis}
  et~al.}{2016}]{vazdekis_etal_2016}
{Vazdekis} A.,  {Koleva} M.,  {Ricciardelli} E.,  {R{\"o}ck} B.,
  {Falc{\'o}n-Barroso} J.,  2016, \mn@doi [\mnras] {10.1093/mnras/stw2231},
  \href {https://ui.adsabs.harvard.edu/abs/2016MNRAS.463.3409V} {463, 3409}

\bibitem[\protect\citeauthoryear{{Whitford}}{{Whitford}}{1977}]{whitford_1977}
{Whitford} A.~E.,  1977, \mn@doi [\apj] {10.1086/154959}, \href
  {https://ui.adsabs.harvard.edu/abs/1977ApJ...211..527W} {211, 527}

\bibitem[\protect\citeauthoryear{{Wilkinson}, {Maraston}, {Goddard}, {Thomas}
  \& {Parikh}}{{Wilkinson} et~al.}{2017}]{wilkinson_etal_2017}
{Wilkinson} D.~M.,  {Maraston} C.,  {Goddard} D.,  {Thomas} D.,   {Parikh} T.,
  2017, \mn@doi [\mnras] {10.1093/mnras/stx2215}, \href
  {https://ui.adsabs.harvard.edu/abs/2017MNRAS.472.4297W} {472, 4297}

\bibitem[\protect\citeauthoryear{{Worthey}, {Faber}, {Gonzalez}  \&
  {Burstein}}{{Worthey} et~al.}{1994}]{worthey_etal_1994}
{Worthey} G.,  {Faber} S.~M.,  {Gonzalez} J.~J.,   {Burstein} D.,  1994,
  \mn@doi [\apjs] {10.1086/192087}, \href
  {http://adsabs.harvard.edu/abs/1994ApJS...94..687W} {94, 687}

\bibitem[\protect\citeauthoryear{{Wu}, {Singh}, {Prugniel}, {Gupta}  \&
  {Koleva}}{{Wu} et~al.}{2011}]{wu_etal_2011}
{Wu} Y.,  {Singh} H.~P.,  {Prugniel} P.,  {Gupta} R.,   {Koleva} M.,  2011,
  \mn@doi [\aap] {10.1051/0004-6361/201015014}, \href
  {https://ui.adsabs.harvard.edu/abs/2011A&A...525A..71W} {525, A71}

\bibitem[\protect\citeauthoryear{{Yan} et~al.,}{{Yan}
  et~al.}{2016}]{yan_etal_2016}
{Yan} R.,  et~al., 2016, \mn@doi [\aj] {10.3847/0004-6256/152/6/197}, \href
  {https://ui.adsabs.harvard.edu/abs/2016AJ....152..197Y} {152, 197}

\bibitem[\protect\citeauthoryear{{Yan} et~al.,}{{Yan}
  et~al.}{2019}]{yan_etal_2019}
{Yan} R.,  et~al., 2019, \mn@doi [\apj] {10.3847/1538-4357/ab3ebc}, \href
  {https://ui.adsabs.harvard.edu/abs/2019ApJ...883..175Y} {883, 175}

\bibitem[\protect\citeauthoryear{{York} et~al.,}{{York}
  et~al.}{2000}]{york_etal_2000}
{York} D.~G.,  et~al., 2000, \mn@doi [\aj] {10.1086/301513}, \href
  {https://ui.adsabs.harvard.edu/abs/2000AJ....120.1579Y} {120, 1579}

\bibitem[\protect\citeauthoryear{{Zhao}, {Zhao}, {Chu}, {Jing}  \&
  {Deng}}{{Zhao} et~al.}{2012}]{lamost2}
{Zhao} G.,  {Zhao} Y.,  {Chu} Y.,  {Jing} Y.,   {Deng} L.,  2012, arXiv
  e-prints, \href {https://ui.adsabs.harvard.edu/abs/2012arXiv1206.3569Z} {p.
  arXiv:1206.3569}

\bibitem[\protect\citeauthoryear{{Zinn} \& {West}}{{Zinn} \&
  {West}}{1984}]{zinn_west_1984}
{Zinn} R.,  {West} M.~J.,  1984, \mn@doi [\apjs] {10.1086/190947}, \href
  {https://ui.adsabs.harvard.edu/abs/1984ApJS...55...45Z} {55, 45}

\makeatother
\end{thebibliography}



\appendix
\section{Parameter space of  spectra removed by the quality flags.}
In Section~\ref{sec:flags} we describe the flags we apply in order to extract the highest quality sub-sample of spectra. Here we show the parameter space covered by the {\it excluded spectra}, in order to assess qualitatively which part of the parameter space is removed as a consequence of applying flags.  We stress that this assessment is qualitative because the reason for excluding spectra is that we judge the calculated parameters not fully trustworthy. 
\begin{figure*}
	\includegraphics[width=0.49\textwidth]{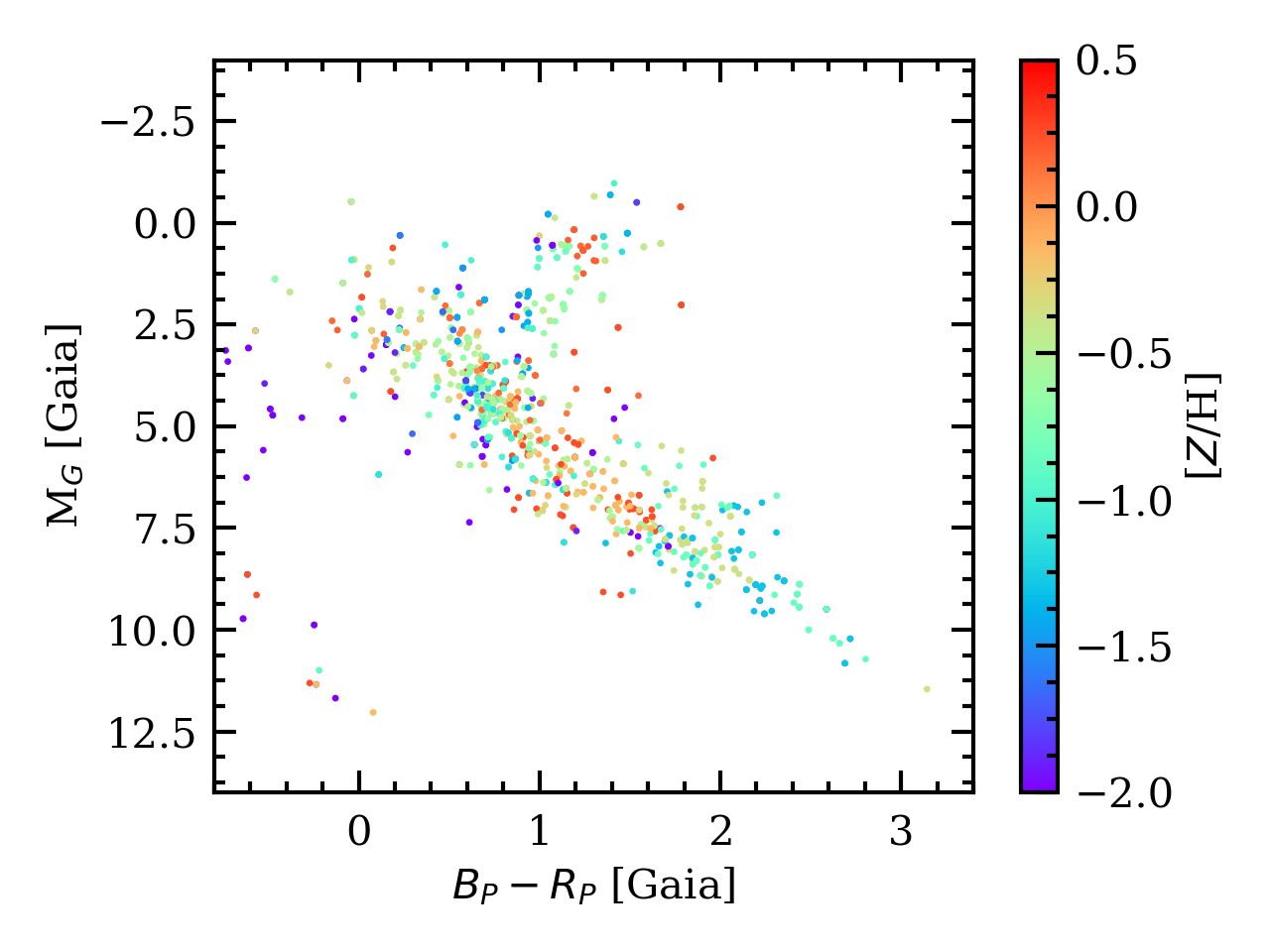}
	\includegraphics[width=0.49\textwidth]{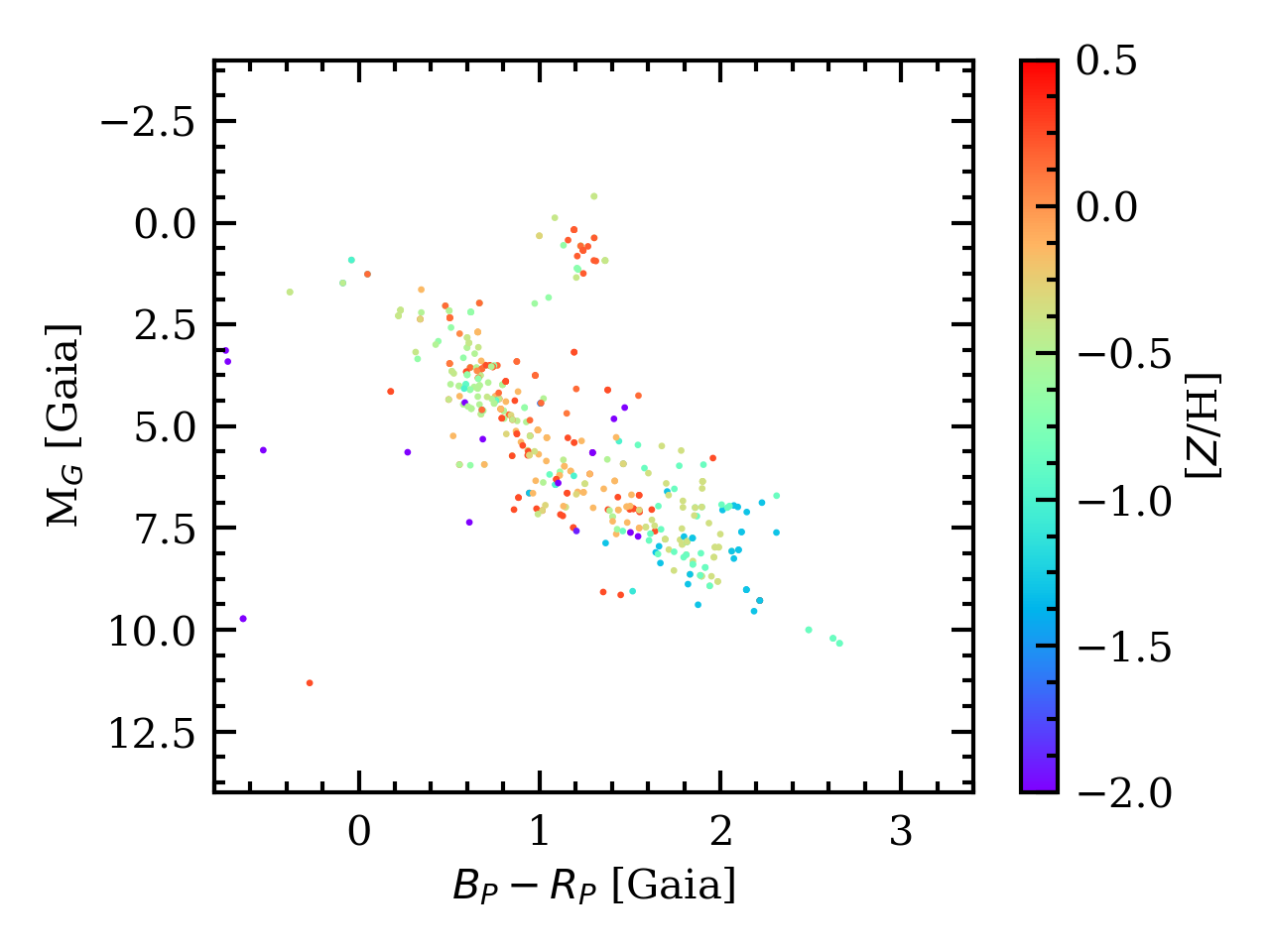}
\caption{Colour-magnitude diagram (CMD) in GAIA filters $M_{G}$ vs. ($B_{\rm p}-R_{\rm p}$) as in Figure~\ref{fig:cmd}, but here for the {\it removed} spectra, after application of all flags (left-hand panel) and of only the 'uncertain extinction' one (right-hand panel). Spectra are colour-coded by total metallicity [Z/H]. Parameters refer to the Th-set.}
 \label{fig:a1}
\end{figure*}

Figure~\ref{fig:a1} shows - in a CMD similar to Figure~\ref{fig:cmd} - the parameter space of the removed spectra colour-coded by total metallicity. The left-hand panel shows the total removed spectra, i.e. after application of all flags listed in Section~\ref{sec:flags}, whilst the right-hand panel focuses on the effect of the sole 'high-extinction' flag. The 'high-extinction' flag mostly removes MS stars with no bias in metallicity (although a small cluster of high-metallicity supergiants is visible). The parameter space of all removed stars after applying all flags (left-hand panel) resembles the actual parameter space of the good stars (Figure~\ref{fig:cmd}), being just broader due to the inaccuracy of the estimated parameters.

\section{Model SEDs of stellar phases.}
\label{app:phases}

Here we include all plots showing the model SEDs integrated over the various stellar evolutionary phases, which are analogues to Figures~\ref{fig:phasesvpoor} and \ref{fig:phases2sun}.
\begin{figure*}
 \includegraphics[width=\textwidth]{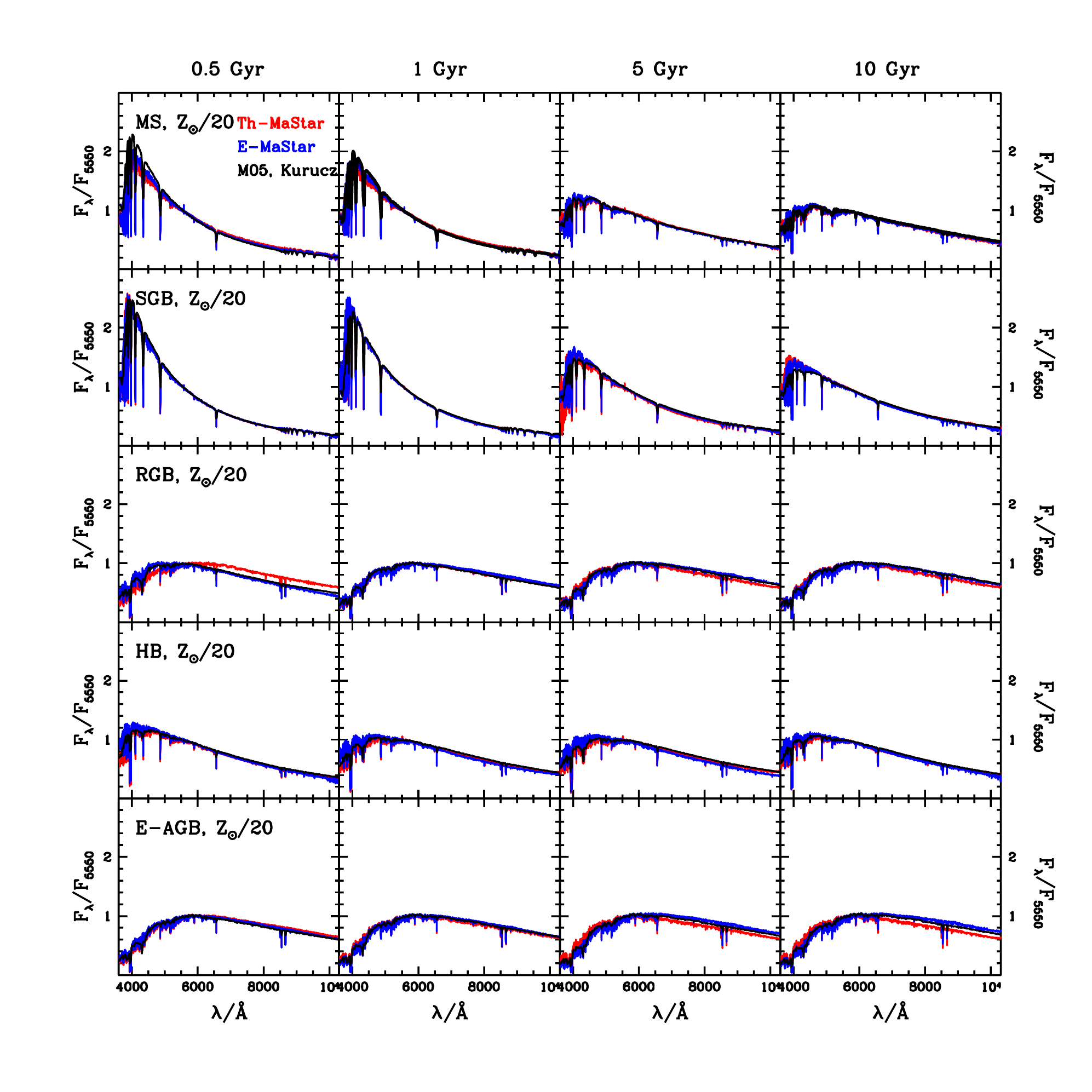}
  \caption{Comparison of integrated SEDs of individual stellar phases for E-MaStar and Th-MaStar models (blue and red colours, respectively), as in Figure~\ref{fig:phasesvpoor}, here for $[Z/H]=-1.35$}
  \label{fig:phasespoor}
\end{figure*}

Figure~\ref{fig:phasespoor} shows that the model SEDs integrated over stellar phases for a metallicity $[Z/H]=-1.35$ are well consistent between the two model flavours and the correspondent ones in M05. 

\begin{figure*}
  \includegraphics[width=\textwidth]{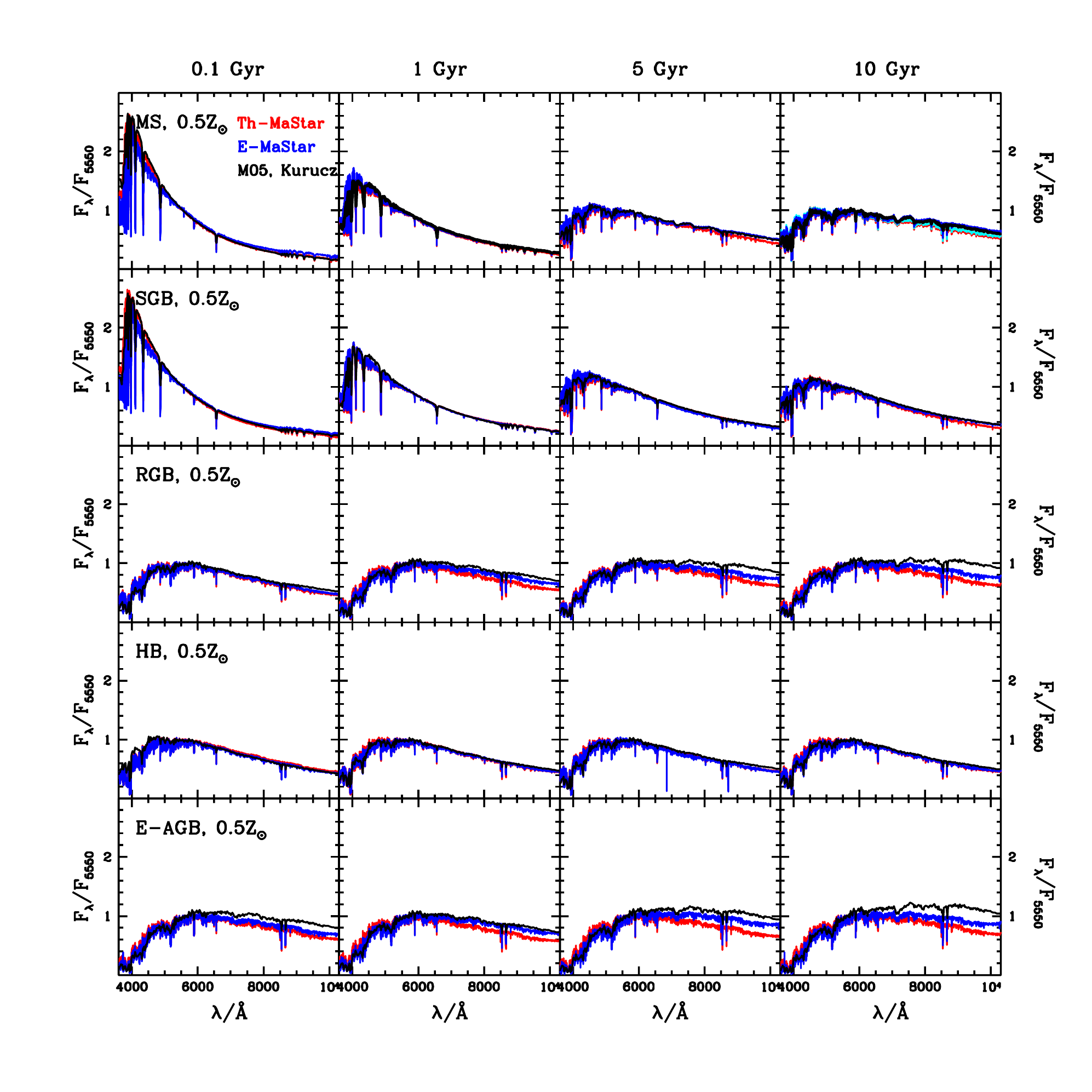}
  \caption{Comparison of integrated SEDs of individual stellar phases for the two flavours of MaStar-models, as in Figure~\ref{fig:phasesvpoor}, here for half-solar metallicity $[Z/H]=-0.33$.}
  \label{fig:phaseshalf}
\end{figure*}

At half-solar metallicity (Figure~\ref{fig:phaseshalf}), the integrated MS SEDs are consistent within MaStar models, which is an excellent achievement considering the importance of this stellar phases for the optical part of the SED. The same consistency is found for the SGB and HB. The coldest phases of the RGB and E-AGB reveal a discrepant flux long-ward $\sim7000$~\AA, where both MaStar models display a lower flux with respect to M05, Th-MaStar in particular. As the MaStar spectra do reach close to the RGB tip and there is no bias in the metallicity of the giant spectra, the slightly lower flux maybe an intrinsic feature of the empirical spectra. M11-STELIB does show the same integrated behaviour as already noted. We should note however that the fuel consumption at the tip-RGB is small due to the rapid evolutionary timescale ($< 2\%$ of the bolometric, cfr. M05) as is the total contribution of the E-AGB ($\sim 4\%$, cfr. M05). This explains why the integrated SEDs do not differ much after all (cfr. Figure~\ref{fig:ssp_sun}). Extensive comparisons with data which we plan for the future maybe be able to shed light on the models.

\begin{figure*}
  \includegraphics[width=\textwidth]{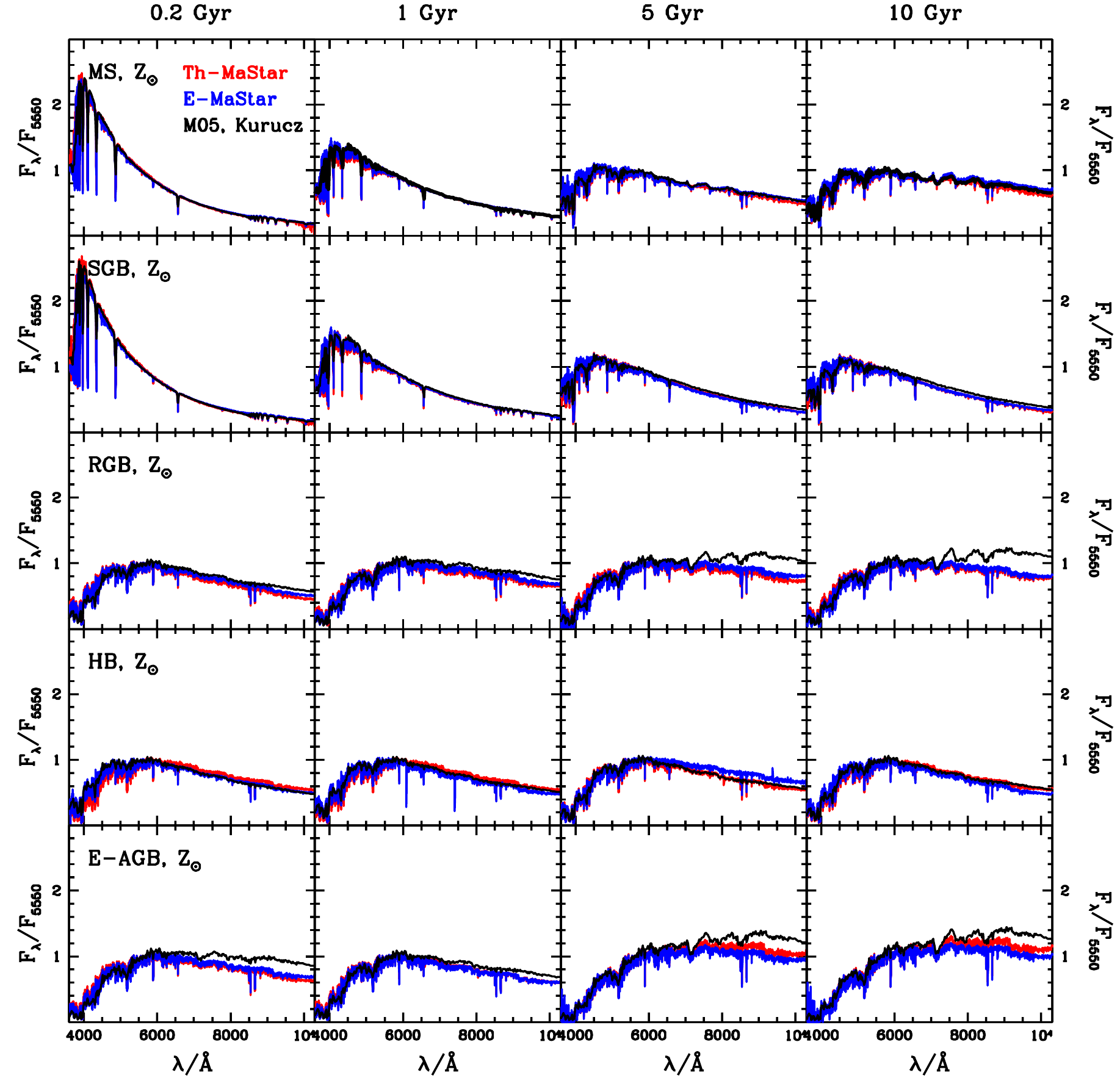}
  \caption{Comparison of integrated SEDs of individual stellar phases for the two flavours of MaStar-models, as in Figure~\ref{fig:phaseshalf}, here for solar metallicity.}
  \label{fig:phasessun}
\end{figure*}

Unlike the half-solar metallicity case, for solar metallicity (Figure~\ref{fig:phasessun}) both cold dwarf and giants are modelled consistently in the two MaStar models. In this case it is Th-MaStar having slightly more near-IR flux over E-MaStar in the giant phases due to a better coverage of the tip-RGB (cfr. Figure~\ref{fig:hr_sol}). The higher flux noted in the E-MaStar 5 Gyr population model spectrum (cfr. Figure~\ref{fig:ssp_sun}) is not due to the RGB, rather to a higher flux in the HB phase, which partly compensates the lower flux from the RGB. This example reinforces the value of inspecting each evolutionary phase integrated SED and not just the total. 

Finally, the plot showing the SEDs of evolutionary phases for the super-solar metallicity case is in the main text (Figure~\ref{fig:phases2sun}).

\section{Integrated Lick indices of stellar population models.}
\label{app:lick}

The integrated Lick indices of our new MaStar stellar population models are visualised in the following plots, each one referring to a different metallicity. These plots are analogues to Figure~\ref{fig:licksun}. 

\begin{figure*}
  \includegraphics[width=\textwidth]{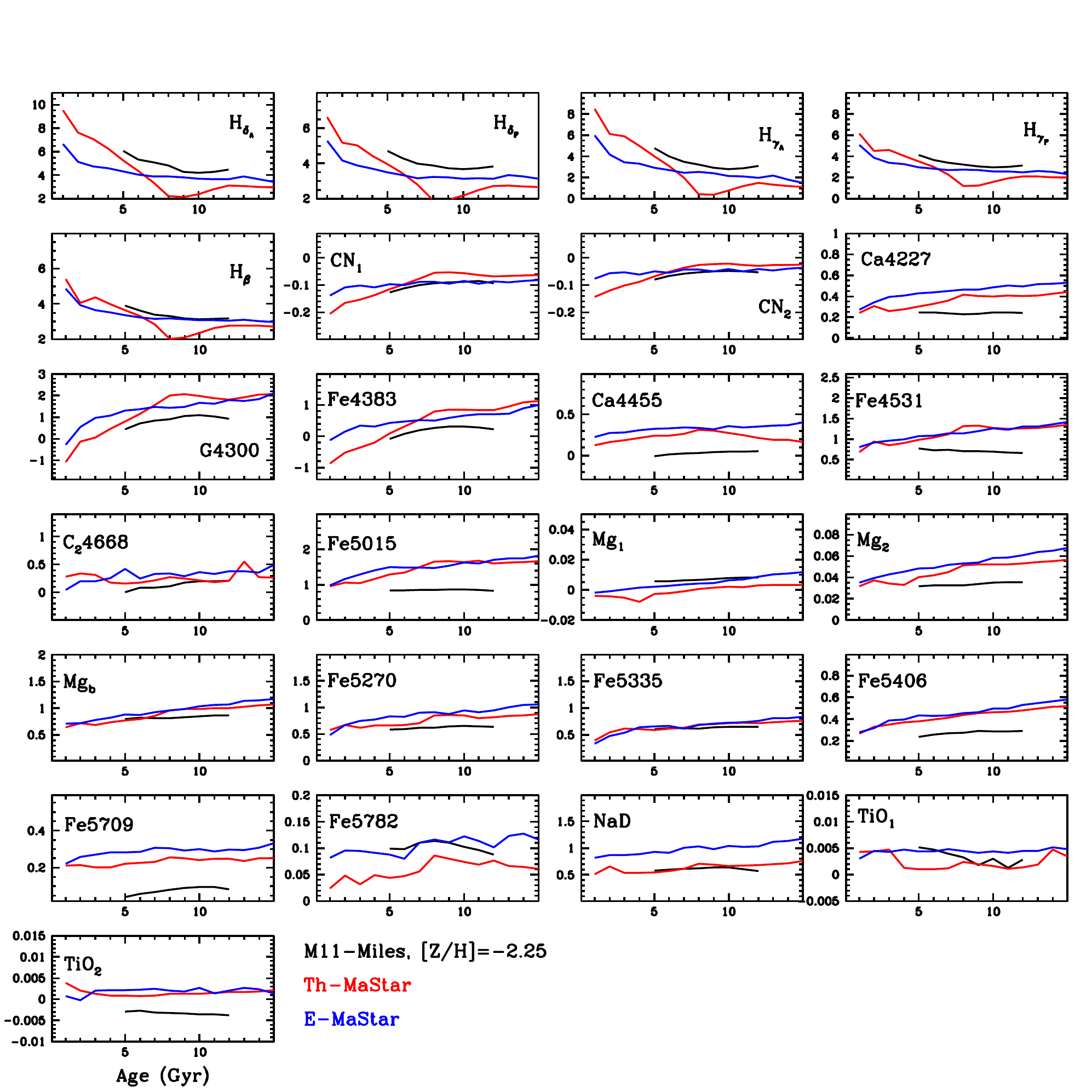}
  \caption{Comparison between the Lick line indices calculated on the two flavours of MaStar-based models (Th- and E-, depicted in red and blue as in previous plots) and those for the M11-MILES models (black).}
  \label{fig:lickvpoor}
\end{figure*}

\begin{figure*}
  \includegraphics[width=\textwidth]{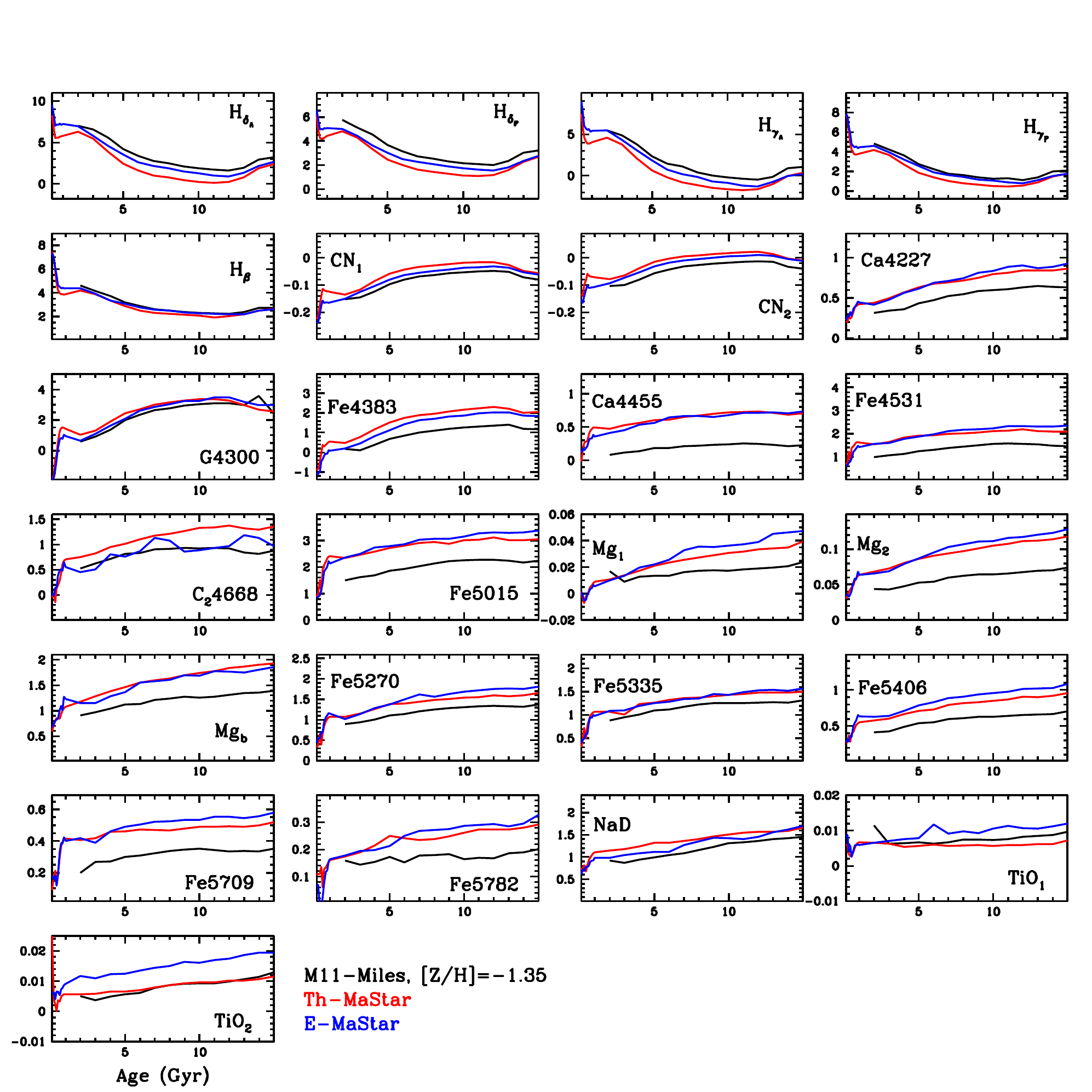}
  \caption{Lick indices of MaStar-based stellar population models as in Figure~\ref{fig:lickvpoor} here for $[Z/H]=-1.35$.}
  \label{fig:lickpoor}
\end{figure*}

\begin{figure*}
  \includegraphics[width=\textwidth]{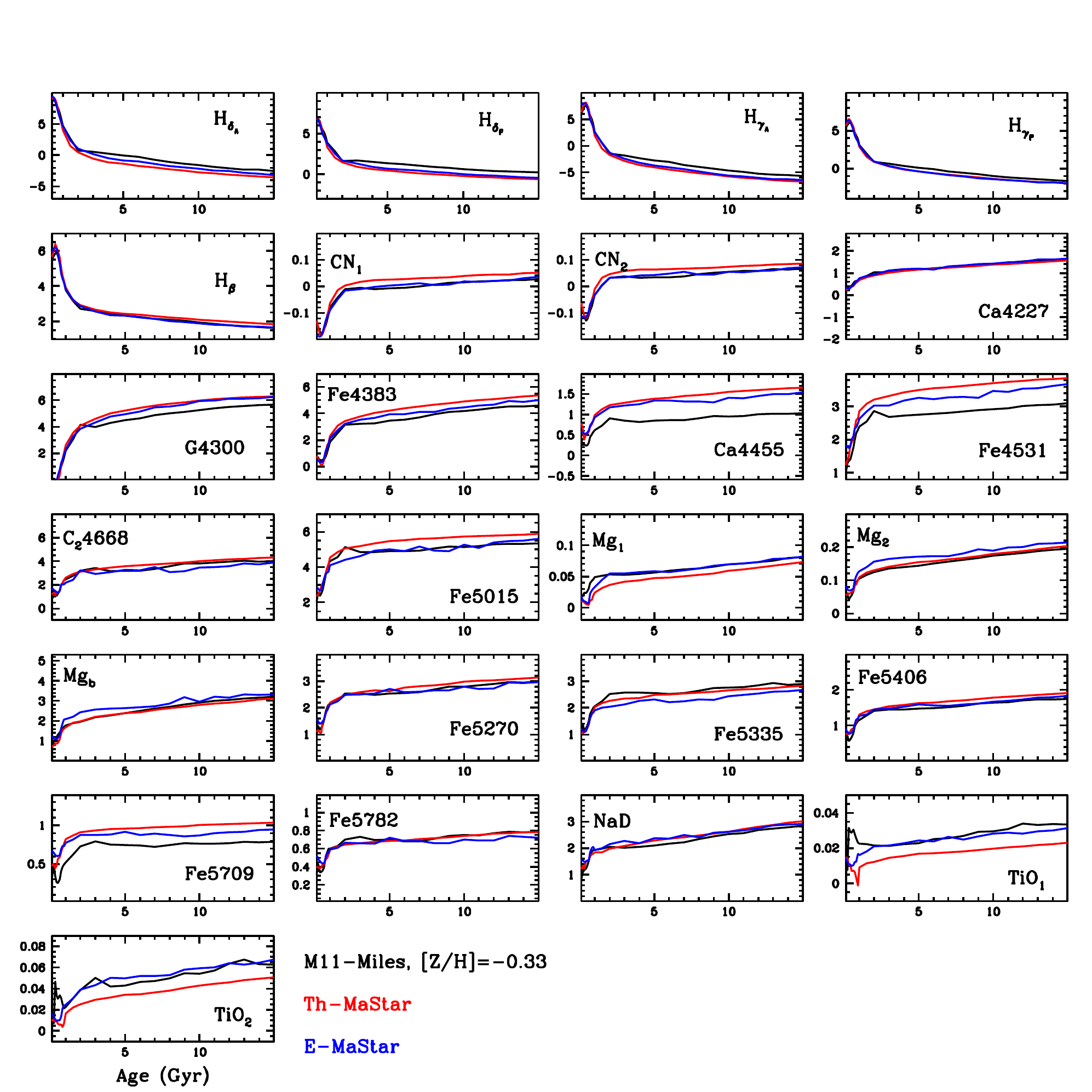}
  \caption{Lick indices of MaStar-based stellar population models as in Figure~\ref{fig:lickvpoor} here for $[Z/H]=-0.33$.}
  \label{fig:lickhalf}
\end{figure*}

\begin{figure*}
		\includegraphics[width=1.\textwidth]{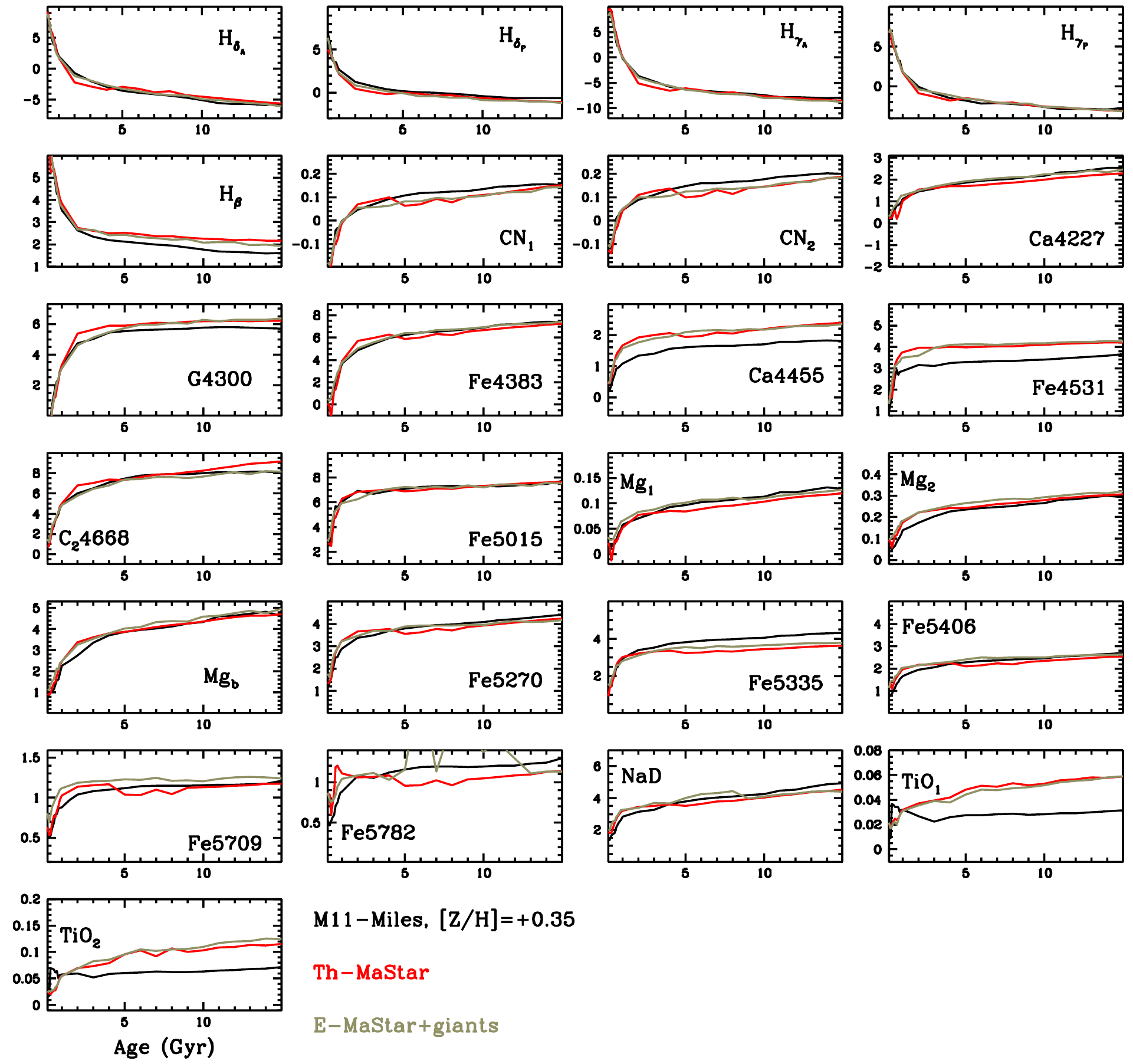}
  \caption{Lick indices of MaStar-based stellar population models as in Figure~\ref{fig:lickvpoor} here for super solar metallicity $[Z/H]=0.35$. 'E-MaStar + giant' models refer to E-MaStar models after addition of cool giants (cfr. Figures~\ref{fig:ssp_2sun} and \ref{fig:phases2sun}).}
    \label{fig:lick_2sun}
\end{figure*}

At the lowest metallicity of the models ($[Z/H]=-2.25$, Figures~\ref{fig:lickvpoor}) the effect of including different (empirical) stellar spectra due to a different assignment of stellar parameters becomes evident in the spectral features.
Lick indices calculated on the two MaStar-model-flavours agree reasonably well, with some noticeable exceptions, such as Fe5782 and NaD. Also the Balmer indices are discrepant, which may be reflective of the different metallicity distribution of MS and HB spectra. The comparison with M11-MILES indices is not straightforward to interpret, with subtle differences overall. Reassuringly, widely used indices like Mgb and Fe5335 are well consistent between all models. From Figure~\ref{fig:lickvpoor} and the similar ones for other metallicities (Figures~\ref{fig:lickpoor}, \ref{fig:lickhalf}, \ref{fig:licksun} and \ref{fig:lick_2sun}) we note that the scatter in Mg- and
Ca-based indices depends somewhat on the metallicity of stars. It is larger at low
metallicity and disappears at high metallicity. Since many metal-poor
stars have enhanced $\alpha$-element abundances and their percentage is variable, this may explain the deviation of the $\alpha$-sensitive indices.
In addition, it should be noted that at this very low metallicity metal lines are weak hence it may be difficult to distinguish the best model if observational uncertainties are not smaller than the models. A quantitative comparison of these models with observed line indices should be able to shed some light, which is the scope of future work. 


Unlike the situation at the lowest metallicity bin, the integrated line strengths for $[Z/H]\sim-1.3$ are consistent between the two MaStar models for the majority of indices (Figure~\ref{fig:lickpoor}).
The only exceptions are the $C_{2}4668$~and the $TiO_{2}$~indices. Discrepancies are found with respect to indices calculated on M11-MILES models. It will be interesting to understand the effect these discrepancies have on more complex models of Lick indices including element-abundance ratio effects (as in Thomas, Maraston \& Johansson~2012). 

The integrated Lick indices of half-solar metallicity MaStar models and M11-MILES models are shown in Figure~\ref{fig:lickhalf}. Indices from the two MaStar models are consistent in the majority of cases and are consistent to those from M11-MILES, with the exception of Ca4455, Fe4531 and Fe5709. The TiO indices are lower for Th-MaStar as a consequence of the lower flux from the giant phases. 

Finally, Figure~\ref{fig:lick_2sun} shows the Lick indices obtained on the super metal-rich MaStar models. As already noticed for the solar metallicity case, but here to a greater extent, the TiO indices are stronger in the new MaStar models over MILES-based models, which will have an impact on the analysis of galaxy spectra. Note that here we have plotted only the indices referred to the 'E-MaStar + giant' models (grey lines in Figures~\ref{fig:ssp_2sun} and \ref{fig:phases2sun}), which due to the addition of cool giants, agree well with the TiO indices of Th-MaStar models. The E-MaStar without the addition of cool giants release lower TiO indices, though still larger than M11-MILES. No other Lick index is affected.

For the rest of indices, it is amusing to note that the models are more consistent in the regime of high-metallicity which is the most relevant to galaxies, rather than at lower metallicity. \footnote{The Fe5782 of E-MaStar models shows a sudden increase at ages around 5 Gyr possibly due to stellar SEDs corrupted in the index region and contributing to those ages. It is very hard to trace back the exact spectra, but we should not worry as the accurate modelling of line indices we plan for future work will mostly be accomplished by calculating fitting functions \citep[e.g.][]{worthey_etal_1994} exactly in order to overcome star-to-star variations. For full spectral fitting the anomalous behaviour of just one line strength does not affect the results.}. %
\bsp	
\label{lastpage}
\end{document}